# Semi-Empirical Parameterization of Interatomic Interactions and Kinetics of the Atomic Ordering in Ni–Fe–C Permalloys and Elinvars

V.A. Tatarenko[1,a], S.M. Bokoch[2,b], V.M. Nadutov[1,c],
T.M. Radchenko[1,d] and Y.B. Park[3,e]

[1]G.V. Kurdyumov Institute for Metal Physics, N.A.S.U.,
Department of Solid State Theory, 36 Academician Vernadsky Boulevard,
UA-03680 Kyyiv-142,
Ukraine

[2]Taras Shevchenko Kyyiv National University, Physics Department,
2 Academician Glushkov Avenue,
UA-03022 Kyyiv-22,
Ukraine

[3]Department of Materials Science and Metallurgical Engineering,
Nano Materials Research Centre, Sunchon National University,
315 Maegok, Sunchon, Jeonnam 540-742,
Korea

[a]tatar@imp.kiev.ua, [b]sergiy.bokoch@gmail.com, [c]nadvl@imp.kiev.ua, [d]tarad@imp.kiev.ua,
[e]ybpark@sunchon.ac.kr

## Abstract

Within the framework of the lattice-statics and static fluctuation-waves' methods, the available energies of strain-induced interaction of interstitial–interstitial, interstitial–substitutional and substitutional–substitutional impurity atomic pairs are collected and analysed for f.c.c.-(Ni,Fe)–C solutions allowing for discrete atomic structure of the host-crystal lattice. The lattice spacings, elasticity moduli and/or quasi-elastic force parameters of the host-crystal lattice, and concentration coefficients of the dilatation of solid-solution lattice due to the respective solutes are selected as the input numerical experimental data used. The above-mentioned interaction energies prove to have non-monotonically decreasing ('quasi-oscillating') and anisotropic dependences on discrete interatomic radius-vector, and themselves are strong and long-range. In all f.c.c.-(Ni,Fe)-base solutions, there is strain-induced attraction in many co-ordination shells. In general, the strain-induced interaction between impurity atoms in γ-Fe is weaker than in α-Ni (but in some solid solutions, it may prove to be of the same order). The verification of applicability of the approximation of strain-induced interaction of impurities for f.c.c.-(Ni,Fe)–C alloys (by means of analysis of thermodynamic C activity and 'short-range order' parameters of C-atoms' distribution revealed by Mössbauer spectroscopy) showed that it must be supplemented with additional short-range ('electrochemical') repulsion in the first co-ordination shell. Nevertheless, in any case, the strain-induced interaction of impurity atoms must be taken into account for analysis of structure and properties of f.c.c.-(Ni,Fe)-base solutions. The Monte Carlo simulation procedures applied for constitution of a nanoscale Fe–C-austenite crystallite and based on analysis of the dependences of numbers of the different atomic configurations on C–C interatomic-interaction en-ergies reveal correlation between the potential energy of such a modelling system and the numbers of iterations as well as Monte Carlo steps for the approach to constrained equilibrium. As shown by the example of austenite, for adequate representation of the experimental data on thermodynamic C activity, one needs to take into account for computations the 'electrochemical' (direct) and strain-induced (indirect) contributions to C–C interaction. The estimated sets of energies of such a total interaction within the first several interstitial co-ordination shells with rated radii are presented, and optimal set is selected, which optimally corresponds to experimental concentration and temperature dependences of C activity and the Mössbauer-spectroscopy data on the nearest neighbourhoods of Fe atoms with octahedral C interstitials. The 'equilibrium' relative parts of the different atomic Fe–C and C–C configurations (depending on C–C interaction energies) are determined. The Khachaturyan–Cook microscopic approach is considered to relate the time dependence of the long-range order (LRO) or short-range order to atomic diffusion. It enables to use the data of measurements of time dependence of radiation diffraction or diffuse-scattering intensity for a Ni–Fe solid solution for calculation of both probabilities of elementary atomic-migration jumps to different lattice sites per unit time and 'exchange' or vacancy-controlled diffusion coefficients, respectively. Using the quantitative experimental information about the Curie temperatures, $T_C$, and neutron diffuse-scattering intensities for disordered f.c.c.-Ni–Fe alloys, it is possible to evaluate the Fourier components, $\widetilde{w}_{tot}(\mathbf{k})$, of effective Ni–Fe atomic 'mixing' energies (inclusive the competing exchange interactions of respective permanent magnetic moments) taking into account both long-range 'paramagnetic' ('electrochemical' + 'strain-induced') and Ising-type magnetic contributions that drive the long-range ordering. Magnetism and 'chemical' (atomic) LRO in f.c.c.-Ni$_{1-c_{Fe}}$Fe$_{c_{Fe}}$ alloys are analysed within the self-consistent field approximation, in which the statistical thermodynamics of the non-stoichiometric $L1_2$(Ni$_3$Fe)-type permalloy (as well as $L1_0$(NiFe)-type elinvar) is determined by several energy parameters $\{\widetilde{w}_{tot}(\mathbf{k})\}$. There is a revealed interplay of magnetism and long-range atomic ordering with the order–disorder transformation temperatures, $T_K$, (below $T_C$) appreciably different from the corresponding isolated $T_K$ values (above $T_C$). The interplay of these two phenomena is examined along two lines, *i.e.* through the estimation of both temperature–$c_{Fe}$ dependence of spatial LRO parameter and magnetisations of Ni and Fe subsystems. Not only the temperature-dependent phase states of such binary f.c.c. alloys can be reproduced, but also the dependence of $T_K$ *vs.* $c_{Fe}$, including the observed asymmetry of phase-diagram curves due to the *T*- and $c_{Fe}$-dependent magnetic contribution to the effective interatomic interactions *etc.* As revealed for f.c.c.-Ni–Fe alloy with the use of single relaxation-time kinetics approximation for calculation of equilibrium intensity values, the magnetic contribution to the 'mixing' energy of atoms (in low-spin states) of this alloy facilitates its atomic ordering, and the presence of atoms with essentially different spins may cause the virtually abrupt phase transition from paramagnetic state into magnetic one. The optimal sets of exchange-interaction energy parameters for f.c.c. Ni–Fe alloy are selected. As shown, the doping with small amounts of interstitial C impurities most likely increases ferromagnetic component of bond of Ni spins with Fe spins, reduces ferromagnetic component of bond of Ni spins with Ni spins, and increases antiferromagnetic component of bond of Fe spins with Fe spins in an f.c.c.-Ni–Fe alloy.

## Introduction

In spite of a long history of researches of the objects considered in offered article, namely Ni–Fe–(C) permalloys and elinvars, there is still their topicality inasmuch as many interesting and important effects, which are revealed in them and are widely discussed in a plenty of important theoretical and especially experimental literature, till now have not received an adequate explanation. (However, the set of prestigious prizes was awarded with revelations of some of them.) In particular, such circumstance has been conditioned by absence of adequate parameterization of those mechanisms and factors, which govern one or another phenomenon at the atomic level. Not solving a formidable task to survey all literature concerning the properties of interstitial elements in such transition-metal alloys, their paramagnetic properties, phase stability *etc.* because of the limited length of article, authors hope that the problem of accumulation of corresponding parameters of both the interatomic interactions and the kinetics of atomic redistribution in these alloys is solved in part (nevertheless, appreciably) in a given article, for the first time in the state-of-the-art review literature. Actually, it is one of the main means of achievement of the goal, for the sake of which a given article was written, *viz.* the presented results of reasonable comparison and analysis of own original knowledge and the compiled suitable data ought to find the further application in atomistic modelling of the processes, which occur in the materials at issue and the invars closely related to them.

As a matter of fact, the interactions of impurity atoms in cubic metals have been the subjects of numerous experimental and theoretical investigations because this information is of topical interest for understanding numerous physical phenomena like short-range order, long-range ordering, segregation, diffusion *etc.* The interatomic-interaction energies are also necessary for calculations of the phase equilibria and phase diagrams, crystallophysical and mechanical properties of alloys [1–4]. For instance, the thermodynamic data (like concentration and temperature dependences of the activity of impurities *etc.*) were traditionally used to get information about the interactions of impurities in f.c.c. solid solutions (see examples in [5–9] and Refs in [4]). Presently, there are commonly accepted physical concepts and reliable experimental data (see, *e.g.*, [4, 10–13 *etc.*]), but their analysis was performed within the framework of the existing approximations of impurity–impurity interaction in the range of the I-st or I-st and II-nd co-ordination shells only. However, the long-range strain-induced interaction of solute atoms is well known to be important for numerous interstitial and substitutional solid solutions in metals [1–3].

The theory of strain-induced (and 'elastic') interaction of impurity atoms in a cubic metal has been developed [1, 2] within the framework of the approximation of the discrete and elastically anisotropic crystal lattice. The energies of 'effectively-pairwise' strain-induced interactions of substitutional (s–s), interstitial (i–i), and interstitial–substitutional (i–s) impurity atoms have been estimated for many solid solutions based on cubic metals [1, 3, 14–25, 28–31 *etc.*].

In spite of the physical limitations of the approximation of pairwise interatomic interactions, it is useful for many problems' solution in solid-state physics. For instance, in case of interstitial–substitutional alloys based on cubic metals, the parameters of i–s-pairwise interaction appeared to be essential for analysis of Mössbauer spectra [3, 4, 29], tetragonality [19], and residual electrical resistivity under tempering [18] or diffraction patterns [21, 29–31 *etc.*].

On the other hand, in cases of α-Ni and γ-Fe, energies of strain-induced i–i interactions have also been calculated [17, 21, 22, 24, 25, 28–31], but for all of them, the different simplified forms of the dynamic matrix of f.c.c. host crystal (with a limited extent and unpaired components of ion–ion interactions) were used [16, 37, 38]. These phonon-spectrum forms are allowed for the experimental data (*e.g.* [33–36]) of the dispersion curves for only some symmetry directions in reciprocal space. Energies of strain-induced s–s and i–s interactions in α-Ni and γ-Fe were calculated in [16, 23, 24] only.

As shown within the analysis of i–i interaction in cubic metals (see, *e.g.*, [1, 17–19, 21, 22, 24–32 *etc.*]), the long-range strain-induced i–i interaction must be supplemented with additional short-range ('electrochemical') repulsion inside of the nearest co-ordination shells. As postulated in Refs [17, 24–27, 32], a screened electrostatic (Coulomb) interaction of charged interstitials may be the possible origin of this repulsion. But in general, the 'direct' ('electrochemical') interaction of point defects inside of the crystal arises because of both an interaction of electrical charges, which form these defects, on conditions of their screening by the free charge carriers and the effect of the Pauli exclusion principle for the formation of the state distribution of the identical particles [1, 3].

Approaches, which do not include both effects due to the lattice displacements and the intrinsic 'electrochemical' interactions on the equal footing, will be discarded.

The theory of 'direct' interactions of solute atoms is still not perfectly developed, and it is so far impossible to estimate microscopically the respective energies for many alloys. The spatial range (or 'radius of blocking') of such i–i repulsion was particularly estimated through comparison of the results of Monte Carlo simulation of tracer diffusion [17, 32] and calculation of Snoek-type internal friction effect ('short-range diffusion') [26, 27] with respective experimental data or analysis of the structure of long-range ordered solid solutions [1] (see also Refs in [24]). As shown, the repulsive i–i interaction in b.c.c. metals extends beyond the III-rd or IV-th co-ordination shell.

However, as shown by means of approximate calculation with the use of available semi-empirical

atom–atom H–H potentials [22, 30, 31] or by diffusion simulation [17] in case of H in f.c.c. metals ($\alpha$-Ni, $\gamma$-Fe *etc.*), the additional repulsion of interstitial H atoms is weak (as compared with strain-induced H–H interaction [17, 22, 30, 31]) or absent almost everywhere. One can explain this by the great distance between two interstitial H atoms located in two neighbouring octahedral interstices of a typical f.c.c. lattice [17, 22, 24, 30, 31]. This distance of common occurrence in the I-st co-ordination shell within a typical f.c.c. lattice is longer than both the revealed 'effective radius' of an H–H screened electrostatic repulsion in a typical b.c.c. metal lattice [17, 26, 27, 32] and the range between H nuclei and/or between the 'overlapping' filled electron shells, which surround these nuclei in $H_2$ molecules.

On the other hand, the applicability of the approximation of C–C strain-induced interaction for description of $\alpha$-Ni–C solid solution or C austenite was studied in [3, 17, 24, 25, 28], and the necessity is presently known to supplement this long-range interaction with the short-range repulsion, as in interstitial solid solutions based on the b.c.c. metals (*e.g.*, $\alpha$-Fe) [1, 17–19, 27, 32] or other isomorphous f.c.c. solid solutions such as N in $\alpha$-Ni or $\gamma$-Fe [21, 28, 29, 31] *etc.* One can verify it, for instance, by means of calculation and comparison with experimental data concerning thermodynamic impurity activity (*e.g.*, [4, 10–13]) or the relative part of the Fe sites with different interstitial (or Ni) atomic neighbourhoods estimated with the use of Mössbauer spectroscopy (*e.g.*, [4, 7, 8, 53, 54, 57]), as performed in Refs [3, 24, 25, 28, 29, 56].

The aims of this work are as follows:

(i) to compare the known energies of i–i strain-induced interaction in f.c.c.-(Ni,Fe) alloys [3, 17, 22, 24, 25, 28, 30, 31] and those calculated with the use of the more precise dependence of lattice parameter of solid solution on the interstitial-impurity concentration [49] (see Refs in [22, 51]) with available energies of 'electrochemical' interaction of respective interstitial non-metallic atoms [40–42];

(ii) to compare the known [16, 23, 24] and recalculated (adequately with regard to Refs [49, 50, 51]) energies of s–s or i–s strain-induced interactions in solid solution based on f.c.c.-(Ni,Fe) crystal with available energies of 'electrochemical' interactions of respective substitutional metal atoms with each other or with interstitial non-metallic atoms [42, 43–47];

(iii) to investigate applicability of the approximation of strain-induced interaction of impurity atoms supplemented with their relatively short-range real-space repulsion in $\gamma$-Fe (or $\alpha$-Ni), for instance, by means of comparison of the results of computer simulation of the thermodynamic impurity activity and indirect parameters of their spatial distribution taken from Mössbauer-spectroscopy with available experimental database, *e.g.*, [4, 10–13, 53, 54, 57].

Besides, there are some dedicated aspects of the effect of experimentally well revealed magnetic ordering on phase stability of Ni–Fe–(C) alloys formed with two magnetic components, which have to be considered in phase-diagram and kinetics calculations:

(iv) first, magnetic ordering can be considered as a magnetic stabilization effect [58, 59, 62, 95–99];

(v) the second aspect is the interdependence of magnetism and spatial LRO or decomposition (segregation) reactions, which occur in these alloys [2, 59, 61, 62, 95].

The need to incorporate the both 'chemical' and magnetic interactions (with a strain-induced contribution) within any complete model of f.c.c.-$Ni_{1-c_{Fe}}Fe_{c_{Fe}}$ alloy ($c_{Fe}$—relative concentration of Fe) has been indicated in many earlier studies. Despite the success of the formalistic treatments [62–67, 91, 92, 96] developed previously for magnetic and atomic ordering (or segregation) in f.c.c. alloys jointly with a method restriction by nearest-neighbour interactions only, it is not worth to abstain from any improved physical model, particularly if the interplay of atomic and magnetic order has to be considered and reveals itself on the modified temperatures of structural and magnetic order–disorder transitions (introduced into the formalism), which can considerably be raised or lowered. These models should be lucid for physical interpretation and sufficiently simple to be applicable to *practical* cases. This means [62]: the use of the interatomic and magnetic interactions as effectively-pairwise ones; the use of the magnetic-exchange-interaction energies, $\{J_{\alpha\beta}(\mathbf{r})\}$ [2, 61], of spins ($\alpha$, $\beta$ = Ni, Fe) spaced with a position vector $\mathbf{r}$ as almost independent on the temperature and $c_{Fe}$ (unlike the magnetic contribution to the pairwise 'interchange' ('mixing') energies, $\{w_{tot}(\mathbf{r})\}$ [1, 2, 61], of substitutional atoms) and the 'para-magnetic' ('electrochemical' [3] + 'strain-induced' [1–3]) contribution, $\{w_{prm}(\mathbf{r})\}$, to $\{w_{tot}(\mathbf{r})\}$ as almost independent on $c_{Fe}$ and perhaps weakly dependent on temperature (like the elastic moduli).

As regards the methods, which may enable properly account for the long-range contributions to interatomic interactions within the framework of the statistical-thermodynamic theory of ordering of solid solutions at issue, one of them is significant and has been proposed by Khachaturyan [1, 2]. Even for the long-range interactions, all thermodynamic quantities may be evaluated within the self-consistent field approximation (SCFA), if we know just a few energy parameters—$\{\widetilde{w}_{tot}(\mathbf{k})\}$, which are the Fourier components of the pairwise 'interchange' ('mixing') energies, $\{w_{tot}(\mathbf{r})\}$, taken in the non-equivalent points $\{\mathbf{k}\}$ [1–3, 100, 102, 109]. By definition,

$$\widetilde{w}_{tot}(\mathbf{k}) = \sum_{\mathbf{r}} w_{tot}(\mathbf{r}) e^{-i\mathbf{k}\cdot\mathbf{r}},$$

where $\mathbf{r} = \mathbf{R} - \mathbf{R}'$ ($\mathbf{R}$, $\mathbf{R}'$—radius-vectors of sites in f.c.c. crystal lattice at issue), the wave vectors $\mathbf{k}$ refer to the positions of fundamental ($\mathbf{k} = \mathbf{k}_0 \equiv \mathbf{0}$) and superstructural ($\mathbf{k} = \mathbf{k}_1, \mathbf{k}_2, \ldots, \mathbf{k}_s, \ldots, \mathbf{k}_{m-1}$) reciprocal-space points—the 'stars', $\{\mathbf{k}_s\}$, of wave vectors in the first Brillouin zone ($m$—a number of sublattices formed during the atomic ordering of alloy), and the sum is taken over all vector differences $\{\mathbf{R} - \mathbf{R}'\}$ between the all $N_{u.c}$ crystal-lattice sites $\{\mathbf{R}, \mathbf{R}' \ etc.\}$. The pertinent number of the energy parameters, $\{\widetilde{w}_{tot}(\mathbf{k})\}$, is determined by the structure of the ordered phase. For example, the statistical thermodynamics of the $L1_2$- and $L1_0$-type f.c.c. substitutional binary alloys [1, 2] is determined, within the SCFA, by two energy parameters:

$$\widetilde{w}_{tot}(\mathbf{k}_X) = -4w_I + 6w_{II} - 8w_{III} + 12w_{IV} - 8w_V + 8w_{VI} + \ldots$$

and

$$\widetilde{w}_{tot}(\mathbf{0}) = 12w_I + 6w_{II} + 24w_{III} + 12w_{IV} + 24w_V + 8w_{VI} + \ldots,$$

where $\mathbf{k}_X = 2\pi \mathbf{a}_1^* (2\pi \mathbf{a}_2^* \text{ or } 2\pi \mathbf{a}_3^*)$ and $\mathbf{k}_0 = \mathbf{0}$ are the wave vectors corresponding to the (1 0 0) and (0 0 0) reciprocal-space points of such alloy, respectively; $\mathbf{a}_1^*, \mathbf{a}_2^*, \mathbf{a}_3^*$ are the basis vectors of the reciprocal lattice along [1 0 0], [0 1 0], [0 0 1] directions [1, 2]; $w_I, w_{II}, w_{III}, w_{IV}, w_V, w_{VI}, \ldots$ are the 'interchange' energies of atomic pairs for the 1-st, 2-nd, 3-rd, 4-th, 5-th, 6-th, ... co-ordination shells with radii $r_I = a/(2)^{1/2}$, $r_{II} = a$, $r_{III} = a(6)^{1/2}/2$, $r_{IV} = a(2)^{1/2}$, $r_V = a(10)^{1/2}/2$, $r_{VI} = a(3)^{1/2}$, ..., respectively, in f.c.c. lattice with parameter $a$.

On the other hand, quantitative information about the radiation diffuse-scattering intensity, $I_{diff}(\mathbf{q})$, associated with the (correlation) short-range order effects, *i.e.* configuration fluctuations as thermodynamic perturbations, in the disordered f.c.c.-$Ni_{1-c_{Fe}}Fe_{c_{Fe}}$ alloy at issue ($c_{Fe}$—the atomic fraction of Fe) in the *equilibrium* state (after 'infinitely' long annealing) at the absolute temperature $T$, may be estimated by the extrapolation of the available intensity values measured *in situ* at a point $\mathbf{q}$, which is disposed at a distance $\mathbf{k}$ from $2\pi\mathbf{B}$—nearest reciprocal-space 'site' ($\mathbf{B}$—reciprocal-lattice vector) of the f.c.c. lattice of disordered alloy. Only such ('asymptotic') information can be used to evaluate adequately the Fourier component, $\widetilde{w}_{tot}(\mathbf{k})$, of 'interchange' energies of 'solvent' ('Ni') and 'solute' ('Fe') atoms with coherent-scattering lengths $b_{Ni}$ and $b_{Fe}$, respectively, that may be found, *e.g.*, on the known Krivoglaz–Clapp–Moss formula for the intensity of elastic neutron diffuse scattering by the short-range atomic order [1, 2, 61, 100, 102]:

$$I_{diff}(\mathbf{q}) \cong N_{u.c}(b_{Ni} - b_{Fe})^2 \frac{Dc_{Fe}(1 - c_{Fe})}{1 + c_{Fe}(1 - c_{Fe})\widetilde{w}_{tot}(\mathbf{k})/(k_B T)} e^{-2M-L}.$$

Here $e^{-2M-L}$ is the so-called overall (static and dynamic [102]) Debye–Waller factor of intensity interference-maximum attenuation used in the kinematic theory of single scattering; $D = D(T, c_{Fe})$ ($\cong 1 \pm 0.2$) is normalizing factor [2]; $k_B$—the Boltzmann constant. Thus, one has unique possibility of direct com-putation of energy parameters, $\{\widehat{w}_{\text{tot}}(\mathbf{k})\}$, of the alloy at issue using experimental data (see, *e.g.*, [80]). In this way, one has to use available data on the neutron, x-ray or electron diffuse-scattering intensities at the respective reciprocal-space points $\{\mathbf{k}\}$ for the disordered phase and then determine these desired parameters, $\{\widehat{w}_{\text{tot}}(\mathbf{k})\}$, by using the Krivoglaz–Clapp–Moss formula (to within $\pm 15\%$).

To work out in detail the binary sections of phase diagrams of $\text{Ni}_{1-c_{\text{C}}}\text{Fe}_{c_{\text{Fe}}}\text{C}_{c_{\text{C}}}$ systems ($c_{\text{C}}$—relative C content) based on f.c.c. crystal lattice and estimate the influence of interstitial C on the exchange-interaction energies for Ni and Fe atoms, the other aim of a given work was the comparison of derived analytical relationships (within the linear approximation of dependence of exchange-interaction parameters on C content) with both literary experimental data on the Curie temperature $T_C = T_C(c_{\text{Fe}}, c_{\text{C}})$ of disordered f.c.c. Ni–Fe–C alloys and evaluation data obtained in the measurements of temperature-dependent magnetic susceptibility. Due to the optimisation procedure of fitting parameters of a content-dependent $T_C$ for Ni–Fe–(C) alloys without of long-range atomic order, the suitable statistical-thermodynamic parameters of such dependence may be estimated. These parameters are defined by combinations of values of Ni and Fe atomic spins and their exchange-interaction energies. The prognostic conclusions related to the interaction between spins of Ni–Fe, Ni–Ni, Fe–Fe atomic pairs were the desirable results. The magnetic-transition temperatures for disordered low-C f.c.c.-Ni–Fe–C alloys may be calculated allowing for new information on evaluated exchange-interaction energies. In this way, the disposition of paramagnetic and magnetic fields in phase diagram of the f.c.c. substitutional subsystem of Ni–Fe–C system can be detailed for such C-concentration range.

However, for all that, the criteria of absence of any nearest-neighbour exchange-interaction energies' temperature-dependence influence on $T_C$ must be suggested, particularly, for f.c.c.-Ni–Fe alloys, taking into account a comparatively-weak dependence of their lattice parameter on the temperature (*i.e.* low thermal expansion) within the temperature range at issue.

The influence of interstitial C on the Kurnakov's temperature $T_K = T_K(c_{\text{C}})$ of long-ranged atomic ordering in substitutional f.c.c.-Ni–Fe alloys (in accordance with $L1_2$ or $L1_0$ superstructural types) must also be analysed at least within the framework of the fourth-order approximation of the dependence of free energy on the C content and long-range order parameter (for $c_{\text{C}} \ll 1$) taking into account the long-range interatomic interactions.

With all this going on, the lattice-statics method, atom–atom potential methodology, static concentration-wave method, SCFA, Ising lattice statistics, simplified Heisenberg-approximation methodology *etc.* are used in analysis of the multifarious independent experimental data.

## Part I. Paramagnetic Interactions of Atoms in F.C.C.-Ni–Fe–C Alloys

### 1. Strain-Induced Interaction Energies of Impurity Atoms

**1.1. Calculation of the Strain-Induced Interaction Energies.** Many physical effects are known to depend rather sensitively on the volume (or as good as on the lattice-parameter) changes, but reliable experimental data are not often available, even for binary alloys. The correct predictions of the variation of the mean interatomic distance with composition and constraints for that are the fundamental problems within the theory of alloys. Various empirical 'rules' have been proposed for this purpose, for instance, Végard's rule (1921) calls for a linear variation of the lattice parameter with concentration. Examples, in which this 'rule' is strictly observed, are very scarce (see Ref. [51]) so one wonders why it enjoys such long-lived popularity and continues to inspire so many attempts and calculations to explain its failure. Actually, concentration dependences of lattice spacings of binary α-Ni–Fe, γ-Fe–Ni, γ-Fe–C or α-Ni–C alloys, including those obtained through the extrapolation from high temperatures, are 'irregular' outside the following concentration ranges: $0 \le c_{\text{Fe}} < 0.55$, $0 \le c_{\text{Ni}} \equiv 1 - c_{\text{Fe}} < 0.2$, $0 \le c_{\text{C}} < 0.09$ or $0 \le c_{\text{C}} < 0.02$, respectively, and deviate from Végard's rule quite appreciably, especially for various Invar alloys within the temperature range from room temperature up to 300°C [49–51]. The success of this simple relation implies that 'elastic' (strain-induced) effects dominate the changes of lattice parameter on alloying; as presupposed within the first approximation, one may ig-nore all charge transfer effects. Although this simple model cannot be used to make the predictions on the stability of the alloy, the 'elastic law' expressed by simple Végard's rule should be very useful for estimation of the volume and energy of formation in crystalline solids.

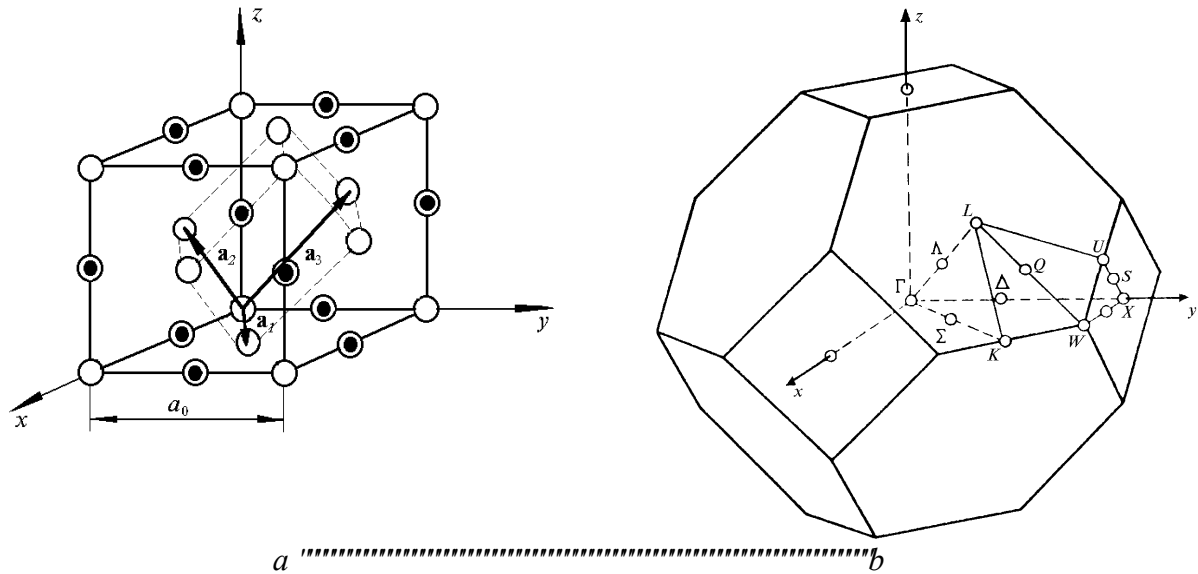

Fig. I.1. Conditional cubic (large cube) and primitive (interior small parallelepiped) unit cells (*a*), and the first Brillouin zone (*b*) of the Bravais f.c.c. lattice (in Fig. I.1*a*, ○—sites, ◉—octahedral intersti­ces, {$\mathbf{a}_1, \mathbf{a}_2, \mathbf{a}_3$}—the basis vectors; in Fig. I.1*b*, {$\Gamma, X, W, L, K(U)$}—high-symmetry points, {$\Delta, \Lambda, \Sigma, Q, S$}—high-symmetry directions in *BZ*).

The calculation technique considered in a given work is based on the corresponding theory formu­lated by Khachaturyan [1, 2] within the reciprocal-space approach and Matsubara–Kanzaki–Krivoglaz lattice-statics method (see bibliography in Refs [1–3]).

Let interstitial atoms are located in octahedral interstices of an f.c.c. lattice amongst its sites. Each unit cell of the f.c.c. lattice (see Fig. I.1*a*) has one octahedral site. Let vectors $\mathbf{R}$ and $\mathbf{R}'$ describe the po­sitions of primitive unit cells with atoms of $\alpha$ or $\beta$ sorts located in lattice sites or interstices. The strain-induced interaction energies $V^{\alpha\beta}(\mathbf{R}-\mathbf{R}')$ of $\alpha$ and $\beta$ atoms ($\alpha\beta=$(i–i) or (s–s) or (i–s)) in primitive unit cells separated by the vector $\mathbf{r}=\mathbf{R}-\mathbf{R}'$ are calculated by means of the Fourier-series expansion,

$$V^{\alpha\beta}(\mathbf{R}-\mathbf{R}') = N_{\text{u.c}}^{-1} \sum_{\mathbf{k}'} \tilde{V}^{\alpha\beta}(\mathbf{k}') e^{i\mathbf{k}'\cdot\mathbf{r}} , \qquad (\text{I.1})$$

where the summation is carried out over all $N_{\text{u.c}}$ points of quasi-continuum {$\mathbf{k}'$} within the first Bril­louin zone (*BZ*) of the f.c.c. lattice (see Fig. I.1*b*), allowed according to the cyclic boundary conditions, $\mathbf{k}'$ is the wave vector, $\tilde{V}^{\alpha\beta}(\mathbf{k}')$ is the Fourier component of the strain-induced interaction energies.

The function $\tilde{V}^{\alpha\beta}(\mathbf{k})$ may be expressed through such material parameters as concentration coeffi­cients of the host-lattice dilatation—$L^{\text{i}}$ (for interstitial atoms) and $L^{\text{s}}$ (for substitutional atoms), elastic­ity moduli, and/or frequencies of natural quasi-harmonic vibrations of the host crystal [1–3].

Within the framework of both superposition approximation [2] and quasi-harmonic one [1–3], this (*non-analytic*) function looks like the algebraic sum of scalar products,

$$\tilde{V}^{\alpha\beta}(\mathbf{k}) \approx -\tilde{\mathbf{F}}^{\alpha+}(\mathbf{k}) \cdot \tilde{\mathbf{u}}^{(\beta)}(\mathbf{k}) + \delta_{\alpha\beta} N_{\text{u.c}}^{-1} {\sum_{\mathbf{k}'}}' \tilde{\mathbf{F}}^{\alpha+}(\mathbf{k}') \cdot \tilde{\mathbf{u}}^{(\beta)}(\mathbf{k}') \qquad (\text{I.2})$$

for $\mathbf{k} \neq \mathbf{0}$ (see expression for $\tilde{V}^{\alpha\beta}(\mathbf{0})$ at $\mathbf{k}=\mathbf{0}$ in Refs [1–3]), where $\delta_{\alpha\beta}$ is the Kronecker delta-symbol,

$$\tilde{\mathbf{F}}^{\alpha}(\mathbf{k}) = \sum_{\mathbf{r}} \mathbf{F}^{\alpha}(\mathbf{r}) e^{-i\mathbf{k}\cdot\mathbf{r}} \qquad (\text{I.3})$$

is the Fourier transform of the 'coupling' force $\mathbf{F}^{\alpha}(\mathbf{r})$ (the so-called Kanzaki force) 'acting' on the undisplaced host atom (at the site $\mathbf{R} = \mathbf{r}$) by $\alpha$-type impurity atom (at the position within the primitive unit cell corresponding to $\mathbf{R}' = \mathbf{0}$). The vectors $\tilde{\mathbf{F}}^{i}(\mathbf{k})$ and $\tilde{\mathbf{F}}^{s}(\mathbf{k})$ are material parameters at the given conditions. In the primed sum in Eq. (I.2), the term, which corresponds to $\mathbf{k}' = \mathbf{0}$, is omitted.

$$\tilde{\mathbf{u}}^{(\beta)}(\mathbf{k}) = \sum_{\mathbf{R}} \mathbf{u}^{(\beta)}(\mathbf{R}) e^{-i\mathbf{k}\cdot\mathbf{R}} \qquad (I.4)$$

is a Fourier transform of host-atom displacement-vector component $\mathbf{u}^{(\beta)}(\mathbf{R})$ (at a site $\mathbf{R}$) caused by the $\beta$-type atom (at $\mathbf{R}' = \mathbf{0}$). Within the framework of the superposition approximation [2] and lattice-statics approach, the value of $\tilde{\mathbf{u}}^{(\beta)}(\mathbf{k})$ (for $\mathbf{k} \neq \mathbf{0}$) may be obtained from the equation of mechanical equilibrium,

$$\tilde{\mathbf{A}}(\mathbf{k})\tilde{\mathbf{u}}^{(\beta)}(\mathbf{k}) \approx \tilde{\mathbf{F}}^{\beta}(\mathbf{k}), \qquad (I.5)$$

where the tensor $\tilde{\mathbf{A}}(\mathbf{k})$ is a dynamical matrix $\|\tilde{A}^{ij}(\mathbf{k})\|$ of a host crystal; $i, j = x, y, z$ are the Cartesian indices.

The final purpose of calculations of the dynamical matrix $\|\tilde{A}^{ij}(\mathbf{k})\|$ is determination of the matrix elements $\tilde{A}^{ij}(\mathbf{k})$ at any $\mathbf{k} \in BZ$, if, for $\mathbf{k}$ with the ends lying along the symmetry directions, $\|\tilde{A}^{ij}(\mathbf{k})\|$ are known ($\|\tilde{A}^{ij}(\mathbf{k})\|$ along symmetry directions is directly determined from inelastic slow-neutron scattering data).

The dynamical matrix $\|\tilde{A}^{ij}(\mathbf{k})\|$ may also be calculated by using the quasi-elastic force parameters of the host-crystal lattice. Introduction of these Born–von Kármán coupling parameters is the most common way to parameterise the experimental dispersion curves of host crystals; however, the Born–von Kármán approximation for the dynamical matrix may not be accurate enough. The major limitation of this approach is the large increase of the number of fitting parameters with increasing interionic-interaction range; the disadvantage of the simple force approximation is that a large number (20–30) of unknown expansion coefficients is required if the interionic potential is not extremely short range. It ensures impossibility of unambiguous determination of these coefficients from the available experimental data. The numerical values of these coupling parameters determined by a least-squares fit do not necessarily have any direct physical meaning. They provide a merely semi-phenomenological description of the spatial dispersion. A determination of the real physical couplings involves not only exact knowledge of the phonon frequencies but requires also information on the off-symmetry phonons' polarizations (see bibliography in Ref. [34]).

The representation of the dynamical matrix in terms of the Born–von Kármán parameters for the f.c.c. crystal lattice is given in Ref. [16] according to [37]. (For $\gamma$-Fe at 1428 K, the applicable Born–von Kármán parameters may also be taken from Ref. [33], and that of $\alpha$-Ni at room temperature from Refs [35] or [36].)

In other semi-phenomenological models of the crystal-lattice dynamics, which satisfy the conditions of the invariance of the 'potential' energy of f.c.c. crystal with respect to translations and rotations of the crystal as a rigid body, the dynamical matrices can, in regard to the phonon properties, be expressed by relatively short-ranged real-space force parameters. The dynamical-matrix expression presented in Ref. [3] according to [38] through the parameters of both the centrally symmetric pairwise and non-pairwise (three-particle) short-range contributions to the interaction between the host ions for the f.c.c. crystal was approximately obtained in Refs [21, 22] in terms of the values of the experimentally measurable elasticity moduli ($C_{11}$, $C_{12}$, $C_{44}$ in Voigt's designations) (see Refs [33, 39]) and the vibration normal-mode frequencies of the acoustic-phonon branches with longitudinal ($\omega_{LA}$) and doubly degenerate transverse ($\omega_{TA}$) polarizations (e.g. see Refs [33, 34, 36]) for $\mathbf{k}$ (along the [1 0 0] direction), which correspond to the $BZ$-boundary symmetry point $X$ (see Fig. I.1b).

The inverse Fourier transformations similar to (I.1) and summation of series in (I.2) over $\mathbf{k}' \in BZ$ may be carried out with the use of a computer, for instance, by summing over $\sim 10^{5}$–$10^{7}$ uniformly-distributed points [16, 17, 22, 24, 25] or over only some 'special' points of the 'principal' value ([21, 28]; see bibliography in Ref. [3]) within the irreducible wedge in the first Brillouin zone ('1/48 $BZ$') of the f.c.c. host lattice.

When the Kanzaki 'coupling' forces, $\mathbf{F}^i(\mathbf{r})$ and $\mathbf{F}^s(\mathbf{r})$, do not vanish for the nearest I-st co-ordination shell around the interstitial or substitutional impurity atom only and are directed along the straight line from the dissolved impurity atom to the host atom, the vectors $\tilde{\mathbf{F}}^i(\mathbf{k})$ and $\tilde{\mathbf{F}}^s(\mathbf{k})$ have the following forms [1–3]:

$$\tilde{\mathbf{F}}^i(\mathbf{k}) \cong -i\frac{a_0^2}{2}(C_{11}+2C_{12})L^i \left\| \begin{matrix} \sin\left(\dfrac{a_0}{2}k_x\right) \\[2mm] \sin\left(\dfrac{a_0}{2}k_y\right) \\[2mm] \sin\left(\dfrac{a_0}{2}k_z\right) \end{matrix} \right\| e^{-i(k_x+k_y+k_z)\frac{a_0}{2}} \tag{I.6}$$

and

$$\tilde{\mathbf{F}}^s(\mathbf{k}) \cong -i\frac{a_0^2}{4}(C_{11}+2C_{12})L^s \left\| \begin{matrix} \sin\left(\dfrac{a_0}{2}k_x\right)\left[\cos\left(\dfrac{a_0}{2}k_y\right)+\cos\left(\dfrac{a_0}{2}k_z\right)\right] \\[2mm] \sin\left(\dfrac{a_0}{2}k_y\right)\left[\cos\left(\dfrac{a_0}{2}k_z\right)+\cos\left(\dfrac{a_0}{2}k_x\right)\right] \\[2mm] \sin\left(\dfrac{a_0}{2}k_z\right)\left[\cos\left(\dfrac{a_0}{2}k_x\right)+\cos\left(\dfrac{a_0}{2}k_y\right)\right] \end{matrix} \right\|, \tag{I.7}$$

respectively; $k_x$, $k_y$, $k_z$—Cartesian components of $\mathbf{k}$. The value $K \equiv (C_{11}+2C_{12})/3$ is (elastic) bulk compression modulus of the host crystal, $a_0$ is the length of the edge of the smallest conventional cubic unit cell of the pure host f.c.c. lattice.

It is clear that there is an urgent need of a simple rule to estimate (predict) the $L^i$ or $L^s$ with a known variation of the lattice parameter (average change of atomic volume) with concentration in solid solutions.

The concentration coefficients of the dilatation of the host lattice, $L^i$ or $L^s$, may be obtained according to the dependence of the parameter of crystal lattice of solid solution, $a = a(\{c_\alpha\}, T)$, on relative concentration $c_\alpha$ ($\alpha = $ i, s),

$$L^\alpha \cong \{\partial \ln a / \partial c_\alpha\}|_{c_\alpha = 0}, \tag{I.8}$$

where $c_\alpha$ is the number of interstitially dissolved solute atoms *per* solvent atom (for $L^i$, *e.g.*, in f.c.c.-Ni(Fe)C$_{c_c}$) or the atomic fraction of substitutionally dissolved solute atoms (for $L^s$, *e.g.*, in f.c.c.-Ni$_{1-c_{Fe}}$Fe$_{c_{Fe}}$).

To determine the $L^i$ and $L^s$ values, it is necessary to use the concentration dependence of $a$ in the f.c.c. phase field, *i.e.* at high temperatures. The limited data are known for C in f.c.c.-Ni(Fe) and for f.c.c.-Ni$_{1-c_{Fe}}$Fe$_{c_{Fe}}$ only (see, *e.g.*, [48–51, 114, 115]). The $L^i$ and $L^s$ values calculated according to these data weakly depend on temperature. This means that one can use available x-ray data at room or elevated temperatures (corrected subsequently for thermal expansion), *e.g.*, for α-Ni-base equilibrium solid solutions in case of doping by Fe [50, 51, 115] and C [51, 114] atoms, for retained austenite [24] or unstrained one (*i.e.* fully austenitic and thus strain-free specimen in equilibrium) [21, 28] in case of C additions [48–50, 114], and for γ-Fe-base solid solutions in case of Ni alloying [50, 51, 115]. (It is important that, for other additions of substitutional atoms like Mn, Mo, Al, Cu *etc.* to α-Ni-base alloys containing a fixed amount of Fe, $L^{\mathrm{s}}$ values may also be obtained [24] according to changes of a lattice parameter of α-Ni–Fe solid solutions [115] related to the third alloying element. Nevertheless, for all that, the lattice parameter of the binary α-Ni–Fe solution [115] with the same Ni content as in the ternary solid solution should be used as $a_0$.)

Table I.1. Crystallographic and mechanical parameters used in the calculations.

| | Elastic moduli [GPa] | | | $a_0$ [Å] | $L^{\mathrm{C}}$ | $L^{\mathrm{H}}$ | $L^{\mathrm{Ni}}$ | $L^{\mathrm{Fe}}$ |
| | $C_{11}$ | $C_{12}$ | $C_{44}$ | | | | | |
|---|---|---|---|---|---|---|---|---|
| γ-Fe at 1428 K [33]: | 154 | 122 | 77 | 3.666 | 0.196 0.210 | 0.080 | −0.007 0.001 | 0 |
| α-Ni at 298 K [39]: | 245.3 | 146.1 | 124.7 | 3.5243 | 0.210 | | 0 | 0.034 |
| α-Ni at 500 K [39]: | 239.6 | 149.0 | 115.9 | 3.52 | 0.210 | 0.077 | 0 | 0.033 |
| Lattice parameters from: | | | | [52], [51], [22] | [49], [50], [114] | [30], [22], [114] | [50], [51], [115] | [50], [115] |

The numerical values of the coefficients $L^{\mathrm{i}}$ and $L^{\mathrm{s}}$ displayed in Table I.1 cannot be considered highly reliable, especially, for α-Ni–C and γ-Fe–H solid solutions, and need to be determined more accurately in future. That is why the energies must be presented in functional form [1, 3, 16, 17, 23–25], which is suitable for calculation with arbitrary values of the coefficients $L^{\mathrm{i}}$ and $L^{\mathrm{s}}$. The results of such calculations easily enabled to estimate the numerical values of strain-induced interaction energies for any dissolved atoms in α-Ni or γ-Fe without additional computing. In order to do that, it was sufficient to use the numerical values of $L^{\mathrm{i}}$ and $L^{\mathrm{s}}$ inherent to the relevant solid solution and the results of numeric computations obtained in Refs [16, 17, 21–25, 28, 30] and listed in the tables below.

The energies $V^{\alpha\beta}(\mathbf{R} - \mathbf{R}')$ calculated for γ-Fe-base solid solutions at 1428 K (e.g. in Ref. [24] with the Born–von Karman parameters obtained in Ref. [33]) are strictly suitable at this temperature. Nevertheless, one will use these energies in statistical-thermodynamic analysis at different temperatures and for comparison with the interaction energies of impurities in α-Ni calculated for other temperatures (e.g., for the room temperature and 500 K). It is important that the strain-induced interaction of impurity atoms is about the same in some solid solutions based on γ-Fe as those for some solid solutions based on α-Ni (Figs. I.2–4) [24, 25] despite the differences of temperatures (and compositions). (It is usually assumed that the interatomic-interaction energies weakly depend on temperature.)

**1.2. Computational Results and Discussion.** Since $\mathbf{F}^{\mathrm{i}}(\mathbf{k})$ and $\mathbf{F}^{\mathrm{s}}(\mathbf{k})$ are linear functions of $L^{\mathrm{i}}$ and $L^{\mathrm{s}}$, respectively, it follows from (I.2) and (I.5) that the energies $V^{\alpha\beta}(\mathbf{R} - \mathbf{R}')$ are proportional to $L^{\alpha}L^{\beta}$:

$$V^{\alpha\beta}(\mathbf{R} - \mathbf{R}') = \mathrm{A}^{\alpha\beta}(\mathbf{R} - \mathbf{R}')L^{\alpha}L^{\beta}. \tag{I.9}$$

Three sets of the 'universal' coefficients $\{\mathrm{A}^{\alpha\beta}(\mathbf{R} - \mathbf{R}')\}$—$\{\mathrm{A}^{\mathrm{ii}}(\mathbf{R} - \mathbf{R}')\}$, $\{\mathrm{A}^{\mathrm{ss}}(\mathbf{R} - \mathbf{R}')\}$, $\{\mathrm{A}^{\mathrm{is}}(\mathbf{R} - \mathbf{R}')\}$— are the same for all solid solutions based on the generic f.c.c. host-crystal lattice (α-Ni or γ-Fe). The coefficients $\{\mathrm{A}^{\alpha\beta}(\mathbf{R} - \mathbf{R}')\}$ may be determined through computer calculations. However, they enable direct evaluation (without involving numerical methods) of the 'pairwise' strain-induced interaction energies provided the coefficients of the concentration dilatation of the host-crystal lattice (concerning the relevant substitutional and interstitial atoms) are known.

Tables I.2–4 give the numerical values of the respective energies $\{V^{\alpha\beta}(\mathbf{R} - \mathbf{R}')\}$ for various distances $\mathbf{r} = \mathbf{R} - \mathbf{R}'$. These tables can also be used directly for calculation of the values of interaction energies for some other solid solutions (e.g., α-Ni–H and γ-Fe–H presented in Table I.2).

**1.2.1. Interstitial–Interstitial Strain-Induced Interaction (Table I.2 and Fig. I.2).** The sign of $V^{\mathrm{ii}}(\mathbf{R} - \mathbf{R}')$ for identical kinds of interstitial atoms is determined by the sign of $\mathrm{A}^{\mathrm{ii}}(\mathbf{R} - \mathbf{R}')$. (A positive interaction energy means repulsion and the negative one—attraction.)

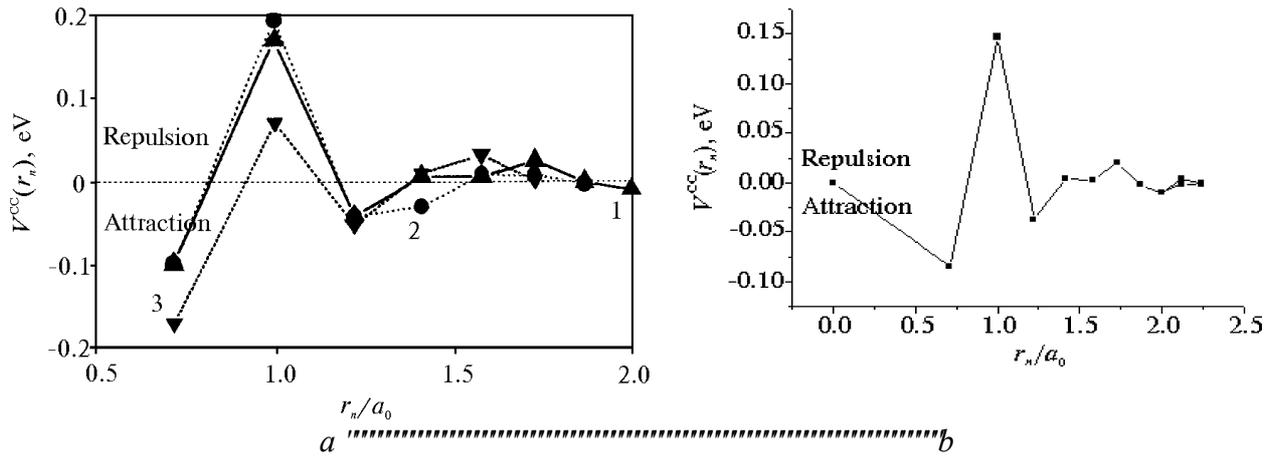

Fig. I.2. Dependence of the energies of C–C strain-induced interaction on discrete interstice–interstice reduced distance $r_n/a_0$ ($n$ = I, II, …) between C atoms within γ-Fe [24, 25] (1), α-Ni [17, 24, 25] (2), γ-Fe [28] (3)—(*a*), and within γ-Fe (present work)—(*b*).

One can see that these energies are anisotropic and quasi-oscillating. There is 'strong' attraction in the I-st and III-rd co-ordination shells, 'strong' repulsion in the II-nd shell and no weak interaction in the VI-th and VIII-th shells. One can see that the C–C interaction extend over two shells taken usually into account for analysis the thermodynamic properties or Mössbauer spectra of austenite [2, 7, 53].

The energies of C–C (Fig. I.2) and H–H interactions differ from data in Refs [28] and [30], respectively. As mentioned in [24, 25], it is possibly connected with the simplified form of the dynamical matrix $\|\tilde{A}^{ij}(\mathbf{k})\|$ used in Refs [21, 28, 30]. Another cause of difference consists in 'rough' estimation [21, 28, 30] using the assumption of i–i interatomic interaction in α-Ni and γ-Fe within only six coordination interstitial shells.

Calculations of the Fourier components, $\{\tilde{A}^{ii}(\mathbf{k})\}$, of the coefficients, $\{A^{ii}(\mathbf{R} - \mathbf{R}')\}$, for interstitial-impurity atoms in some f.c.c. metals [3, 17, 21, 22, 24, 25, 28, 30, 31], which have been performed with the use of different forms of $\|\tilde{A}^{ij}(\mathbf{k})\|$, gave the results (*e.g.*, the energies of H–H strain-induced interaction in α-Pd [17] and [31]), which are satisfactorily fitted with each other (and other studies [20 *etc.*]) for respective solid solutions.

This fact and calculation of H–H interactions in α-Ni [22] give grounds to treat results presented in

Table I.2. Energies [eV] of pairwise strain-induced interaction between interstitial impurity atoms in γ-Fe at 1428 K (the values below the horizontal split bar were estimated in [24]) or in α-Ni at 500 K.

| $2(\mathbf{R} - \mathbf{R}')/a_0$ | 110 | 200 | 211 | 220 | 310 | 222 | 321 | 400 | 330 | 411 | 420 |
|---|---|---|---|---|---|---|---|---|---|---|---|
| Shell | I | II | III | IV | V | VI | VII | VIII | IXa | IXb | X |
| $\|\mathbf{R} - \mathbf{R}'\|/a_0$ | $\cong 0.71$ | 1 | $\cong 1.22$ | $\cong 1.41$ | $\cong 1.58$ | $\cong 1.73$ | $\cong 1.87$ | 2 | $\cong 2.12$ | $\cong 2.12$ | $\cong 2.24$ |
| In γ-Fe: $A^{ii}(\mathbf{R} - \mathbf{R}')$ [24, 25] | −2.209 | +3.827 | −0.957 | +0.095 | +0.085 | +0.523 | −0.046 | −0.269 | +0.110 | −0.034 | −0.037 |
| $V^{CC}(\mathbf{R} - \mathbf{R}')$ | −0.085 | +0.147 | −0.037 | +0.004 | +0.003 | +0.020 | −0.002 | −0.010 | +0.004 | −0.001 | −0.001 |
| | −0.097 | +0.169 | −0.042 | +0.004 | +0.004 | +0.023 | −0.002 | −0.012 | +0.005 | −0.001 | −0.002 |
| $V^{HH}(\mathbf{R} - \mathbf{R}')$ | −0.014 | +0.024 | −0.006 | +0.001 | +0.001 | +0.003 | $-3 \cdot 10^{-4}$ | −0.002 | +0.001 | $-2 \cdot 10^{-4}$ | $-2 \cdot 10^{-4}$ |
| In α-Ni: $V^{CC}(\mathbf{R} - \mathbf{R}')$ | −0.115 | +0.139 | −0.038 | +0.029 | +0.005 | +0.007 | −0.003 | −0.009 | +0.007 | −0.001 | $+9 \cdot 10^{-4}$ |
| $V^{HH}(\mathbf{R} - \mathbf{R}')$ | −0.015 | +0.019 | −0.005 | +0.004 | $+6 \cdot 10^{-4}$ | $+9 \cdot 10^{-4}$ | $-4 \cdot 10^{-4}$ | −0.001 | $+9 \cdot 10^{-4}$ | $-2 \cdot 10^{-4}$ | $+1 \cdot 10^{-4}$ |

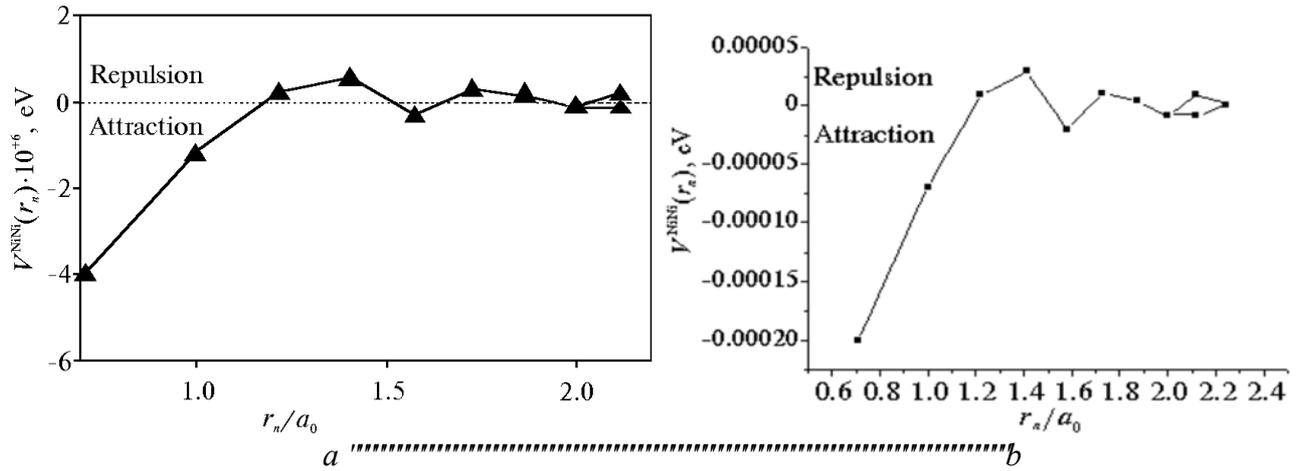

Fig. I.3. Dependence of the Ni–Ni strain-induced interaction energies on discrete site–site reduced distance $r_n/a_0$ ($n$ = I, II, …) between Ni atoms within $\gamma$-Fe: based on [24] ($a$) and present work ($b$).

Table I.2 as more precise than the real-space data in Refs [21, 28, 30].

Comparison of C–C and H–H interactions in $\alpha$-Ni presented, on the one hand, in a given article (see Table I.2) and Ref. [22], and, on the other hand, in Fig. I.2$a$ and Refs [17, 24, 25], shows that the respective interaction-energy functions have essentially different dependences on distance (see, *e.g.*, signs of them in the IV-th co-ordination shell) due to the distinctions of forms of the f.c.c. host-crystal dynamic matrix $\|\tilde{A}^{ij}(\mathbf{k})\|$ used in these Refs.

Nevertheless, for high-symmetry points $\{\Gamma, X, W, L, K(U)\}$ in $BZ$, always-truth is as follows:

$$0 < \tilde{V}^{ii}(\mathbf{k}_X) > \tilde{V}^{ii}(\mathbf{k}_K) \neq \tilde{V}^{ii}(\mathbf{k}_W) \neq \tilde{V}^{ii}(\mathbf{k}_L) \neq \tilde{V}^{ii}(\mathbf{k}\rightarrow\mathbf{0}) \neq \tilde{V}^{ii}(\mathbf{k}_\Gamma=\mathbf{0}) \; (< 0 \text{ usually})$$

that is significant for the priority rating of driving motives in the statistical thermodynamics of interstitial solid solutions based on the f.c.c. lattice.

**1.2.2. Substitutional–Substitutional Strain-Induced Interaction (Table I.3 and Fig. I.3).** The sign of $V^{ss}(\mathbf{R}-\mathbf{R}')$ for substitutional atoms of the same type is also determined by the sign of the coefficient $A^{ss}(\mathbf{R}-\mathbf{R}')$, *i.e.* there is essential attraction in the first two shells. Because the values of these coefficients, $\{A^{ss}(\mathbf{R}-\mathbf{R}')\}$, are of the same order in $\alpha$-Ni and $\gamma$-Fe [16, 24], the differences of respective interactions are determined by $L^s$ values for specific solid solutions.

Table I.3. Energies [eV] of pairwise strain-induced interaction of substitutional impurity atoms in $\gamma$-Fe at 1428 K (the values below the horizontal split bar were estimated on the basis of data taken from Refs [24, 51]) or in $\alpha$-Ni at room temperature.

| $2(\mathbf{R}-\mathbf{R}')/a_0$ | 110 | 200 | 211 | 220 | 310 | 222 | 321 | 400 | 330 | 411 |
|---|---|---|---|---|---|---|---|---|---|---|
| Shell | I | II | III | IV | V | VI | VII | VIII | IXa | IXb |
| $|\mathbf{R}-\mathbf{R}'|/a_0$ | $\cong 0.71$ | 1 | $\cong 1.22$ | $\cong 1.41$ | $\cong 1.58$ | $\cong 1.73$ | $\cong 1.87$ | 2 | $\cong 2.12$ | $\cong 2.12$ |
| In $\gamma$-Fe: $A^{ss}(\mathbf{R}-\mathbf{R}')$ [24] | −4.007 | −1.230 | +0.171 | +0.500 | −0.365 | +0.222 | +0.075 | −0.175 | +0.161 | −0.166 |
| $V^{NiNi}(\mathbf{R}-\mathbf{R}')$ | $-2\cdot10^{-4}$ | $-7\cdot10^{-5}$ | $+9\cdot10^{-6}$ | $+3\cdot10^{-5}$ | $-2\cdot10^{-5}$ | $+1\cdot10^{-5}$ | $+4\cdot10^{-6}$ | $-9\cdot10^{-6}$ | $+9\cdot10^{-6}$ | $-9\cdot10^{-6}$ |
| | $-4\cdot10^{-6}$ | $-1\cdot10^{-6}$ | $+2\cdot10^{-7}$ | $+5\cdot10^{-7}$ | $-4\cdot10^{-7}$ | $+2\cdot10^{-7}$ | $+8\cdot10^{-8}$ | $-2\cdot10^{-7}$ | $+2\cdot10^{-7}$ | $-2\cdot10^{-7}$ |
| In $\alpha$-Ni: $A^{ss}(\mathbf{R}-\mathbf{R}')$ [16] | −4.16 | −1.10 | +0.22 | +0.60 | −0.30 | +0.05 | +0.07 | −0.14 | +0.18 | −0.11 |
| $V^{FeFe}(\mathbf{R}-\mathbf{R}')$ | −0.0048 | −0.0013 | +0.0003 | +0.0007 | −0.0003 | $+6\cdot10^{-5}$ | $+8\cdot10^{-5}$ | −0.0002 | +0.0002 | −0.0001 |

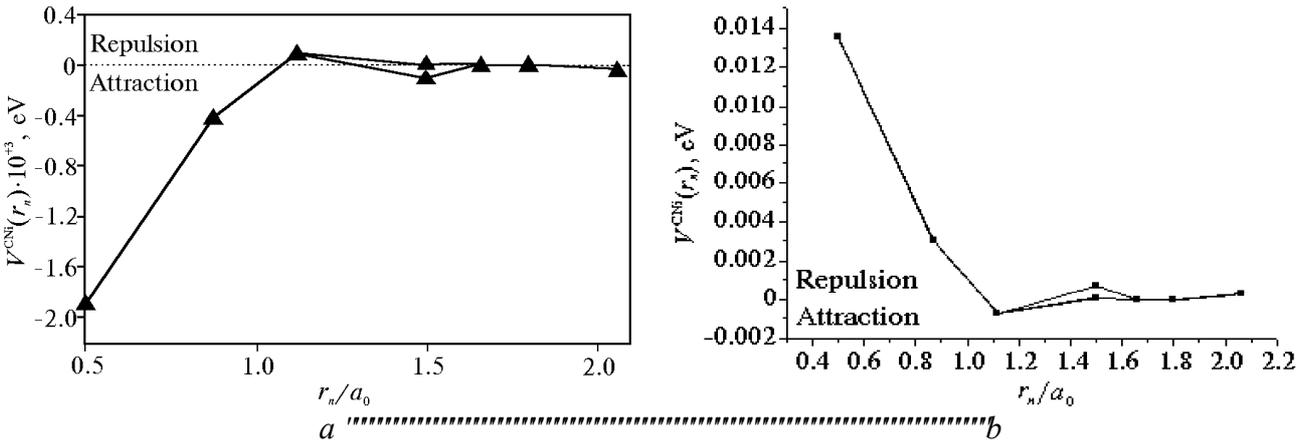

Fig. I.4. Dependence of the strain-induced interaction energies between C and Ni atoms on discrete interstice–site reduced distance $r_n/a_0$ ($n$ = I, II, …) within γ-Fe: from [24] (*a*) and present work (*b*).

interactions are determined by $L^s$ values for specific solid solutions.

The calculations of both the Fourier components, $\{\tilde{A}^{ss}(\mathbf{k})\}$, of the coefficients $\{A^{ss}(\mathbf{R}-\mathbf{R}')\}$ for substitutional impurity atoms in many f.c.c. metals [16, 24] and the energies of strain-induced Mn–Mn interaction in γ-Fe [3] or vacancy–vacancy in α-Pd [31], which have been performed with the use of different forms of the dynamical matrix, $\|\tilde{A}^{ij}(\mathbf{k})\|$, gave results, which are satisfactorily fitted with each other (and other studies [15 *etc.*]) for respective solid solutions.

It is significant for the priority rating of governing factors in the statistical thermodynamics of substitutional solutions based on the f.c.c. lattice that, for high-symmetry points $\{\Gamma, X, W, L, K(U)\}$ in $BZ$,

$$\tilde{V}^{ss}(\mathbf{k}_\Gamma{=}\mathbf{0}) \neq \tilde{V}^{ss}(\mathbf{k}{\to}\mathbf{0}) < 0 < \tilde{V}^{ss}(\mathbf{k}_X) = \tilde{V}^{ss}(\mathbf{k}_L) = \tilde{V}^{ss}(\mathbf{k}_W) = \tilde{V}^{ss}(\mathbf{k}_K)$$

(usually within the framework of approximations (I.7) and Refs [3, 38]; for reference, see also [124]).

**1.2.3. Interstitial–Substitutional Strain-Induced Interaction (Table I.4 and Fig. I.4).** Since all the interstitials expand a crystal lattice, one has $L^i > 0$, and the sign of $V^{is}(\mathbf{R}-\mathbf{R}')$ is determined by the sign of the product of $A^{is}(\mathbf{R}-\mathbf{R}')$ and $L^s$. In the first two co-ordination shells one has $A^{is}(\mathbf{R}-\mathbf{R}') < 0$. In the far shells, the interaction energy is anisotropic and quasi-oscillating.

Table I.4. Energies [eV] of pairwise strain-induced interaction of interstitial–substitutional impurity atoms in γ-Fe at 1428 K (the numbers below the horizontal split bar are estimated on basis of data taken in part from Ref. [24]) or in α-Ni at room temperature.

| $2(\mathbf{R}-\mathbf{R}')/a_0$ | 100 | 111 | 210 | 221 | 300 | 311 | 320 | 410 |
|---|---|---|---|---|---|---|---|---|
| Shell | I | II | III | IVa | IVb | V | VI | VII |
| $\|\mathbf{R}-\mathbf{R}'\|/a_0$ | 0.5 | $\cong 0.87$ | $\cong 1.12$ | 1.5 | 1.5 | $\cong 1.66$ | $\cong 1.8$ | $\cong 2.06$ |
| In γ-Fe: | | | | | | | | |
| $A^{is}(\mathbf{R}-\mathbf{R}')$ [24] | −9.448 | −2.061 | +0.469 | −0.064 | −0.465 | +0.001 | +0.007 | −0.186 |
| | +0.0136 | +0.0030 | −0.0007 | +9·10⁻⁵ | +0.0007 | −1·10⁻⁶ | −1·10⁻⁵ | +0.0003 |
| $V^{CNi}(\mathbf{R}-\mathbf{R}')$ | +0.0146 | +0.0032 | −0.0007 | +0.0001 | +0.0007 | −2·10⁻⁶ | −1·10⁻⁵ | +0.0003 |
| | −0.0019 | −0.0004 | +9·10⁻⁵ | −1·10⁻⁵ | −9·10⁻⁵ | +2·10⁻⁷ | +1·10⁻⁶ | −4·10⁻⁵ |
| | −0.0020 | −0.0004 | +0.0001 | −1·10⁻⁵ | −0.0001 | +2·10⁻⁷ | +1·10⁻⁶ | −4·10⁻⁵ |
| In α-Ni: | | | | | | | | |
| $A^{is}(\mathbf{R}-\mathbf{R}')$ [23] | −10.5 | −2.2 | +0.5 | −0.07 | −0.5 | +0.002 | +0.008 | −0.19 |
| $V^{CFe}(\mathbf{R}-\mathbf{R}')$ | −0.066 | −0.016 | +0.004 | 0 | −0.004 | 0 | 0 | −0.001 |

When a substitutional atom expands the host-crystal lattice ($L^s > 0$), attraction in the I-st and the II-nd co-ordination shells is observed with the maximal attraction in the I-st shell. On the contrary, when a substitutional atom contracts the host-crystal lattice ($L^s < 0$) there is repulsion in the first two shells and weak attraction in the III-rd one. The attraction is weak since the value $|A^{is}(\mathbf{R} - \mathbf{R}')|$ in the III-rd shell is about 20 times less than in the I-st one. This is why the i–s attraction may be essential in α-Ni and γ-Fe only if a substitutional atom expands the crystal lattice ($L^s > 0$). In the latter case, the i–s attraction in the I-st co-ordination shell may be of the same order or stronger than the i–i attraction (see energies of C–C interactions in Table I.2 as well as energies of C–Fe and C–Ni within α-Ni and γ-Fe, respectively, in Table I.4) because the value $|A^{is}(\mathbf{R} - \mathbf{R}')|$ in the I-st shell is about twice the coefficient $|A^{ii}(\mathbf{R} - \mathbf{R}')|$.

For comparison, the co-ordination shell distribution of the C–Ni interaction energies for a case of location of the interstitial C atoms in octahedral interstices in γ-Fe are calculated in this work and [24], and presented in Table I.4. Apparently, the interaction calculated and presented here is strongest in all shells, unlike a case of Ref. [24]. For $L^s < 0$, this interaction is repulsive. The main difference between our results and Ref. [24] is as follows. When a substitutional atom contracts the crystal lattice ($L^s < 0$; see, for instance, Ref. [50]), the i–s attraction in γ-Fe is quite weak (in the III-rd co-ordination shell), in contrast to much more weak repulsion in the III-rd shell in γ-Fe from [24]. One also has to take into consideration that the strain-induced interaction is not the only contribution to i–s interaction, and must be supplemented by the 'electrochemical' interaction, like a case of b.c.c. metals ([27] *etc.*).

## 2. Parameterization of the Direct 'Electrochemical' Interaction Energies

**2.1. 'Electrochemical' Interaction of Solute Atoms.** To specify the energies of direct 'electrochemical' interactions of substitutional and/or interstitial atoms in f.c.c.-(Ni,Fe)–C solutions, the most essential potentials of interactions, which can be compared with each other and used, are as follows:

(i) energy expression [43],

$$\varphi^{NiNi}(r) = -U_0(1+\Re)\exp(-\Re), \quad \Re = B(r/r_0 - 1) + C(r/r_0 - 1)^2 + D(r/r_0 - 1)^3 \quad (r < r_q),$$

where $U_0$—separation energy (cohesive energy with a sign '−') of pairwise 'bond', $r_0$—equilibrium distance for pairwise 'bond', $r_q$—effective range of potential for Ni–Ni interaction with $U_0 \approx 0.6150$ eV, $r_0 \approx 2.48$ Å, $r_q \approx 3.35$ Å, and $B \approx 5.21$, $C \approx -2.38$, $D \approx 2.38$ [44];

(ii) Lennard-Jones potential, $\varphi^{NiNi}(r) = 4\varepsilon\left\{(\sigma/r)^{12} - (\sigma/r)^6\right\}$, $\varepsilon = 0.518914994$ eV, $\sigma = 2.28$ Å [45];

(iii) generalized Lennard-Jones potential,

$$\varphi^{NiNi}(r) = D\left\{\left(\frac{r_0}{r}\right)^{m_1} - \frac{m_1}{m_2}\left(\frac{r_0}{r}\right)^{m_2}\right\},$$

$D = 30.61803872$ eV/atom, $r_0 = 2.26010$ Å, $m_1 = 9.9315$, $m_2 = 9.4481$ for Ni–Ni-pair of atoms [46, 47];

(iv) generalized Lennard-Jones potentials,

$$\varphi^{\alpha\beta}(r) = D^{\alpha\beta}\left\{\left(\frac{r_0^{\alpha\beta}}{r}\right)^{m_1} - \frac{m_1}{m_2}\left(\frac{r_0^{\alpha\beta}}{r}\right)^{m_2}\right\},$$

where, in accordance with phenomenological calculation of phase equilibria in the Ni–Fe system [117], $D^{NiNi} = 0.7391$ eV, $r_0^{NiNi} = 2.486$ Å for Ni–Ni-pair of atoms, $D^{FeFe} = 0.7007$ eV, $r_0^{FeFe} = 2.517$ Å for Fe–Fe-pair of atoms, $D^{NiFe} = 0.7919$ eV, $r_0^{NiFe} = 2.509$ Å for Ni–Fe-pair of atoms; $m_1 = 7.0$, $m_2 = 3.5$;

(v) 'angular-dependent' potentials (ADP) [118] for pair interactions are postulated in the form of a generalized Lennard-Jones function of the interatomic distances, $r$,

$$\varphi^{\alpha\alpha}(r) = \psi\left(\frac{r - r_c}{h}\right)\left[\frac{V_0}{b_2 - b_1}\left(\frac{b_2}{z^{b_1}} - \frac{b_1}{z^{b_2}}\right) + \delta\right] + m\rho(r),$$

which is parameterized with respective fitting parameters: $b_1$, $b_2$, $r_1$, $V_0$, $\delta$, $m$ for $\alpha$ = Ni or Fe, $z = r / r_1$, $\rho(r)$—the electron density function, which is chosen in the form

$$\rho(r) = \psi\left(\frac{r - r_c}{h}\right)\left[A_0 z^y e^{-\gamma z}(1 + B_0 e^{-\gamma z}) + C_0\right],$$

where $z = r - r_0$, $\psi(x)$ is a cut-off function defined by

$$\psi(x) \equiv \begin{cases} \dfrac{x^4}{1 + x^4}, & \text{if } x < 0, \\ 0, & \text{if } x \geq 0, \end{cases}$$

with respective fitting parameters $B_0$, $C_0$, $r_0$, $y$, $\gamma$, $r_c$, $h$ for $\alpha$ = Ni or Fe; the cross-interaction function $\varphi^{\mathrm{NiFe}}(r)$ is chosen in the form of a mixture of the pair-interaction functions of Fe and Ni with exponential weights,

$$\varphi^{\mathrm{NiFe}}(r) = t_1 e^{-t_2 r} \varphi^{\mathrm{FeFe}}(r) + t_3 e^{-t_4 r} \varphi^{\mathrm{NiNi}}(r)$$

with four parameters $t_1$, $t_2$, $t_3$, $t_4$; the all optimized parameters are reported in Ref. [118] after investigation by first-principles calculations and atomistic simulations of Ni–Fe-system phase stability by means of the fitting database of experimental values of the 'equilibrium' value of the respective cubic lattice parameters ($a_0$), the 'equilibrium' cohesive energy ($-U_0$), three elastic moduli ($C_{11}$, $C_{12}$, $C_{44}$), the vacancy formation ($E_v^{\mathrm{f}}$) and migration ($E_v^{\mathrm{m}}$) energies, and the surface energy ($\gamma_s$) for Fe and Ni;

(vi) Born–Mayer-type potential, $\varphi^{\alpha\beta}(r) = Ae^{-br}$ ($R_l \leq r \leq R_u$) ($\alpha$, $\beta$ = Ni, Fe), where for Ni–Ni interaction, $A \approx 13271$ eV, $b \approx 3.56819$ Å$^{-1}$, $R_l \approx 0.79376559$ Å, $R_u \approx 1.85211971$ Å or $R_u \approx (3.17506236–4.23341648)$ Å; for Fe–Fe interaction, $A \approx 11931$ eV, $b \approx 3.57730$ Å$^{-1}$, $R_l \approx 0.79376559$ Å, $R_u \approx 1.85211971$ Å or $R_u \approx (3.17506236–4.23341648)$ Å; for Ni–Fe interaction, $A \approx 12583.17531$ eV, $b \approx 3.572745$ Å$^{-1}$, $R_l \approx 0.79376559$ Å, $R_u \approx 1.85211971$ Å or $R_u \approx (3.17506236–4.23341648)$ Å [42].

As is essential to note in the statistical-thermodynamic analysis of alloy phase stability, it is necessary not only to confirm a consistency or an inconsistency of application of those or other potentials of 'electrochemical' interatomic interactions, $\varphi^{\alpha\beta}(r)$, by quantitative characteristics of both these values and their derivatives with respect to distance between atoms, $r$, but also to supplement the same with the analysis and the account of symmetry of distribution of extrema of their Fourier components all along proper reciprocal space of an alloy lattice (see [1–3, 18, 19, 21, 22, 28–31, 61, 100, 102, 109]).

The potential of Born–Mayer type (vi) within the scope of Abrahamson's parameterization [42] rather well reproduces (with a margin of errors to 3%) values of energy of repulsive (*non-bonded*) interaction of pair atoms within the limits of extension in space ($R_l \leq r \leq R_u$) of its effective action. However, combinations corresponding to it such as 'interchange' ('mixing') energy of atoms of Fe and Ni on sites already outside of the first site co-ordination shell of an f.c.c. lattice of a Ni–Fe alloy have extremely underestimated values that causes quantitative insufficiency of the energy characterization of thermodynamics of an f.c.c.-Ni–Fe alloy, for example, by means of Fourier components of such 'interchange' energies. Nevertheless, distribution of the latter values along reciprocal space of an f.c.c. lattice can properly represent the configuration-symmetrical description of ordering structures of a considered substitutional alloy.

As to the generalized potentials of Lennard-Jones type, for example, within the framework of the parameterization (iv) of Horiuchi *et al.* [117], they rather well reproduce values of energies of 'electro-chemical' interactions of pairs of atoms within the range of the first site co-ordination sphere and also reasonable values of combinations corresponding to them such as 'interchange' energy of atoms of Fe and Ni on sites of an f.c.c. lattice of a Ni–Fe alloy. However, unfortunately, on sites outside the first site co-ordination sphere, such 'interchange' energies are already characterized by too overestimated values. In that way, they do not provide the unconditionally-adequate statistical-thermodynamic description of an f.c.c.-Ni–Fe alloy, for example, by means of Fourier components of such 'interchange' energies, and distribution of extrema of the latter values along reciprocal space of an f.c.c. lattice at all does not correspond to the configuration-symmetrical description of well-known structures of a considered alloy with the long-range atomic order.

At the same time, both specified potentials can be used for order-of-magnitude estimation of macroscopic characteristics of an alloy (such as elastic modules and this sort of values) determined mainly by magnitudes (and character) of derivatives of potentials with respect to distance between atoms. Especially, it is true for the ADP of generalized Lennard-Jones type (v) within the scope of the parameterization of Mishin *et al.* [118]. Due to suitable numerical values of corresponding energies of 'electrochemical' interactions of pairs of atoms in several (usually, 4) site co-ordination shells (with radii $r_\mathrm{I} = a/(2)^{1/2}$, $r_\mathrm{II} = a$, $r_\mathrm{III} = a(6)^{1/2}/2$, and $r_\mathrm{IV} = a(2)^{1/2}$ in f.c.c. lattice with parameter $a$) and also their combinations such as 'interchange' energy of atoms of Fe and Ni on sites of a lattice of an f.c.c.-Ni–Fe alloy, which in reasonable essentially-non-monotone ('quasi-oscillating') manner depend on interatomic distances in pairs, global extrema of Fourier components of such 'interchange' energies are distributed along reciprocal space of an f.c.c. lattice in tolerably-acceptable remoteness from those high-symmetry points of this space, which actually can 'generate' well-known ordered structures of a considered alloy. In particular, some lack of coincidence of position of a global minimum of Fourier component of 'interchange' energy made up only on basis of corresponding 'electrochemical' interactions of pairs of atoms of Fe and Ni with the centre $X$ of square faces of a Brillouin zone (Fig. I.1*b*) can be quite compensated by other adjustable contributions, for instance, magnetic and/or strain-induced ones, in total interaction of substitutional atoms in a lattice of an f.c.c.-Ni–Fe alloy. Thus, with this kind of potentials of 'electrochemical' interaction of atoms of Ni and Fe supplemented with magnetic and strain-induced contributions, it is possible to expect also satisfactory results in statistical thermodynamics of both the disordered state, but with the short-range atomic order, and the ordered states of a considered alloy in conformity with available results of data management in both physicochemical or thermodynamic testing and diffraction experiments (see, *e.g.*, [58, 70, 71, 73, 76, 80, 84, 86, 99, 103, 105–111, 121, 122, 124]).

As ascertained, atom–atom potentials for interactions of interstitial atoms should be calculated as follows:

(vii) by the Born–Mayer-type formula, $\varphi^{C\beta}(r) = Ae^{-br}$ ($R_l \le r \le R_u$) ($\beta$ = C, Ni, Fe), where for 'contact' C–C repulsion, $A \approx 1316.1$ eV, $b \approx 3.80959$ Å$^{-1}$, $R_l \approx 0.52917706$ Å, $R_u \approx 1.58753118$ Å or $R_u \approx \approx (3.17506236–4.23341648)$ Å; for 'contact' C–Fe repulsion, $A \approx 3962.624016$ eV, $b \approx 3.693445$ Å$^{-1}$, $R_l \approx (0.52917706–0.79376559)$ Å, $R_u \approx (1.58753118–1.85211971)$ Å or $R_u \approx (3.17506236–4.23341648)$ Å; for 'contact' C–Ni repulsion, $A \approx 4179.229964$ eV, $b \approx 3.68889$ Å$^{-1}$, $R_l \approx (0.52917706–0.79376559)$ Å, $R_u \approx (1.58753118–1.85211971)$ Å or $R_u \approx (3.17506236–4.23341648)$ Å [42];

Table I.5. Energies [eV] of central-force 'electrochemical' interaction of octahedral interstitial C atoms on discrete interstice–interstice reduced distance $r_n/a_0$ ($n$ = I, II, …) in γ-Fe ($a_0 = 3.666$ Å [52]).

| $2(\mathbf{R} - \mathbf{R}')/a_0$ | 110 | 200 | 211 | 220 | 310 | 222 |
|---|---|---|---|---|---|---|
| Shell | I | II | III | IV | V | VI |
| $\lvert\mathbf{R} - \mathbf{R}'\rvert/a_0 = r_n/a_0$ | $1/\sqrt{2} \cong 0.71$ | 1 | $\sqrt{6}/2 \cong 1.22$ | $\sqrt{2} \cong 1.41$ | $\sqrt{5}/\sqrt{2} \cong 1.58$ | $\sqrt{3} \cong 1.73$ |
| $\varphi^{CC}(\mathbf{R} - \mathbf{R}')$ [41] | +0.163 | −0.003 | −0.002 | −0.001 | −0.0005 | −0.0003 |
| $\varphi^{CC}(\mathbf{R} - \mathbf{R}')$ [40] | +0.240 | −0.003 | −0.003 | −0.001 | −0.0007 | −0.0004 |

(viii) by the Buckingham formula, $\varphi^{CC}(r) \approx B_{CC}\exp(-C_{CC}r) - A_{CC}/r^6$, for non-bonded interatomic C–C interaction with parameters $B_{CC} = 3105$ eV, $C_{CC} = 3.68$ Å$^{-1}$, $A_{CC} = 18.26$ eV·Å$^6$ [41] or $B_{CC} = 3627$ eV, $C_{CC} = 3.6$ Å$^{-1}$, $A_{CC} = 24.63$ eV·Å$^6$ [40] (respective parameters of potentials $\varphi^{NN}(r)$ and $\varphi^{HH}(r)$ for non-bonded interatomic N–N and H–H interactions can be found *ibidem*).

The calculated values of the 'electrochemical' interaction energies $\varphi^{CC}(r_n)$ for C–C atomic pairs within the nearest six co-ordination shells ($n = I, II, ..., VI$) in γ-Fe [28] are presented in Table I.5.

**2.2. Calculation of the Thermodynamical C activity in Fe–C austenite.** To verify both the applicability of the approximation of strain-induced interaction between interstitial C atoms for description of their solutions based on γ-Fe (or α-Ni) and the necessity to supplement it with short-range repulsion due to a screened Coulomb interaction (following [24, 25]) *etc.*, one will simulate the reliable available experimental data of both thermodynamical C-activity in austenite and abundance of the Fe sites with different C neighbourhoods (as indirect parameters of the short-range order) obtained by means of Mössbauer spectroscopy, taking into account the C–C interaction within the six or eight shells.

**2.2.1. Statistical-Thermodynamics Method of Calculation.** To simulate the concentration and temperature dependences of the thermodynamical C-activity, the Murch and Thorn approach [9] to the Fe–C austenite can be used. This approach was selected because it can easily be extended for the long-range C–C interaction. The expression for the C activity can be presented in the following form [7–9]:

$$a_C \approx a_{conf}\exp\left(\frac{\Delta G}{k_B T}\right),\tag{I.10}$$

where $a_{conf}$ is a configurational factor that depends directly on the C-atomic distribution; $\exp\{\Delta G/(k_B T)\}$ is a factor accounting for the non-configurational expression; $\Delta G = \Delta H - T\Delta S$, where $\Delta H$ and $\Delta S$ are the relative partial enthalpy and non-configurational partial entropy of C atoms, respectively, *i.e.* the difference between the isobaric-isothermal thermodynamic potentials *per* C atom in a 'standard' state (*e.g.*, graphite) and infinitely dilute Fe–C solid solution.

If the C–C 'blocking' effect is supposed, the satisfactory approximation is ensured by the representation of relative probability to find the 12 empty nearest interstices around the certain 'accessible' interstice through the 4 statistically independent identical 'triplets' composed of three correlatively unoccupied interstices; within this supposition [55], $a_{conf} \approx \Gamma_C^{(4)}c_C(1-c_C)^3/(1-4c_C)^4$ (where the coefficient of proportionality may only be temperature dependent— $\Gamma_C^{(4)} = \Gamma_C^{(4)}(T)$).

Taking into account that, for infinitely dilute binary solid solution, $a_{conf} \cong c_C/(1-c_C)$ [5], the non-configurational coefficient can be estimated from experimental $a_C$ data. For small C concentration,

$$a_C \cong \frac{c_C}{1-c_C}\exp\left(\frac{\Delta G}{k_B T}\right),\tag{I.11}$$

where $c_C = N_C/N_{Fe} \ll 1$, and $N_C$ and $N_{Fe}$ are the total numbers of C and Fe atoms in γ-Fe–C solution, respectively.

The experimental values of $a_C$ at $c_C \leq 0.02$ ($\leq 0.03$ at 1073 K) and the 'theoretical' point ($c_C = 0$, $a_C = 0$) were used in [24, 25] to calculate the factor $\exp\{\Delta G/(k_B T)\}$ at several temperatures by means of the least-squares method. The estimated values of $\exp\{\Delta G/(k_B T)\}$ [24, 25] were smoothed through the fitting curve $\exp\{\Delta G/(k_B T)\} \cong 0.1434\exp\{5297.4 \text{ K}/T\}$ in Ref. [56]. These values are proved somewhat different from those determined in Ref. [6] while using slightly different assortment of experimental activity data ([10, 11, 12]; see also bibliography in Ref. [4]).

The algorithm of calculation of a configurational factor $a_{conf}$, which uses the Monte Carlo procedure, is explained in detail in Refs [7–9]. The idea of this method is as follows. For every member of statistical ensemble generated during simulation, only a part of the statistical sum of the system is calculated, namely the one specified by a virtual input of a single new C atom with fixed positions of the others. This partition function is averaged through the usual Monte Carlo procedure. The expression for $a_{conf}$ [7–9] is as follows:

$$a_{conf} = \frac{M(N_C + 1)}{\sum\limits_{K=1}^{M}\left[\sum\limits_{J=1}^{N_{Fe}-N_C} \exp\left(-\frac{\Delta E_J^K}{k_B T}\right)\right]},$$  (I.12)

where the summation over $J$ and $K$ is performed for all empty octahedral interstices and Monte Carlo steps, respectively; $\Delta E_J^K$ is a change of the system energy when one 'virtual' C atom is embedded [9]; $M$ is the number of Monte Carlo steps, and $(N_{Fe} - N_C)$ is the number of empty octahedral interstices in the model-based crystal lattice.

The Monte Carlo simulation is carried out in the following way. Within the framework of the Végard's approximation, the configurational Hamiltonian of a system composed of interacting with each other C atoms and non-interacting empty octahedral interstices, is equal to the sum of pairwise interaction energies $W^{CC}(\mathbf{R} - \mathbf{R}')$ [1–3],

$$U \approx \frac{1}{2}\sum_{\mathbf{R}}\sum_{\mathbf{R}'} W^{CC}(\mathbf{R} - \mathbf{R}')c_C(\mathbf{R})c_C(\mathbf{R}').$$  (I.13)

Here vectors $\mathbf{R}$ and $\mathbf{R}'$ indicate the positions of primitive unit cells with C atoms located in octahedral interstices; $c_C(\mathbf{R})$ are the occupation numbers of these interstices ($c_C(\mathbf{R}) = 1$ if there is a C atom in the octahedral interstice within the primitive unit cell $\mathbf{R}$; otherwise $c_C(\mathbf{R}) = 0$). A certain amount of mobile C atoms is placed randomly in octahedral interstices of model f.c.c. crystallite, which has the dimensions $12{\times}12{\times}12a_0^3$ [24, 25] and $22.5{\times}22.5{\times}22.5a_0^3$ or $52.5{\times}52.5{\times}52.5a_0^3$ (in present work and [56]) with periodic boundary conditions.

One of 12 octahedral interstices nearest to the occupied one is selected in a random way. When the chosen interstice is empty, the Hamiltonian difference $\Delta U$ for the respective jump of the selected C atom into that interstice is calculated. The jump is allowed if either $\Delta U$ proves to be negative or the probability of the jump, $\propto \exp\{-\Delta U/(k_B T)\}$, is greater than a certain random number $\zeta \in (0, 1)$. After multiple repetitions of the process, the equilibrium spatial distribution of interstitial C atoms for certain temperature and concentration is achieved.

Upon receiving this equilibrium distribution, after a certain number of cycles ($M \cong 10$), the configurational energy change $\Delta E_J^K$ after a 'virtual' embedding of the C atom into every $J$-th empty interstice is determined as follows:

$$\Delta E_J^K = \sum_{\mathbf{R}'(\neq \mathbf{R}_J)} W^{CC}(\mathbf{R}_J - \mathbf{R}')c_C(\mathbf{R}')$$  (I.14)

where the summation over $\mathbf{R}'$ is performed for all other octahedral interstices (within primitive unit cells with radius-vectors $\{\mathbf{R}'(\neq \mathbf{R}_J)\}$).

**2.2.2. Monte Carlo Simulation Results and Comparison with Mössbauer Spectroscopy Data.**
The $a_C$ concentration dependence has been calculated in [24, 25] at six temperatures either taking into account only the strain-induced C–C interaction within eight co-ordination shells (first way) or with additional (fitting) repulsion in the I-st co-ordination shell (second way). For the first case, any calculated $a_C(c_C)$ line (at $T$ = const) did not coincide with set of experimental data. It means that the strain-induced interaction can solely not describe the thermodynamic properties of austenite. Because the approximation of strain-induced i–i interaction was previously shown [1, 17–19, 26, 27, 32] to be applicable for description of b.c.c.-metal solid solutions only if it is supplemented with 'electrochemical' repulsion (*e.g.*, a screened Coulomb repulsion) in about three co-ordination shells, the strain-induced C–C interaction in austenite has to be also supplemented with a short-range repulsion. The three co-ordination shells distance of octahedral interstices is equal to 2.49 Å in α-Fe, and this matches the dis-tance within the first shell in γ-Fe (2.59 Å). This is why the additional C–C interatomic repulsion ($\varphi^{CC}(r_1) > 0$) may be taken into consideration in the first shell only, and the energies in other shells are preserved as strain-induced in [24, 25]. At least for the first co-ordination shell ($r_1 = |\mathbf{R} - \mathbf{R'}| \cong 0.71a_0$), the 'total' energy $W^{CC}(r_1) = V^{CC}(r_1)$ ('strain-induced' part) + $\varphi^{CC}(r_1)$ ('electrochemical' part) must be used in the Eqs (I.13) and (I.14).

The $W^{CC}(r_1)$ values were fitted in [24, 25] through the least-squares method, which showed the best correlation of simulated $a_C$ values with the experimental ones at every temperature: these $a_C$ values calculated with the use of fitting energies, $W^{CC}(r_1)$, manifest good agreement of computed and experimental C thermodynamic activity values in austenite.

The average $W^{CC}(r_1)$ value estimated in [24, 25] is +0.115 eV, *i.e.* the short-range C–C repulsion overrides the strain-induced C–C attraction in the first shell. Therefore, there is strong C–C repulsion in the first two shells as in case of the well-known approach of C–C interaction in two shells only [53]. However, the 'total'-interaction energies fitted and computed in Refs [24, 25], $W^{CC}(r_1)$ ($\approx$ +0.115 eV) and $W^{CC}(r_{II})$ ($\approx$ +0.169 eV), respectively (as well as estimated independently, *i.e.* semi-empirically, in this work and [56]), are much greater than the respective values in Ref. [53]—+0.036 eV and +0.075 eV, because the long-range interaction within many shells was not taken into account in [53].

The $a_C$ concentration dependence simulated within the six co-ordination shells, taking into account both verified energies of both strain-induced C–C interaction (the optimal set above the horizontal split bar in Table I.2 of this work; see Fig. I.2*b*) and 'electrochemical' C–C interaction (the optimal sets in Table I.5 based on Refs [28, 40, 41]), is presented in Fig. I.5 with solid lines at several temperatures. One can see that these curves do not coincide with experimental data (circles ● [13]). Nevertheless, with increase of the dimensions of a simulated crystallite [56], the computed lines may approach the experimental data. As a whole, the activity calculation shows that the approximation of strain-induced interaction supplemented with 'electrochemical' repulsion in the first shell is mostly applicable for the description of thermodynamic properties of austenite.

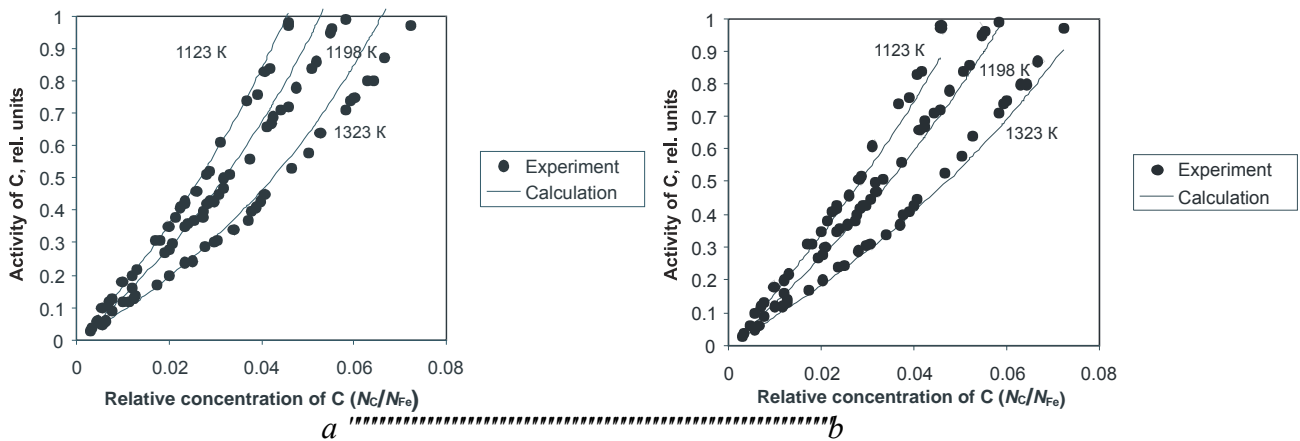

Fig. I.5. The concentration dependence of C activity in austenite simulated by f.c.c. crystallite that has dimensions $52.5 \times 52.5 \times 52.5 a_0^3$ with periodic boundary conditions and calculated within the framework of the Monte Carlo procedure for $M = 5$ at the end of 50 dovetailing iterations at $T = 1123, 1198, 1323$ K (solid lines [56] provided that $\exp\{\Delta G/(k_B T)\} \cong 0.1434\exp\{5297.4\,\text{K}/T\}$ based on data of [10–12] cited in [24, 25] and also referred in [4]) taking into account the roughened strain-induced C–C interaction within the six co-ordination shells (see Table I.2 and Fig. I.2b) supplemented with additional 'electrochemical' repulsion in the I-st co-ordination shell and a weak 'polarization' attraction within the next five co-ordination shells from Table I.5 [28]. Accordingly, the respective 'total' energies $\{W^{CC}(r_n),\ n = \text{I, II,} \ldots, \text{VI}\}$ are specified as follow: $W^{CC}(r_1) \approx$ +0.155 eV, $W^{CC}(r_{II}) \approx$ +0.144 eV, $W^{CC}(r_{III}) \approx$ −0.040 eV, $W^{CC}(r_{IV}) \approx$ +0.003 eV, $W^{CC}(r_V) \approx$ +0.0023 eV, $W^{CC}(r_{VI}) \approx$ +0.0196 eV (*a*); $W^{CC}(r_1) \approx$ +0.078 eV, $W^{CC}(r_{II}) \approx$ +0.144 eV, $W^{CC}(r_{III}) \approx$ −0.039 eV, $W^{CC}(r_{IV}) \approx$ +0.003 eV, $W^{CC}(r_V) \approx$ +0.0025 eV, $W^{CC}(r_{VI}) \approx$ +0.0197 eV (*b*). Experimental data [13] are shown by circles ●.

As regards the Mössbauer data for f.c.c.-Fe–C alloy, they are available only in the narrow concentration range $c_C = 0.05$–$0.09$ [3, 4, 53, 54, 57] due to the limited storability of the metastable f.c.c. structure during the cooling to room temperature.

According to interpretation [7, 8, 53, 54, 57] one can determine the following parameters of the distribution of octahedral C atoms in the vicinity of Fe atoms depending on the C–C interaction energies: $P_0$, $P_1$, $P_{21}^{90°}$ and $P_{22}^{180°}$. $P_0$ and $P_1$ are the atomic fractions of Fe that has no or one C atom in the nearest octahedral interstices, respectively. $P_{21}^{90°}$ is the atomic fractions of Fe atoms that have two C atoms in the nearest octahedral interstices separated by the vector $\mathbf{r}_I = (1/2, 1/2, 0)a_0$ (*i.e.* within the I-st co-ordination shell). $P_{22}^{180°}$ is the fraction of Fe atoms that have two C atoms in the nearest octahedral interstices separated by the vector $\mathbf{r}_{II} = (1, 0, 0)a_0$ (*i.e.* in the II-nd co-ordination shell).

Calculation of the short-range order and average values of the ten fractions $\{P_k; k = 0, 1, 21^{90°}, 22^{180°}, 31, 32, 41, 42, 5, 6\}$ was performed by using Monte Carlo computer simulation of spatial distribution of C atoms described above, taking into account the energies of long-range C–C strain-induced interaction within eight or six co-ordination shells without or with additional parameter of 'electro-chemical' repulsion at least in the first shell. Table I.6 presents the experimental data [53, 54] and the results of respective simulations [24, 25, 56]. (For comparison, see also [57].)

Table I.6. Abundances of the different Fe-sites and respective sets of energies, $\{W^{CC}(r_n) = V^{CC}(r_n) + \varphi^{CC}(r_n)\}$ [eV], of C–C interaction within the first eight co-ordination shells ($n = I, II, ..., VIII$) [24] or within the first six co-ordination shells ($n = I, II, ..., VI$) [56] according to the fitting of Mössbauer-spectroscopy data obtained in different works [53, 54] or independent semi-empirical estimation (see Tables I.2, I.5, and [56]), respectively.

| $c_C$ | | $T_f$, K | $P_0$ | $P_1 + P_{21}^{90°}$ | $P_{22}^{180°}$ |
|---|---|---|---|---|---|
| 0.087 | Experiment [53] | | 0.545 | 0.455 | < 0.006 |
| | 'Total'—$W^{CC}(r_I) \approx +0.020$, and strain-induced only—$W^{CC}(r_{II}) \approx +0.169$, $W^{CC}(r_{III}) \approx -0.042$, $W^{CC}(r_{IV}) \approx +0.004$, $W^{CC}(r_V) \approx +0.004$, $W^{CC}(r_{VI}) \approx +0.023$, $W^{CC}(r_{VII}) \approx -0.002$, $W^{CC}(r_{VIII}) \approx -0.012$ (Table I.2 and [25]) [24]: | 773 | 0.5443 | 0.4536 | 0.0021 |
| | 'Total'—$W^{CC}(r_I) \approx +0.089$, and strain-induced only—$W^{CC}(r_{II}) \approx +0.169$, $W^{CC}(r_{III}) \approx -0.042$, $W^{CC}(r_{IV}) \approx +0.004$, $W^{CC}(r_V) \approx +0.004$, $W^{CC}(r_{VI}) \approx +0.023$, $W^{CC}(r_{VII}) \approx -0.002$, $W^{CC}(r_{VIII}) \approx -0.012$ (Table I.2 and [25]) [24]: | 1600 | 0.54 | 0.4530 | 0.0070 |
| 0.083 | Experiment [54] | | 0.575± ±0.01 | 0.425± ±0.01 | < 0.01 |
| | 'Total'—$W^{CC}(r_I) \approx +0.004$, and strain-induced only—$W^{CC}(r_{II}) \approx +0.169$, $W^{CC}(r_{III}) \approx -0.042$, $W^{CC}(r_{IV}) \approx +0.004$, $W^{CC}(r_V) \approx +0.004$, $W^{CC}(r_{VI}) \approx +0.023$, $W^{CC}(r_{VII}) \approx -0.002$, $W^{CC}(r_{VIII}) \approx -0.012$ (Table I.2 and [25]) [24]: | 773 | 0.5743 | 0.4239 | 0.0018 |
| | Total—$W^{CC}(r_I) \approx +0.155$, $W^{CC}(r_{II}) \approx +0.144$, $W^{CC}(r_{III}) \approx -0.040$, $W^{CC}(r_{IV}) \approx +0.003$, $W^{CC}(r_V) \approx +0.0023$, $W^{CC}(r_{VI}) \approx +0.0196$ (Tables I.2, I.5, and [40]) [56]: | 1123–1323 | 0.5190± ±0.0002 | 0.4774± ±0.0004 | 0.0035± ±0.0001 |
| | Total—$W^{CC}(r_I) \approx +0.078$, $W^{CC}(r_{II}) \approx +0.144$, $W^{CC}(r_{III}) \approx -0.039$, $W^{CC}(r_{IV}) \approx +0.003$, $W^{CC}(r_V) \approx +0.0025$, $W^{CC}(r_{VI}) \approx +0.0197$ (Tables I.2, I.5, and [41]) [56]: | 1123–1323 | 0.5416± ±0.0005 | 0.4538± ±0.001 | 0.0041± ±0.00015 |
| | 'Total'—$W^{CC}(r_I) \approx +0.036$, and strain-induced only—$W^{CC}(r_{II}) \approx +0.169$, $W^{CC}(r_{III}) \approx -0.042$, $W^{CC}(r_{IV}) \approx +0.004$, $W^{CC}(r_V) \approx +0.004$, $W^{CC}(r_{VI}) \approx +0.023$, $W^{CC}(r_{VII}) \approx -0.002$, $W^{CC}(r_{VIII}) \approx -0.012$ (Table I.2 and [25]) [24]: | 1600 | 0.5718 | 0.4228 | 0.0054 |

The simulation results are dependent on the 'freezing' temperature, $T_f$, of the thermodynamically equilibrium atomic distribution during quenching of the austenite specimens. $T_f$ was estimated as 600 K in Ref. [53], and 773 K in Ref. [54]. Because no estimations are trustworthy, the dominating $\{P_k; k = 0, 1, 21^{90°}, 22^{180°}\}$ were calculated for several temperatures ($T_f = 773, 1123, 1198, 1323, 1600$ K), which enclose the temperature range of $a_C(c_C)$ simulations (1073–1573 K in [24, 25, 56]).

The strain-induced C–C interaction can solely not describe the experimental data, and this is necessary to supplement it with repulsion at least within the first shell, as in case of the thermodynamic activity. According to Mössbauer data (see Table I.6), the optimal independently (*semi-empirically*) estimated value $W^{CC}(r_1) \approx 0.078$ eV [56] is located within the range of *fitting* evaluations from +0.004 eV to +0.089 eV (see [24]). This result may only be treated as qualitative but not quantitative due to the problems of both the detailed interpretation of Mössbauer spectra and correct $T_f$ determination.

## Part II. Study of Magnetic Interactions of Substitutional Atoms in F.C.C.-Ni–Fe–(C) Alloys

### 3. Calculations of the Exchange Interaction Energies

**3.1. Magnetic Contribution into the Interaction Energies of Ni and Fe Atoms.** The f.c.c.-$Ni_{1-c_{Fe}}Fe_{c_{Fe}}$ system is a metallic ordering alloy with magnetic Curie temperatures higher than the spatial disorder–order transformation temperatures [58]. As soon as between 50 and 75 at.% Fe, its magnetism may be of the very weak itinerant-electron type [58]. It is therefore quite evident that a description of this system by a distinguishable-spin-$\geq\frac{1}{2}$ Ising model Hamiltonian is practice-relevant enough. Of course, it is difficult to explain all the mechanisms of multifarious phenomena within the scope of this model. Nevertheless, it is interesting to consider the model, which is a simple interpolation scheme for this case and to see what kind of interaction parameters and observable quantities one obtains with this procedure. It should be kept in mind that such a comparison is only semi-quantitative and can yield only a rough description of the phenomena in f.c.c.-$Ni_{1-c_{Fe}}Fe_{c_{Fe}}$ alloys ($0.5 < c_{Fe} < 0.75$): we by no means attempt to claim that suggested simple model applies directly to the metallic, itinerant and complex situation of these alloys [119, 120].

The mathematical Heisenberg model of the system of randomly located arbitrary 'spins' concerning 'slowly'-relaxing arrangement of their carriers in a lattice may be updated for binary solid solutions, which consist of both magnetic components. The system Hamiltonian of transition-metal (Ni, Fe) atomic-'spin' operators $\{\hat{\mathbf{s}}_\alpha(\mathbf{R})\}$ at sites $\{\mathbf{R}\}$ of f.c.c. lattice can be constructed, where the contributions of exchange-interaction energies of 'spins' of Ni and Fe atoms in Fe–Fe, Ni–Ni, and Ni–Fe pairs, $\{J_{\alpha\beta}(\mathbf{R}-\mathbf{R}'); \alpha, \beta = Ni, Fe\}$, at arbitrary distances $|\mathbf{R}-\mathbf{R}'|$ are taken into account. Namely, with the neglect of the anisotropic and antisymmetric components of the 'spin'–'spin' exchange interactions in a Ni–Fe alloy, it can be written down as follows:

$$\hat{H}_{sp} = \frac{1}{2}\sum_{\mathbf{R},\mathbf{R}'}\Big\{J_{FeFe}(\mathbf{R}-\mathbf{R}')c_{Fe}(\mathbf{R})c_{Fe}(\mathbf{R}')\big(\hat{\mathbf{s}}_{Fe}(\mathbf{R})\cdot\hat{\mathbf{s}}_{Fe}(\mathbf{R}')\big) +$$

$$+ J_{NiNi}(\mathbf{R}-\mathbf{R}')c_{Ni}(\mathbf{R})c_{Ni}(\mathbf{R}')\big(\hat{\mathbf{s}}_{Ni}(\mathbf{R})\cdot\hat{\mathbf{s}}_{Ni}(\mathbf{R}')\big) + J_{FeNi}(\mathbf{R}-\mathbf{R}')c_{Fe}(\mathbf{R})c_{Ni}(\mathbf{R}')\big(\hat{\mathbf{s}}_{Fe}(\mathbf{R})\cdot\hat{\mathbf{s}}_{Ni}(\mathbf{R}')\big) +$$

$$+ J_{NiFe}(\mathbf{R}-\mathbf{R}')c_{Ni}(\mathbf{R})c_{Fe}(\mathbf{R}')\big(\hat{\mathbf{s}}_{Ni}(\mathbf{R})\cdot\hat{\mathbf{s}}_{Fe}(\mathbf{R}')\big)\Big\};$$

here $c_\alpha(\mathbf{R})$ ($\alpha = Ni, Fe$) are the occupation numbers for sites $\{\mathbf{R}\}$ ($c_\alpha(\mathbf{R}) = 1$ if there is an $\alpha$-atom at the origin site of the primitive cell $\mathbf{R}$; otherwise $c_\alpha(\mathbf{R}) = 0$). After ensemble averaging over orientations of all atomic 'spin' moments of both constituents amenably to the statistical-mechanics formula

$$\langle\ldots\rangle_{sp} = \text{Tr}\left\{(\ldots)\exp\left(-\hat{H}_{sp}\big/(k_B T)\right)\right\}\bigg/\text{Tr}\left\{\exp\left(-\hat{H}_{sp}\big/(k_B T)\right)\right\},$$

we have a following relation:

$$
\begin{aligned}
\left\langle\hat{H}_{sp}\right\rangle_{sp} = \frac{1}{2}\sum_{\mathbf{R},\mathbf{R}'}\Big\{ & J_{\text{FeFe}}(\mathbf{R}-\mathbf{R}')\left\langle\left(\hat{\mathbf{s}}_{\text{Fe}}(\mathbf{R})\cdot\hat{\mathbf{s}}_{\text{Fe}}(\mathbf{R}')\right)\right\rangle_{sp}c_{\text{Fe}}(\mathbf{R})c_{\text{Fe}}(\mathbf{R}') + \\
& + J_{\text{NiNi}}(\mathbf{R}-\mathbf{R}')\left\langle\left(\hat{\mathbf{s}}_{\text{Ni}}(\mathbf{R})\cdot\hat{\mathbf{s}}_{\text{Ni}}(\mathbf{R}')\right)\right\rangle_{sp}c_{\text{Ni}}(\mathbf{R})c_{\text{Ni}}(\mathbf{R}') + \\
& + J_{\text{FeNi}}(\mathbf{R}-\mathbf{R}')\left\langle\left(\hat{\mathbf{s}}_{\text{Fe}}(\mathbf{R})\cdot\hat{\mathbf{s}}_{\text{Ni}}(\mathbf{R}')\right)\right\rangle_{sp}c_{\text{Fe}}(\mathbf{R})c_{\text{Ni}}(\mathbf{R}') + \\
& + J_{\text{NiFe}}(\mathbf{R}-\mathbf{R}')\left\langle\left(\hat{\mathbf{s}}_{\text{Ni}}(\mathbf{R})\cdot\hat{\mathbf{s}}_{\text{Fe}}(\mathbf{R}')\right)\right\rangle_{sp}c_{\text{Ni}}(\mathbf{R})c_{\text{Fe}}(\mathbf{R}')\Big\}.
\end{aligned}
$$

Let us

$$\left\langle\left(\hat{\mathbf{s}}_\alpha(\mathbf{R})\cdot\hat{\mathbf{s}}_\beta(\mathbf{R}')\right)\right\rangle_{sp} \equiv \left(\left\langle\hat{\mathbf{s}}_\alpha(\mathbf{R})\right\rangle_{sp}\cdot\left\langle\hat{\mathbf{s}}_\beta(\mathbf{R}')\right\rangle_{sp}\right).$$

In other words, let us guess weakness of correlation between 'fast'-relaxing orientations of the 'spin' moments localized on atoms in sites $\mathbf{R}$ and $\mathbf{R}'$. As regards the orbital moments, they are assumed as 'silent', at least, because of a feeble magnetic anisotropy in orientation of magnetic moments of atoms, $\{\boldsymbol{\mu}_\alpha\}$, with reference to crystallographic axes of a lattice of an alloy. Therefore, values of Lande's factors [59], $g_\alpha$, for both constituents ($\alpha = $ Ni, Fe) not strongly differ from 'spin value': $g = 2$. In that case,

$$
\begin{aligned}
\left\langle\hat{H}_{sp}\right\rangle_{sp} \cong \frac{1}{2}\sum_{\mathbf{R},\mathbf{R}'}\Big\{ & J_{\text{FeFe}}(\mathbf{R}-\mathbf{R}')\left(\left\langle\hat{\mathbf{s}}_{\text{Fe}}(\mathbf{R})\right\rangle_{sp}\cdot\left\langle\hat{\mathbf{s}}_{\text{Fe}}(\mathbf{R}')\right\rangle_{sp}\right)c_{\text{Fe}}(\mathbf{R})c_{\text{Fe}}(\mathbf{R}') + \\
& + J_{\text{NiNi}}(\mathbf{R}-\mathbf{R}')\left(\left\langle\hat{\mathbf{s}}_{\text{Ni}}(\mathbf{R})\right\rangle_{sp}\cdot\left\langle\hat{\mathbf{s}}_{\text{Ni}}(\mathbf{R}')\right\rangle_{sp}\right)c_{\text{Ni}}(\mathbf{R})c_{\text{Ni}}(\mathbf{R}') + \\
& + J_{\text{FeNi}}(\mathbf{R}-\mathbf{R}')\left(\left\langle\hat{\mathbf{s}}_{\text{Fe}}(\mathbf{R})\right\rangle_{sp}\cdot\left\langle\hat{\mathbf{s}}_{\text{Ni}}(\mathbf{R}')\right\rangle_{sp}\right)c_{\text{Fe}}(\mathbf{R})c_{\text{Ni}}(\mathbf{R}') + \\
& + J_{\text{NiFe}}(\mathbf{R}-\mathbf{R}')\left(\left\langle\hat{\mathbf{s}}_{\text{Ni}}(\mathbf{R})\right\rangle_{sp}\cdot\left\langle\hat{\mathbf{s}}_{\text{Fe}}(\mathbf{R}')\right\rangle_{sp}\right)c_{\text{Ni}}(\mathbf{R})c_{\text{Fe}}(\mathbf{R}')\Big\}.
\end{aligned}
$$

Furthermore, let us define relative, *i.e.* reduced, spontaneous magnetization of an $\alpha$-th subsystem of an alloy by means of averaging of the operator of the total spin moment of atom of an $\alpha$-th constituent (with spin number of $s_\alpha$) at the arbitrary site $\mathbf{R}$ over orientations of all spin moments of atoms of both constituents in crystallographically equivalent sites of a lattice $\{\mathbf{R}\}$, at least, in conditions of a feeble internal crystal magnetic anisotropy with reference to a certain crystallographic axis along which we shall direct the Cartesian axis $Oz$:

$$\sigma_\alpha = \left\langle\hat{\mathbf{s}}_\alpha(\mathbf{R})\right\rangle_{sp}\big/s_\alpha \cong \left\langle\hat{s}_\alpha^z(\mathbf{R})\right\rangle_{sp}\big/s_\alpha.$$

It is obvious that $-1 \leq \sigma_\alpha \leq +1$ and, *e.g.*, $s_{\text{Fe}}\sigma_{\text{Fe}} \cong \mu_{\text{Fe}}/(g_{\text{Fe}}\mu_B)$; the Lande's factor value of Fe, $g_{\text{Fe}}$, is close to its 'spin value' (2); the same concerns to a subsystem of magnetic moments on Ni atoms, $\{\boldsymbol{\mu}_{\text{Ni}}\}$.

Moreover, let us suppose that

$$\left(\left\langle\hat{\mathbf{s}}_\alpha(\mathbf{R})\right\rangle_{sp}\cdot\left\langle\hat{\mathbf{s}}_\beta(\mathbf{R}')\right\rangle_{sp}\right) \cong s_\alpha s_\beta\sigma_\alpha\sigma_\beta.$$

(Apparently, this guess partly exaggerates both a role and the significance of magnetism of atoms in their interaction, especially, at large distances between them.) In that case,

$$\left\langle \hat{H}_{sp} \right\rangle_{sp} \cong \frac{1}{2} \sum_{\mathbf{R}, \mathbf{R}'} \{ J_{FeFe}(\mathbf{R} - \mathbf{R}') s_{Fe}^2 \sigma_{Fe}^2 c_{Fe}(\mathbf{R}) c_{Fe}(\mathbf{R}') +$$

$$+ J_{NiNi}(\mathbf{R} - \mathbf{R}') s_{Ni}^2 \sigma_{Ni}^2 c_{Ni}(\mathbf{R}) c_{Ni}(\mathbf{R}') + 2 J_{FeNi}(\mathbf{R} - \mathbf{R}') s_{Fe} s_{Ni} \sigma_{Fe} \sigma_{Ni} c_{Fe}(\mathbf{R}) c_{Ni}(\mathbf{R}') \}.$$

Within the scope of the guess of an admissibility of approximation of interaction between the spin moments by means of interaction of each of them with the effective self-consistent ('molecular') field, for system of such spin moments, which do 'not interact' with each other, inside of this effective field, the relative magnetization, $\sigma_\alpha = B_{s_\alpha}(s_\alpha H^\alpha_{eff}/(k_B T))$, is defined by the Brillouin function [58, 59],

$$B_{s_\alpha}(y_\alpha) = \left(1 + \frac{1}{2s_\alpha}\right) \mathrm{cth}\left[\left(1 + \frac{1}{2s_\alpha}\right) y_\alpha\right] - \frac{1}{2s_\alpha} \mathrm{cth}\left[\frac{1}{2s_\alpha} y_\alpha\right], \tag{II.1}$$

in terms of $y_\alpha = s_\alpha H^\alpha_{eff}/(k_B T)$ and spin numbers $s_\alpha$ ($= s_{Fe}$ or $s_{Ni}$). For instance, $\sigma_{Fe} = \langle s_{Fe}^z(\mathbf{r})\rangle_{sp}/s_{Fe}$ is relative thermally averaged ($\langle\ldots\rangle$) 'magnetization' of Fe subsystem of f.c.c.-Ni–Fe alloy with effective (thermally averaged) $z$ projection of spontaneous local magnetic moment *per* Fe ion, $\mu_{Fe} \cong 2\mu_B s_{Fe} \sigma_{Fe}$, at the temperature $T$ and infinitesimal applied magnetic field, where $\mu_B$ is the Bohr magneton. 'Effective' field $H^\alpha_{eff} = -\mu_B \sum_{\beta = Ni, Fe} g_\beta \Gamma_{\alpha\beta} \sigma_\beta$ is proportionate to $\sigma_\beta$ ($g_\beta$ ($\cong 2$)—Lande's factor; $\Gamma_{\alpha\beta}$ is coefficient of the Weiss 'molecular'-field contribution), and hence this relationship, $\sigma_\alpha = B_{s_\alpha}(y_\alpha)$, is not so much expression for $\sigma_\alpha$ as equation with regard to $\sigma_\alpha$ within the self-consistent field approximation [2, 59, 60–62]. (The detailed tables of $B_{s_\alpha}$ and $B'_{s_\alpha}$ values as functions of $y_\alpha \in [0, 10]$ for different values of spin number $s_\alpha = 1/2, 1, 3/2, 2, 5/2, 3, 7/2$, and also $B_{s_\alpha}(y_\alpha)$ graphs against $y_\alpha$ for values of parameter $s_\alpha = 1/2, 3/2, 7/2$ and $\infty$ are given in book [59].)

The respective equations follow from the deduced expression for the total 'configuration-dependent' free energy *per* site,

$$f \cong \frac{1}{2}\left[ c_{Fe}^2 \tilde{w}_{prm}(\mathbf{0}) + c_{Fe}^2 \tilde{J}_{FeFe}(\mathbf{0}) s_{Fe}^2 \sigma_{Fe}^2 + (1 - c_{Fe})^2 \tilde{J}_{NiNi}(\mathbf{0}) s_{Ni}^2 \sigma_{Ni}^2 + 2(1 - c_{Fe}) c_{Fe} \tilde{J}_{NiFe}(\mathbf{0}) s_{Ni} s_{Fe} \sigma_{Ni} \sigma_{Fe} +$$

$$+ \frac{3}{16} \eta^2 \left( \tilde{w}_{prm}(\mathbf{k}_X) + \tilde{J}_{FeFe}(\mathbf{k}_X) s_{Fe}^2 \sigma_{Fe}^2 + \tilde{J}_{NiNi}(\mathbf{k}_X) s_{Ni}^2 \sigma_{Ni}^2 - 2\tilde{J}_{NiFe}(\mathbf{k}_X) s_{Ni} s_{Fe} \sigma_{Ni} \sigma_{Fe} \right) \right] +$$

$$+ \frac{k_B T}{4}\left[ 3\left(c_{Fe} - \frac{\eta}{4}\right)\ln\left(c_{Fe} - \frac{\eta}{4}\right) + 3\left(1 - c_{Fe} + \frac{\eta}{4}\right)\ln\left(1 - c_{Fe} + \frac{\eta}{4}\right) + \left(1 - c_{Fe} - \frac{3\eta}{4}\right)\ln\left(1 - c_{Fe} - \frac{3\eta}{4}\right) + $$

$$+ \left(c_{Fe} + \frac{3\eta}{4}\right)\ln\left(c_{Fe} + \frac{3\eta}{4}\right) \right] - k_B T c_{Fe}\left[ \ln \mathrm{sh}\left(\left(1 + \frac{1}{2s_{Fe}}\right) y_{Fe}(\sigma_{Fe})\right) - \ln \mathrm{sh}\left(\frac{1}{2s_{Fe}} y_{Fe}(\sigma_{Fe})\right) - \sigma_{Fe} y_{Fe}(\sigma_{Fe}) \right] - $$

$$- k_B T (1 - c_{Fe})\left[ \ln \mathrm{sh}\left(\left(1 + \frac{1}{2s_{Ni}}\right) y_{Ni}(\sigma_{Ni})\right) - \ln \mathrm{sh}\left(\frac{1}{2s_{Ni}} y_{Ni}(\sigma_{Ni})\right) - \sigma_{Ni} y_{Ni}(\sigma_{Ni}) \right],$$

$$\tag{II.2}$$

for such f.c.c.-Ni$_{c_{Ni}}$Fe$_{c_{Fe}}$ alloy ($c_{Ni} + c_{Fe} = 1$) ordered in accordance with $L1_2$ (Ni$_3$Fe) superstructural type (see Fig. II.1$a$) with a long-range order parameter, $\eta$ [60] (see also Eq. (III.3) below). Here for the zero wave vector $\mathbf{0}$ ($\Gamma$ point) and the three rays of star of wave vector—$\mathbf{k}_X = 2\pi(1\,0\,0)$, $2\pi(0\,1\,0)$, and $2\pi(0\,0\,1)$—$X$ points (see Fig. I.1$b$), the Fourier components of 'microscopic'-model energy parameters, $\tilde{J}_{\alpha\alpha'}(\mathbf{k}_X)$ and $\tilde{J}_{\alpha\alpha'}(\mathbf{0})$, of the spin-carriers ($\alpha, \alpha'$ = Fe, Ni) exchange interaction,

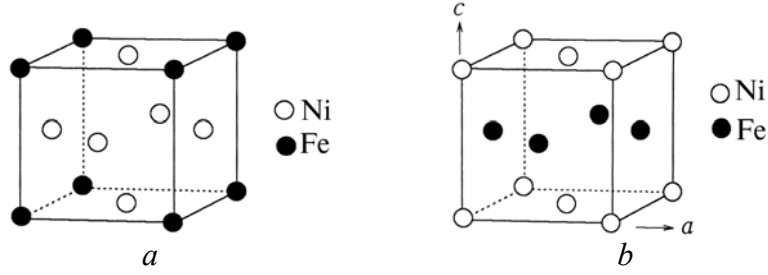

Fig. II.1. Ni$_3$Fe $L1_2$-type (*a*) and NiFe $L1_0$-type (*b*) superstructures based on f.c.c. lattice. At low $T$, the layered NiFe superstructure may have the tetragonally-distorted unit cell with non-equal parameters, $c$ and $a$ (however, $c/a \cong 1$).

$$\tilde{J}_{\alpha\alpha'}(\mathbf{k}) = \sum_{\mathbf{r}} J_{\alpha\alpha'}(\mathbf{r}) e^{-i\mathbf{k}\cdot\mathbf{r}},$$

as well as the Fourier components, $\tilde{w}_{\text{prm}}(\mathbf{k}_X)$ and $\tilde{w}_{\text{prm}}(\mathbf{0})$, of 'mixing' energies of Ni and Fe atoms,

$$w_{\text{prm}}(\mathbf{r}) = W^{\text{NiNi}}(\mathbf{r}) + W^{\text{FeFe}}(\mathbf{r}) - 2W^{\text{NiFe}}(\mathbf{r}),$$

within the f.c.c.-Ni–Fe alloy in paramagnetic state should be determined just after Refs [2, 60, 61] (on the development of models [63, 64, 66], which take into account the interaction of only nearest-neighbour atoms and their spins) for involved Fe concentrations, $c_{\text{Fe}}$, as provided by available 'macroscopic' experimental magnetic measurements as well as radiation diffuse-scattering data.

Using the following conditions, $\partial f/\partial\sigma_{\text{Fe}} = 0$, $\partial f/\partial\sigma_{\text{Ni}} = 0$, and $\partial f/\partial\eta = 0$, the set of transcendental equations for calculation of equilibrium values of $\sigma_{\text{Fe}}$, $\sigma_{\text{Ni}}$, and long-range order parameter, $\eta$, may be obtained:

$$\sigma_{\text{Fe}} \cong B_{s_{\text{Fe}}}\Bigg(-\frac{1}{c_{\text{Fe}}k_B T}\Big\{c_{\text{Fe}}^2 \tilde{J}_{\text{FeFe}}(\mathbf{0})s_{\text{Fe}}^2\sigma_{\text{Fe}} + c_{\text{Fe}}(1-c_{\text{Fe}})\tilde{J}_{\text{NiFe}}(\mathbf{0})s_{\text{Ni}}s_{\text{Fe}}\sigma_{\text{Ni}} +$$

$$+\frac{3}{16}\eta^2\Big[\tilde{J}_{\text{FeFe}}(\mathbf{k}_X)s_{\text{Fe}}^2\sigma_{\text{Fe}} - \tilde{J}_{\text{NiFe}}(\mathbf{k}_X)s_{\text{Ni}}s_{\text{Fe}}\sigma_{\text{Ni}}\Big]\Big\}\Bigg),$$

$$\sigma_{\text{Ni}} \cong B_{s_{\text{Ni}}}\Bigg(-\frac{1}{(1-c_{\text{Fe}})k_B T}\Big\{(1-c_{\text{Fe}})^2 \tilde{J}_{\text{NiNi}}(\mathbf{0})s_{\text{Ni}}^2\sigma_{\text{Ni}} + c_{\text{Fe}}(1-c_{\text{Fe}})\tilde{J}_{\text{NiFe}}(\mathbf{0})s_{\text{Ni}}s_{\text{Fe}}\sigma_{\text{Fe}} +$$

$$+\frac{3}{16}\eta^2\Big[\tilde{J}_{\text{NiNi}}(\mathbf{k}_X)s_{\text{Ni}}^2\sigma_{\text{Ni}} - \tilde{J}_{\text{NiFe}}(\mathbf{k}_X)s_{\text{Ni}}s_{\text{Fe}}\sigma_{\text{Fe}}\Big]\Big\}\Bigg),$$

$$\ln\frac{\left(c_{\text{Fe}} - \dfrac{\eta}{4}\right)\left(1 - c_{\text{Fe}} - \dfrac{3\eta}{4}\right)}{\left(c_{\text{Fe}} + \dfrac{3\eta}{4}\right)\left(1 - c_{\text{Fe}} + \dfrac{\eta}{4}\right)} \cong$$

$$\cong \frac{\eta}{k_B T}\Big[\tilde{w}_{\text{prm}}(\mathbf{k}_X) + \tilde{J}_{\text{FeFe}}(\mathbf{k}_X)s_{\text{Fe}}^2\sigma_{\text{Fe}}^2 + \tilde{J}_{\text{NiNi}}(\mathbf{k}_X)s_{\text{Ni}}^2\sigma_{\text{Ni}}^2 - 2\tilde{J}_{\text{NiFe}}(\mathbf{k}_X)s_{\text{Ni}}s_{\text{Fe}}\sigma_{\text{Ni}}\sigma_{\text{Fe}}\Big],$$

(II.3)

Aforecited expression for $f$ (as well as a set of equilibrium equations) for f.c.c.-Ni$_{c_{\text{Ni}}}$Fe$_{c_{\text{Fe}}}$ alloy possibly ordered in accordance with $L1_0$ (NiFe) superstructural type (described with the only one ray of star of wave vector—$\mathbf{k}_X = 2\pi(1\,0\,0)$, $2\pi(0\,1\,0)$, or $2\pi(0\,0\,1)$; see Fig. II.1*b*) may also be presented through the replacement of the factor 3/16 by 1/4 within the configurational internal-energy contribution, and both factors 3/4 and 1/4 by 1/2 within the configurational-entropy contribution.

As magnetic phase transition is a critical phase transition (or of the 1-st kind transition close to critical one), the bifurcation points of the solutions of a set of equations for equilibrium values of relative magnetisations almost coincide with a Curie point, $T_C(c_{Fe})$. To find bifurcation points of such set of equations, they are linearized over $\sigma_{Fe}$ and $\sigma_{Ni}$ simultaneously. After application of the Brillouin-function asymptotics, $\sigma_\alpha \cong B_{s_\alpha}(y_\alpha) \approx \frac{s_\alpha + 1}{3 s_\alpha}\left[y_\alpha + O(y_\alpha^3)\right]$, which is true for small $y_\alpha = s_\alpha H_{eff}^\alpha/(k_B T) << 1$, and relationship $\tilde{J}_{\alpha\alpha'}(\mathbf{k}_X) \equiv -\tilde{J}_{\alpha\alpha'}(\mathbf{0})/3$ ($\alpha$, $\alpha'$ = Fe, Ni) valid for the Fourier components, which correspond to $\mathbf{k}$ points such as $2\pi(1\,0\,0)$ and $2\pi(0\,0\,0)$ in reciprocal space of f.c.c. lattice, within the approximation of only nearest-neighbour spins' interaction that allows for short-range nature of an exchange interaction, the expression for $T_C(c_{Fe})$ is obtained as follows:

$$T_C \cong -\frac{1}{6k_B}\Big\{A(1-c_{Fe})\tilde{J}_{NiNi}(\mathbf{0})(1+s_{Ni})s_{Ni} + Bc_{Fe}\tilde{J}_{FeFe}(\mathbf{0})(1+s_{Fe})s_{Fe} -$$

$$-\Big[\Big(A(1-c_{Fe})\tilde{J}_{NiNi}(\mathbf{0})(1+s_{Ni})s_{Ni} - Bc_{Fe}\tilde{J}_{FeFe}(\mathbf{0})(1+s_{Fe})s_{Fe}\Big)^2 + \qquad \text{(II.4)}$$

$$+4C^2(1-c_{Fe})c_{Fe}\tilde{J}_{NiFe}^2(\mathbf{0})(1+s_{Ni})s_{Ni}(1+s_{Fe})s_{Fe}\Big]^{1/2}\Big\},$$

where $A$, $B$, and $C$ are defined as

$$A = 1 - \eta_C^2\Big/\Big[16(1-c_{Fe})^2\Big], \quad B = 1 - \eta_C^2\Big/\Big[16c_{Fe}^2\Big], \quad C = 1 + \eta_C^2\Big/\Big[16(1-c_{Fe})c_{Fe}\Big]$$

in case of $L1_2$-type (Ni$_3$Fe) superstructure (Fig. II.1$a$) with long-range order parameter $\eta = \eta_C$ (at Fe content $c_{Fe}$ and temperature $T = T_C(c_{Fe})$), and exchange-interaction parameters of spin carriers ($\alpha$, $\alpha'$ = Fe, Ni), $\tilde{J}_{\alpha\alpha'}(\mathbf{k}_X)$.

These expressions for $L1_0$ (NiFe)-type superstructure (Fig. II.1$b$) may be obtained through the replacement of factor 1/16 by 1/12 in $A$, $B$, and $C$.

For disordered states, when the optional value of long-range order parameter is $\eta = \eta_C \equiv 0$ at Curie temperature that fits to $c_{Fe}$ (at $T = T_C(c_{Fe})$), the expression for $T_C$ is simplified, because $A$, $B$, and $C$ are equal to unit. In that case, there is a possibility to determine $\tilde{J}_{NiNi}(\mathbf{0})$, $\tilde{J}_{FeFe}(\mathbf{0})$, $\tilde{J}_{NiFe}(\mathbf{0})$ (see Tables II.1, II.2) with the use of experimental dependence (see Fig. II.2$a$) of Curie temperature on the alloy content in the wide $c_{Fe}$ range, which corresponds to disordered alloy.

The best fit of measured [67] and computed Curie temperatures (see Fig. II.2$a$) for this alloy (without long-range order) within the involved concentration range is revealed at values

$$\frac{s_{Ni}(1+s_{Ni})\tilde{J}_{NiNi}(\mathbf{0})}{k_B} \cong -1840 \text{ K}, \quad \frac{s_{Fe}(1+s_{Fe})\tilde{J}_{FeFe}(\mathbf{0})}{k_B} \cong 3202 \text{ K},$$

$$\frac{\sqrt{s_{Ni}(1+s_{Ni})s_{Fe}(1+s_{Fe})}\tilde{J}_{NiFe}(\mathbf{0})}{k_B} \cong -4670 \text{ K},$$

obtained by least squares' optimisation procedure. Remark this as $\tilde{J}_{NiNi}(\mathbf{0})s_{Ni}(1+s_{Ni}) \approx -0.159$ eV, $\tilde{J}_{FeFe}(\mathbf{0})s_{Fe}(1+s_{Fe}) \approx 0.280$ eV, $\tilde{J}_{NiFe}(\mathbf{0})\sqrt{s_{Fe}(1+s_{Fe})s_{Ni}(1+s_{Ni})} \approx -0.402$ eV; so, $\tilde{J}_{NiNi}(\mathbf{k}_X)s_{Ni}(1+s_{Ni}) \approx 0.053$ eV, $\tilde{J}_{FeFe}(\mathbf{k}_X)s_{Fe}(1+s_{Fe}) \approx -0.093$ eV, $\tilde{J}_{NiFe}(\mathbf{k}_X)\sqrt{s_{Fe}(1+s_{Fe})s_{Ni}(1+s_{Ni})} \approx 0.134$ eV or

$$\tilde{J}_{NiNi}(\mathbf{k}_X)s_{Ni}(1+s_{Ni})/k_B \cong 613 \text{ K}, \quad \tilde{J}_{FeFe}(\mathbf{k}_X)s_{Fe}(1+s_{Fe})/k_B \cong -1067 \text{ K},$$

$$\tilde{J}_{NiFe}(\mathbf{k}_X)\sqrt{s_{Fe}(1+s_{Fe})s_{Ni}(1+s_{Ni})}/k_B \cong 1557 \text{ K}.$$

Table II.1. Calculated Fourier components of exchange-interaction energy for hypothetical *high*-spin states of Fe and Ni atoms in disordered f.c.c.-Ni–Fe alloy. To estimate $\tilde{J}_{\alpha\alpha'}(\mathbf{k}_X)$ ($\mathbf{k}_X = 2\pi(1\,0\,0)$), the approximation of nearest-neighbour spins interaction ($\tilde{J}_{\alpha\alpha'}(\mathbf{k}_X) \cong -\tilde{J}_{\alpha\alpha'}(\mathbf{0})/3$) was used.

| $s_{Ni}$ | $s_{Fe}$ | $\tilde{J}_{NiNi}(\mathbf{0})$ [eV] | $\tilde{J}_{NiNi}(\mathbf{k}_X)$ [eV] | $\tilde{J}_{FeFe}(\mathbf{0})$ [eV] | $\tilde{J}_{FeFe}(\mathbf{k}_X)$ [eV] | $\tilde{J}_{NiFe}(\mathbf{0})$ [eV] | $\tilde{J}_{NiFe}(\mathbf{k}_X)$ [eV] |
|---|---|---|---|---|---|---|---|
| 1/2 | 3 | −0.211 | 0.070 | 0.023 | −0.008 | −0.134 | 0.045 |
| 1 | 3 | −0.079 | 0.026 | 0.023 | −0.008 | −0.082 | 0.027 |
| 1/2 | 5/2 | −0.211 | 0.070 | 0.032 | −0.011 | −0.157 | 0.052 |
| 1 | 5/2 | −0.079 | 0.026 | 0.032 | −0.011 | −0.096 | 0.032 |
| 1/2 | 2 | −0.211 | 0.070 | 0.046 | −0.015 | −0.190 | 0.063 |
| 1 | 2 | −0.079 | 0.026 | 0.046 | −0.015 | −0.116 | 0.039 |

### 3.2. Estimation of Total (Paramagnetic + Magnetic) 'Interchange' Energies of Ni and Fe Atoms.

The comparison of experimental data on $T_C$ and neutron diffuse-scattering intensity summarized in Refs [67] (Fig. II.2*a*) and [85], respectively, gives the estimation: $\tilde{w}_{prm}(\mathbf{k}_X) \in (-0.303, -0.300)$ eV (depending on the spin states of Fe and Ni atoms), if 773 K < T < 801 K (see, for instance, Table II.3).

As revealed for *low*-spin states in f.c.c.-Ni–Fe alloy by using the one-relaxation-time approximation [85] for a calculation of equilibrium diffuse-scattering radiation intensity values for a diffraction vector $\mathbf{q} = 2\pi\mathbf{B}+\mathbf{k}$ in the vicinity of reciprocal-lattice (Bragg) point $\mathbf{B}$,

$$I_{diff}(\mathbf{q}, T, c_{Fe}, \infty) \propto D \frac{c_{Fe}(1-c_{Fe})}{1 + c_{Fe}(1-c_{Fe})\dfrac{\tilde{w}_{tot}(\mathbf{k})}{k_B T}} \qquad \text{(II.5)}$$

(where $D$ is normalizing factor; see Refs in [2, 60, 61]), the magnetic contribution to the Fourier component of 'mixing' energy of this alloy,

$$\tilde{w}_{tot}(\mathbf{k}_X) = \tilde{w}_{prm}(\mathbf{k}_X) + \tilde{J}_{FeFe}(\mathbf{k}_X)s_{Fe}^2\sigma_{Fe}^2 + \tilde{J}_{NiNi}(\mathbf{k}_X)s_{Ni}^2\sigma_{Ni}^2 - 2\tilde{J}_{NiFe}(\mathbf{k}_X)s_{Ni}s_{Fe}\sigma_{Ni}\sigma_{Fe}, \qquad \text{(II.6)}$$

Table II.2. Calculated Fourier components of exchange-interaction energy for assumed (realistic) *low*-spin states of Ni and Fe atoms in disordered f.c.c.-Ni–Fe alloy. To estimate $\tilde{J}_{\alpha\alpha'}(\mathbf{k}_X)$ ($\mathbf{k}_X = 2\pi(1\,0\,0)$), the approximation of nearest-neighbour spins' interaction ($\tilde{J}_{\alpha\alpha'}(\mathbf{k}_X) \cong -\tilde{J}_{\alpha\alpha'}(\mathbf{0})/3$) was used.

| $s_{Ni}$ | $s_{Fe}$ | $\tilde{J}_{NiNi}(\mathbf{0})$ [eV] ([K]) | $\tilde{J}_{NiNi}(\mathbf{k}_X)$ [eV] ([K]) | $\tilde{J}_{FeFe}(\mathbf{0})$ [eV] ([K]) | $\tilde{J}_{FeFe}(\mathbf{k}_X)$ [eV] ([K]) | $\tilde{J}_{NiFe}(\mathbf{0})$ [eV] ([K]) | $\tilde{J}_{NiFe}(\mathbf{k}_X)$ [eV] ([K]) |
|---|---|---|---|---|---|---|---|
| 1/2 | 3/2 | −0.211 (−2449) | 0.070 (812) | 0.074 (854) | −0.025 (−285) | −0.240 (−2785) | 0.080 (928) |
| 1/2 | 1 | −0.211 (−2449) | 0.070 (812) | 0.138 (1601) | −0.046 (−534) | −0.329 (−3795) | 0.110 (1265) |
| 1/2 | 1/2 | −0.211 (−2449) | 0.070 (812) | 0.368 (4270) | −0.123 (−1427) | −0.537 (−6232) | 0.179 (2077) |
| ≅ 0.3 | ≅ 1.4 | −0.407 (−4709) | 0.136 (1570) | 0.082 (953) | −0.027 (−318) | −0.352 (−4062) | 0.117 (1358) |

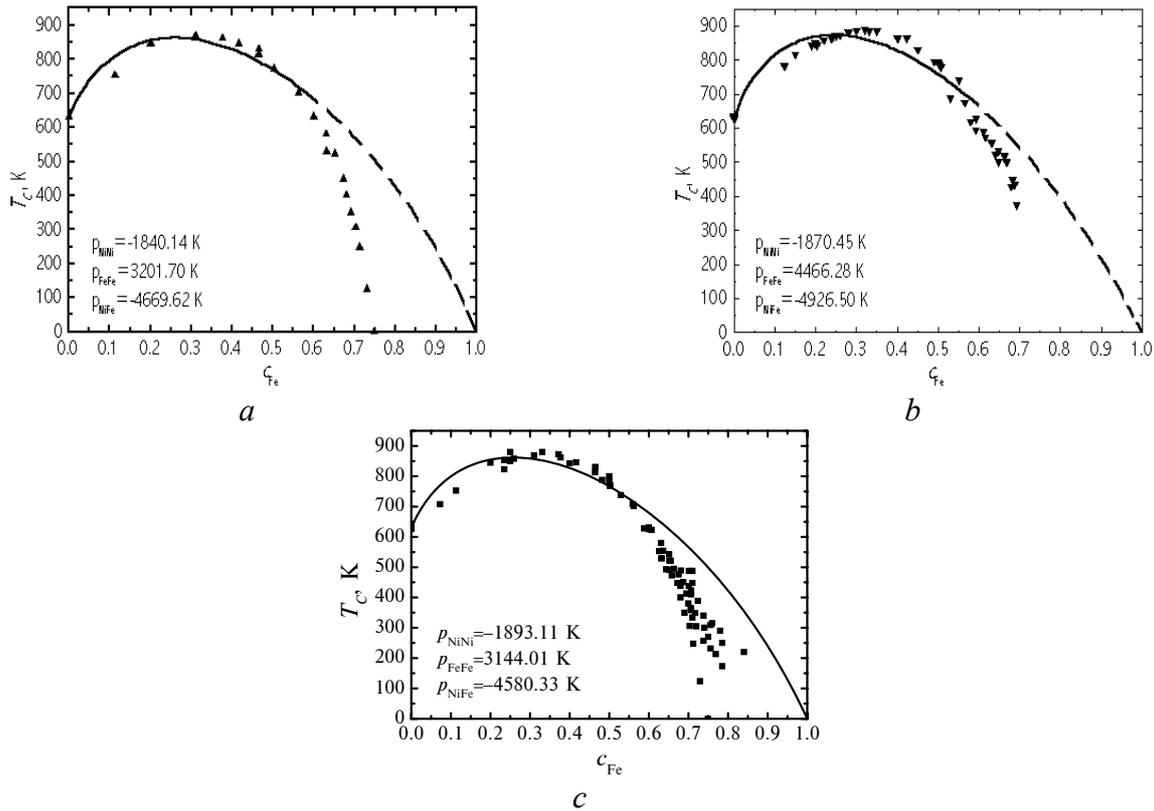

Fig. II.2. Theoretical ('solid' curve) and experimental ('triangles') concentration dependences of Curie temperature of f.c.c.-Ni–Fe alloy: experimental points ('triangles') are taken from [67] (*a*) or [86] (*b*); 'dashed' curve corresponds to the extrapolation of optimising ('solid') curve, obtained by the Levenberg–Marquardt numerical non-linear method (see Refs in [88, 89]), up to right end of concentration segment $0 \le c_{Fe} \le 1$. Theoretical ('solid' curve) and experimental ('small squares') dependences of Curie points, $T_C$ on the Fe concentration $c_{Fe}$, for f.c.c.-Ni–Fe alloy; experimental points ('small squares') are taken from *all* available literary experimental data [67–87] (*c*).

promotes atomic ordering according to the channel $\{\mathbf{k}_X\}$ (see, for instance, Table II.3):

$$T_K \propto -\tilde{w}_{tot}(\mathbf{k}_X)/k_B > -\tilde{w}_{prm}(\mathbf{k}_X)/k_B \ . \tag{II.7}$$

In itself, the presence of atoms with different spin numbers in alloy causes virtually abrupt phase tran-sition from paramagnetic state to the magnetic one (see, for instance, Fig. II.3).

Table II.3. Fourier component of total 'mixing' energy, $\tilde{w}_{tot}(\mathbf{k}_X)$, and 'paramagnetic' component of it, $\tilde{w}_{prm}(\mathbf{k}_X)$, at different temperatures $T \in (T_K, T_C)$ and spins of Ni and Fe atoms for disordered f.c.c.-$^{62}Ni_{0.765}Fe_{0.235}$ solid solution.

| $s_{Ni}$ | $s_{Fe}$ | $T = 773$ K, $\tilde{w}_{tot}(\mathbf{k}_X) \approx -0.341$ eV | | | $T = 776$ K, $\tilde{w}_{tot}(\mathbf{k}_X) \approx -0.338$ eV | | | $T = 783$ K, $\tilde{w}_{tot}(\mathbf{k}_X) \approx -0.334$ eV | | | $T = 801$ K, $\tilde{w}_{tot}(\mathbf{k}_X) \approx -0.325$ eV | | |
|---|---|---|---|---|---|---|---|---|---|---|---|---|---|
| | | $\sigma_{Ni}$ | $\sigma_{Fe}$ | $\tilde{w}_{prm}(\mathbf{k}_X)$ [eV] | $\sigma_{Ni}$ | $\sigma_{Fe}$ | $\tilde{w}_{prm}(\mathbf{k}_X)$ [eV] | $\sigma_{Ni}$ | $\sigma_{Fe}$ | $\tilde{w}_{prm}(\mathbf{k}_X)$ [eV] | $\sigma_{Ni}$ | $\sigma_{Fe}$ | $\tilde{w}_{prm}(\mathbf{k}_X)$ [eV] |
| 1/2 | 1/2 | 0.527 | 0.569 | −0.300 | 0.519 | 0.561 | −0.300 | 0.503 | 0.543 | −0.300 | 0.445 | 0.482 | −0.300 |
| 1/2 | 1 | 0.546 | 0.491 | −0.301 | 0.538 | 0.483 | −0.301 | 0.518 | 0.465 | −0.301 | 0.461 | 0.413 | −0.301 |
| 1/2 | 3/2 | 0.552 | 0.456 | −0.303 | 0.544 | 0.449 | −0.303 | 0.524 | 0.432 | −0.303 | 0.467 | 0.384 | −0.303 |
| '0.3' | 1.4, | 0.580 | 0.411 | −0.302 | 0.571 | 0.404 | −0.302 | 0.551 | 0.389 | −0.302 | 0.491 | 0.345 | −0.302 |

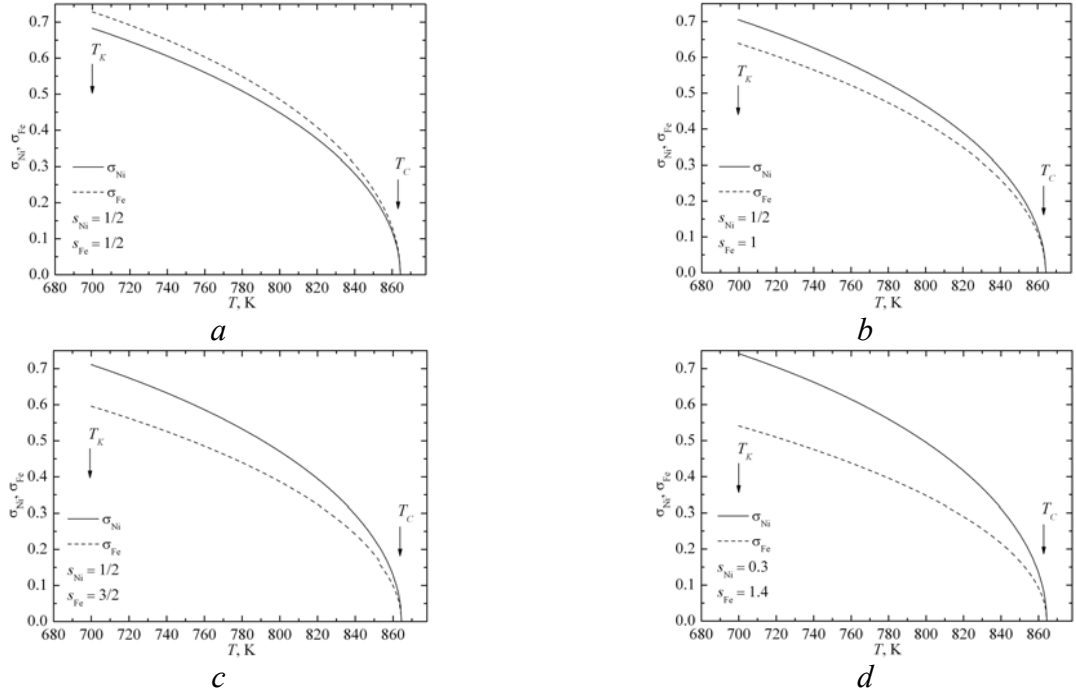

Fig. II.3. Temperature dependences of relative magnetizations, $\sigma_{Ni}$ and $\sigma_{Fe}$, at different spin numbers of Fe and Ni atoms, $s_{Fe}$ and $s_{Ni}$, in disordered f.c.c.-$^{62}Ni_{0.765}Fe_{0.235}$ alloy. For this content, an estimated value of the Curie temperature is as follows: $T_C \cong 865$ K.

As revealed within the framework of approximate representation for $\tilde{w}_{tot}(\mathbf{k}_X)$,

$$\tilde{w}_{tot}(\mathbf{k}_X, T) \cong \tilde{w}_{tot}(\mathbf{k}_X, T_K) + \varpi_1(T - T_K) + \varpi_2(T - T_K)^2 + \varpi_3(T - T_K)^3, \qquad (II.8)$$

in the upper vicinity of $T_K$—Kurnakov's temperature of the order–disorder phase transformation (for f.c.c.-$^{62}Ni_{0.765}Fe_{0.235}$, $T_K \approx 771 \pm 2$ K [85]), $\tilde{w}_{tot}(\mathbf{k}_X, T_K) \approx -0.344$ eV, $\varpi_1 \approx +1.43 \cdot 10^{-3}$ eV/K, $\varpi_2 \approx -7.0 \cdot 10^{-5}$ eV/K$^2$, $\varpi_3 \approx +1.43 \cdot 10^{-6}$ eV/K$^3$. (For reference, see also [124].)

**3.3. Possible Effect of C on the Magnetic-Interaction Energies of Ni and Fe Atoms.** To specify the binary sections of phase diagrams of $Ni_{1-c_{Fe}}Fe_{c_{Fe}}$(C) systems based on f.c.c. crystal lattice, the comparison of derived analytical relationships with available literary experimental data [67–87] (see Fig. II.4) of the Curie temperature of disordered f.c.c.-$Ni_{1-c_{Fe}}Fe_{c_{Fe}}$(C) alloys was carried out ($c_{Fe}$—relative Fe con-tent, $T_K$—the Kurnakov's temperature of atomic order–disorder transformation), *i.e.* for the temperatures, which are higher than upper absolute instability limit $T_c^+$ ($T > T_c^+ > T_K$).

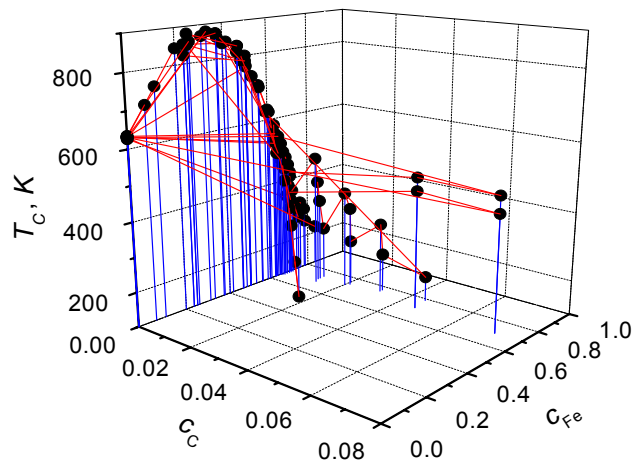

Fig. II.4. Experimental dependence ('circles' [67–87]) of Curie temperature, $T_C$, for f.c.c.-Ni–Fe alloy on the Fe concentration, $c_{Fe}$, and interstitial-impurity (C) concentration, $c_C$.

Table II.4. Predicted (and recommended) Fourier components of exchange-interaction energy for different assumed (realistic) *low*-spin states of Ni and Fe atoms in disordered f.c.c.-Ni–Fe alloy ($c_C = 0$). To estimate $\tilde{J}_{\alpha\alpha'}(\mathbf{k}_X)$ ($\mathbf{k}_X = 2\pi(1\,0\,0)$), the approximation of nearest-neighbour spins' interaction was used ($\tilde{J}_{\alpha\alpha'}(\mathbf{k}_X) \cong -\tilde{J}_{\alpha\alpha'}(\mathbf{0})\,/\,3$).

| $s_{Ni}$ | $s_{Fe}$ | $\tilde{J}_{NiNi}(\mathbf{0})$ [eV] ([K]) | $\tilde{J}_{NiNi}(\mathbf{k}_X)$ [eV] ([K]) | $\tilde{J}_{FeFe}(\mathbf{0})$ [eV] ([K]) | $\tilde{J}_{FeFe}(\mathbf{k}_X)$ [eV] ([K]) | $\tilde{J}_{NiFe}(\mathbf{0})$ [eV] ([K]) | $\tilde{J}_{NiFe}(\mathbf{k}_X)$ [eV] ([K]) |
|---|---|---|---|---|---|---|---|
| 1/2 | 3/2 | −0.218 (−2524) | 0.073 (841) | 0.072 (838) | −0.024 (−279) | −0.235 (−2731) | 0.078 (910) |
| 1/2 | 1/2 | −0.218 (−2524) | 0.073 (841) | 0.361 (4192) | −0.120 (−1397) | −0.526 (−6107) | 0.175 (2036) |

In terms of the Fourier components of exchange-interaction energies for Ni and Fe atoms, the expression for Curie temperature may be presented as follows:

$$T_C \cong -\frac{p_{NiNi}(1-c_{Fe}) + p_{FeFe}c_{Fe} - \sqrt{\left(p_{NiNi}(1-c_{Fe}) - p_{FeFe}c_{Fe}\right)^2 + 4p_{NiFe}^2(1-c_{Fe})c_{Fe}}}{6}, \quad (II.9)$$

where, instead the Fourier components of exchange-interaction energies, $\tilde{J}_{NiNi}(\mathbf{0})$, $\tilde{J}_{FeFe}(\mathbf{0})$, $\tilde{J}_{NiFe}(\mathbf{0})$,

$$p_{NiNi} = \frac{s_{Ni}(1+s_{Ni})\tilde{J}_{NiNi}(\mathbf{0})}{k_B}, \quad p_{FeFe} = \frac{s_{Fe}(1+s_{Fe})\tilde{J}_{FeFe}(\mathbf{0})}{k_B}, \quad p_{NiFe} = \frac{\sqrt{s_{Ni}(1+s_{Ni})s_{Fe}(1+s_{Fe})}\,\tilde{J}_{NiFe}(\mathbf{0})}{k_B}$$

(see Fig. II.2) are the appropriate parameters of exchange interactions of atomic spins.

For f.c.c.-Ni–Fe alloy doped with interstitial C (see Fig. II.4), new parameters ($p'_{NiNi}, p'_{FeFe}, p'_{NiFe}$) of exchange interaction must be used (for $c_C \ll 1$),

$$p_{NiNi} \rightarrow p'_{NiNi} \cong p_{NiNi} + K_1 + K_2 c_C, \quad p_{FeFe} \rightarrow p'_{FeFe} \cong p_{FeFe} + K_4 + K_5 c_C, \quad p_{NiFe} \rightarrow p'_{NiFe} \cong p_{NiFe} + K_7 + K_8 c_C.$$

This analysis enables to get the updated estimations of parameters of exchange-interaction energies for Ni and Fe atoms. If benchmark values of $p_{NiNi}, p_{FeFe}, p_{NiFe}$ are implied as presented in Fig. II.2*a*, the fitting parameters $K_1, K_2, K_4, K_5, K_7, K_8$ are as follows: $K_1 \approx -52.97$ K, $K_2 \approx 3.85 \cdot 10^5$ K, $K_4 \approx -57.69$ K, $K_5 \approx 1.17 \cdot 10^7$ K, $K_7 \approx 89.29$ K, $K_8 \approx -2.16 \cdot 10^6$ K (see also Fig. II.2c with $p'_{NiNi}, p'_{FeFe}, p'_{NiFe}$ at $c_C = 0$).

Table II.5. Forecasted Fourier components of exchange-interaction energy for different assumed (realistic) *low*-spin states of Ni and Fe atoms in disordered low-C f.c.c.-Ni–Fe alloy ($c_C = 0.00071$). To estimate $\tilde{J}_{\alpha\alpha'}(\mathbf{k}_X)$ ($\mathbf{k}_X = 2\pi(1\,0\,0)$), as before, the approximation of nearest-neighbour spins' interaction was used ($\tilde{J}_{\alpha\alpha'}(\mathbf{k}_X) \cong -\tilde{J}_{\alpha\alpha'}(\mathbf{0})\,/\,3$).

| $s_{Ni}$ | $s_{Fe}$ | $\tilde{J}_{NiNi}(\mathbf{0})$ [eV] ([K]) | $\tilde{J}_{NiNi}(\mathbf{k}_X)$ [eV] ([K]) | $\tilde{J}_{FeFe}(\mathbf{0})$ [eV] ([K]) | $\tilde{J}_{FeFe}(\mathbf{k}_X)$ [eV] ([K]) | $\tilde{J}_{NiFe}(\mathbf{0})$ [eV] ([K]) | $\tilde{J}_{NiFe}(\mathbf{k}_X)$ [eV] ([K]) |
|---|---|---|---|---|---|---|---|
| 1/2 | 3/2 | −0.186 (−2159) | 0.062 (720) | 0.262 (3046) | −0.087 (−1015) | −0.314 (−3647) | 0.105 (1216) |
| 1/2 | 1/2 | −0.186 (−2159) | 0.062 (720) | 1.312 (15229) | −0.437 (−5076) | −0.703 (−8155) | 0.234 (2718) |

Table II.6. Exchange-interaction energies, $J_{\alpha\alpha'}(r_1 = a/\sqrt{2}) \cong \tilde{J}_{\alpha\alpha'}(\mathbf{0})/12$, for different assumed nearest-neighbouring *low* spins of Ni and Fe atoms in disordered plain f.c.c.-Ni–Fe alloy ($c_C = 0$), estimated within the framework of the approximation of nearest-neighbour spins' interaction.

| $s_{Ni}$ | $s_{Fe}$ | $J_{NiNi}(r_1)$ [meV] ([K]) | $J_{FeFe}(r_1)$ [meV] ([K]) | $J_{NiFe}(r_1)$ [meV] ([K]) |
|---|---|---|---|---|
| 1/2 | 3/2 | −18 (−210) | 6 (70) | −20 (−228) |
| 1/2 | 1/2 | −18 (−210) | 30 (349) | −44 (−509) |

The appropriate predicted values $\tilde{J}_{\alpha\alpha'}(\mathbf{k}_X)$ and $\tilde{J}_{\alpha\alpha'}(\mathbf{0})$ —Fourier components of the exchange-interaction energies of spin carriers ($\alpha$, $\alpha'$ = Fe, Ni) in assumed *low*-spin states—at two C concentrations ($c_C = 0$ and $c_C = 0.00071$) are presented in Tables II.4 and II.5. (For reference, see also [124].)

Thus, due to the optimisation procedure of fitting parameters of C effect within the framework of the above-mentioned approximation, the following hypothetical conclusion is obtained: because of implicit relation of exchange-interaction parameters to the C concentration, $c_C$, by virtue of relation of these parameters to interspin distance and redistribution of electrons between components, the embedding of small amount of interstitial C atoms results in both decrease of ferromagnetic contribution to interspin Ni–Ni interaction and respective increase of ferromagnetic contribution to interspin Ni–Fe interaction, while the antiferromagnetic contribution to interspin Fe–Fe interaction is increasing.

The spins of nearest-neighbour $\alpha$ and $\alpha'$ atoms (in a direct space of f.c.c. lattice) interact with energy, which may incidentally be estimated as $J_{\alpha\alpha'}(r_1 = a/\sqrt{2}) \cong \tilde{J}_{\alpha\alpha'}(\mathbf{0})/12$ ($a$—the parameter of conditional unit cell of f.c.c. lattice) within the scope of the approximation of only nearest-neighbour spin–spin interaction. Respective values $J_{\alpha\alpha'}(r_1)$ ($\alpha$, $\alpha'$ = Ni, Fe) are presented in Table II.6 and may be compared with available evaluations based on rough and contradictory experimental data (see Table II.7 with an allowance for almost generally accepted definition of signs of 'integrals' $J_{\alpha\alpha'}(r_1)$ of exchange interaction, which, however, does not respond real energies of antiferromagnetic interaction of oppo-

Table II.7. Exchange-interaction 'integrals', $J_{\alpha\alpha'}(r_1)$, for nearest-neighbouring spins of Ni and Fe atoms in disordered plain f.c.c.-Ni–Fe alloy ($c_C = 0$).

| $c_{Fe}$ | $J_{NiNi}(r_1)$ [meV] | $J_{NiFe}(r_1)$ [meV] | $J_{FeFe}(r_1)$ [meV] | References |
|---|---|---|---|---|
| $c_{Fe} \in [0, 1]$ | −52 ± 3 | −39 ± 5 | 9 ± 2.6 | [90] |
| $c_{Fe} \in [0, 1]$ | −34.9 | −24.1 | 1.7 | [64] |
| $c_{Fe} \in [0, 1]$ | −60.3 | −30.6 | 2.2 | [65] |
| $c_{Fe} \in [0, 1]$ | $1 \cdot J_{NiNi}(r_1)$ | $0.93 \cdot J_{NiNi}(r_1)$ | $-0.05 \cdot J_{NiNi}(r_1)$ | [91] |
| $c_{Ni} \in [0, 1]$ | −57 | −34 | 6 | [76] (stress-free films, bulk specimens) |
| $c_{Fe} \approx 0.75$ | −22 | −22 | 5 | [92] |
| $c_{Fe} \approx 0.675$ | −66±13 | 0 | −9.2±1.8 | [93] (strained films) |
| $c_{Fe} \approx 0.5$ | −30 | −30 | 4 | [94] |
| $c_{Fe} \approx 0.5$ | −22 | −42 | 5 | [92] |
| $c_{Fe} \approx 0.25$ | −22 | −45 | 5 | [92] |
| $c_{Fe} \approx 0.2$ | −58.5 | −2.55 | 23.3 | [94] |
| $c_{Fe} \approx 0.0$ | −16.7 | — | — | [77] |
| $c_{Fe} \approx 0.0$ | −17.5 | — | — | [77] |
| $c_{Fe} \approx 0.0$ | −22 | — | — | [92] |

sitely directed and ferromagnetic interaction of like-directed spins [2, 60–62]). (See also Ref. [124].)

**Part III. Study of Atomic Ordering and Magnetic Phase Transitions in F.C.C.-Ni–Fe–(C) Alloys**

## 4. Calculations of the Short-Range and Long-Range Order Parameters

### 4.1. Equilibrium Short-Range and Long-Range Atomic Order Parameters of F.C.C.-Ni–Fe Alloys.
There may be used the Khachaturyan–Cook microscopic approach [1], which relates the annealing-time-dependent short-range order parameter, $\alpha(\mathbf{k}, T, c_{Fe}, t)$, of solid solution to the data of measurements of the dependence of diffuse-scattering intensity of radiations (thermal neutrons, x-rays *etc.*), $I_{diff}(2\pi\mathbf{B}+\mathbf{k}, T, c_{Fe}, t)$, on time $t$ (see also [85, 107]). The kinetics of relaxation of the above mentioned intensity for binary solid solution ($I_{diff}(2\pi\mathbf{B}+\mathbf{k}, c_{Fe}, t) \propto \alpha(T, c_{Fe}, t)$) may be described within discrete kinetic models with one, two or three relaxation times [2, 60, 100]. In this work, we consider explicitly the kinetics of diffuse intensity variation for different temperatures and several reciprocal-space vectors in the vicinity of the special point $\mathbf{k}_X = 2\pi(1\,0\,0)$, where the intensity reaches a peak value.

Within the framework of the one-relaxation-time model [1, 85] and with some simplifying assumptions, the relaxation time $\tau_0(\mathbf{k})$ is approximately proportional to the square of the short-range order correlation length, $\xi \propto a T_K^{1/2}|T_K - T|^{-1/2}$, and inversely proportional to the 'interdiffusion' coefficient, $D_0$: $\tau_0(\mathbf{k}) \propto \xi^2/D_0$. As the temperature decreases, $\xi$ increases, so that $\tau_0(\mathbf{k})$ increases more fast than the one elementary-diffusion jump time, $a^2/(8D_0)$. On the other hand, for f.c.c.-Ni–Fe alloy with a (1 0 0)-type short-range order, the reciprocal-space point (1 0 0) is the node where the weight of the short-range order parameters for the more distant atomic pairs is the largest. Consequently, in the vicinity of (1 0 0), the relaxation time must be longest at $\mathbf{k}_X = 2\pi(1\,0\,0)$ and reach up to tens–hundreds (90–200) minutes.

For the estimation of $I_{diff}(2\pi\mathbf{B}+\mathbf{k}, T, c_{Fe}, \infty)$—equilibrium ($t \to \infty$) values of radiation diffuse-scattering intensity, it is possible to evaluate the equilibrium short-range order parameters [1, 2, 100, 101] of disordered f.c.c.-Ni–Fe phase for different $T$, $c_{Fe}$, and a given wave vector $\mathbf{k}_X = 2\pi(1\,0\,0)$ by using the substitution of numerical relation (II.8) for $\tilde{w}_{tot}(\mathbf{k}_X)$ (taking into account the temperature–concentration dependence of it (due to $\sigma_{Fe}$ and $\sigma_{Ni}$)), *e.g.*, into the Krivoglaz–Clapp–Moss formula (see Refs in [2, 61, 100, 101]),

$$\alpha(\mathbf{k}_X, T, c_{Fe}, \infty) \cong \frac{D}{1 + c_{Fe}(1 - c_{Fe})\dfrac{\tilde{w}_{tot}(\mathbf{k}_X)}{k_B T}}, \qquad (\text{III.1})$$

provided that $D \cong 1 \pm 0.2$, for instance,

$$\alpha(\mathbf{k}_X, T, c_{Fe}, \infty) \cong$$
$$\cong \frac{1}{1 + c_{Fe}(1 - c_{Fe})\dfrac{-0.344 \text{ eV} + \varpi_1(T - 771 \text{ K}) + \varpi_2(T - 771 \text{ K})^2 + \varpi_3(T - 771 \text{ K})^3}{k_B T}} \qquad (\text{III.2})$$

at $T \geq T_K$, where $\varpi_1 \approx +1.43 \cdot 10^{-3}$ eV/K, $\varpi_2 \approx -7.0 \cdot 10^{-5}$ eV/K², $\varpi_3 \approx +1.43 \cdot 10^{-6}$ eV/K³.

It is necessary to note for further comparison of methods that alternating and rather slowly drop-down curve $\alpha(\mathbf{r})$ values (with increasing $|\mathbf{r}|$) [102] is obtained also in rather precise neutron-diffraction study of disordered $^{62}$Ni$_{0.765}$Fe$_{0.235}$ alloy quenched at 808 K [103]. The yielded $\alpha(\mathbf{r})$ values of short-range order parameters for 22 co-ordination shells appeared insufficient for synthesizing the experimental $I_{diff}(\mathbf{q})$ curve. Therefore, it was necessary to take into account all data obtained for 34 shells [103]. The annealing of the $^{62}$Ni$_{0.765}$Fe$_{0.235}$ alloy at temperatures (780 K, 808 K, 958 K), which exceed the critical ordering temperature, $T_K \approx 773$ K, promotes attainment of the equilibrium short-range or-der. The short-term annealing at $T < T_K$ (1800 sec at $T = 745$ K and 3600 sec at $T = 658$ K) promotes formation of the transition short-range order before direct formation of the long-range order [103, 105].

It is significant that the Invar Fe–Ni-alloy was also studied with neutron diffraction in [106]. The obtained values of the short-range order parameters for the first two co-ordination shells are as follows: $\alpha_1 = -0.006$, $\alpha_2 = 0.006$ for Fe–30 at.% Ni with annealing temperature $T_a = 773$ K for 4380 minutes; $\alpha_1 = -0.010$, $\alpha_2 = 0.012$ for Fe–32 at.% Ni annealed at $T_a = 753$ K for 6480 minutes; $\alpha_1 = -0.018$, $\alpha_2 = 0.018$ for Fe–35 at.% Ni ($T_a = 723$ K for 22200 minutes' annealing); $\alpha_1 = -0.022$, $\alpha_2 = 0.024$ for Fe–40 at.% Ni ($T_a = 683$ K for 34800 minutes' annealing).

The negative $\alpha_1 = \alpha(110)$ value (Table III.1) for f.c.c.-Ni–Fe solid solution testifies that such alloy is the system with the short-range order, in which the nearest neighbours are mainly the atoms of different components. However, this tendency is not considerable. The probability that Ni atom is a nearest neighbour of Fe, $P_{Ni|Fe}(110)$, is defined by the formula $P_{Ni|Fe}(\mathbf{r}) = c_{Ni}\{1 - \alpha(\mathbf{r})\}$; thus, $\alpha(110) = 0$, if the probability $P_{Ni|Fe}(110)$ is equal to concentration of Ni—$c_{Ni}$. As $\alpha(110) \cong -0.1$ (Table III.1), the probability that Ni atom is a nearest neighbour of Fe, $P_{Ni|Fe}(110)$, is greater approximately by 10% than in disordered alloy (without the long-range order). When increasing the Fe concentration from 22.5

Table III.1. Short-range order parameters, $\alpha(\mathbf{r})$, for different co-ordination shells $\{n\}$ in the completely ordered f.c.c.-Ni$_3$Fe alloy with $L1_2$-type structure and some disordered (without the long-range order) f.c.c.-Ni–Fe alloys studied with the use of x-ray diffuse scattering [70, 71] (see Refs in [100]).

| | | $\alpha(\mathbf{r})$ | | | |
|---|---|---|---|---|---|
| $n$ | $\mathbf{r}$ | Ordered Ni$_3$Fe [71, 103] | Ni$_{0.775}$Fe$_{0.225}$ 1273 K [71] | Ni$_{0.535}$Fe$_{0.465}$ 1273 K [71] | Ni$_{0.368}$Fe$_{0.632}$ 753 K [70] |
| | 000 | 1 | 1.0012 | 1.0000 | 1.0020 |
| I | 110 | −1/3 | −0.1082 | −0.0766 | −0.0580 |
| II | 200 | 1 | 0.1194 | 0.0646 | 0.0520 |
| III | 211 | −1/3 | −0.0047 | −0.0022 | −0.0030 |
| IV | 220 | 1 | 0.0307 | 0.0037 | 0.0000 |
| V | 310 | −1/3 | −0.0179 | −0.0100 | −0.0060 |
| VI | 222 | 1 | 0.0130 | 0.0037 | |
| VII | 321 | −1/3 | −0.0075 | −0.0030 | |
| VIII | 400 | 1 | 0.0173 | 0.0071 | |
| IXa | 330 | −1/3 | 0.0005 | −0.0021 | |
| IXb | 411 | −1/3 | 0.0046 | 0.0007 | |
| X | 420 | 1 | 0.0048 | 0.0012 | |
| XI | 332 | −1/3 | −0.0032 | −0.0007 | |
| XII | 422 | 1 | 0.0030 | −0.0003 | |
| XIIIa | 510 | −1/3 | −0.0025 | 0.0002 | |
| XIIIb | 431 | −1/3 | −0.0023 | −0.0008 | |
| XIV | 521 | −1/3 | −0.0025 | −0.0011 | |
| XV | 440 | 1 | 0.0023 | 0.0007 | |
| XVIa | 530 | −1/3 | −0.0042 | −0.0002 | |
| XVIb | 433 | −1/3 | −0.0017 | 0.0003 | |
| XVII | 600 | 1 | 0.0087 | −0.0037 | |

at.% up to 46.5 at.%, the short-range atomic order is decreased in the absence of the long-range order (Table III.1): for single crystal $Ni_{0.775}Fe_{0.225}$ $T_K \approx 773$ K [71], for $^{62}Ni_{0.765}Fe_{0.235}$ $T_K \approx 771\pm2$ K [85], for $Ni_3Fe$ $T_K \approx 789$ K (808$\pm$2 K) [108] ([109]), and for $Ni_{0.535}Fe_{0.465}$ $T_K \approx 573$ K [71]. This tendency is expected [71], as temperature of the disorder–order phase transformation is maximal in $Ni_3Fe$ alloy, and for increasing Fe concentration, the $T_K$ is lowered.

After annealing at $T_a = 753$ K, the disordered f.c.c.-$Ni_{0.368}Fe_{0.632}$ alloy was quenched to the room-temperature point [70]. Therefore, the atomic diffusion has been rather slow therein. It means that the solid solution should be in a metastable homogeneous single-phase Invar state [70] (its $T_K$ can, at least, be less than 683–693 K [110] in a state previously unirradiated with neutrons or electrons; see Refs in [71, 111]). The measurement was carried out at 293 K, 60 K, and at three selected x-ray energy values for intensification of the contrast of Fe and Ni scattering to magnify the sensitivity of measurements to local atomic structure [70a, 71]. The first six short-range order parameters in disordered $Ni_{0.368}Fe_{0.632}$ alloy [70] are also presented in Table III.1. The fact that $\alpha(000) \approx 1.0$ means that the data obtained at three x-ray energies are self-consistent, *i.e.* normalizing condition for the short-range order parameter $\alpha_0$ at $\mathbf{r} = \mathbf{0}$ is satisfied. Both values $\alpha(110)$ and $\alpha(200)$ reveal considerable difference of the state of alloy from disordered one, while the other $\alpha(\mathbf{r})$ values tend to 0 (Table III.1). The minor tendency of atoms to redistribution is observed, and by that, the nearest neighbours are the atoms of different components, and the next nearest neighbours are the atoms of the same component that specifies $L1_0$(NiFe)-type ordering and/or $L1_2$($Ni_3Fe$)-one [70]. This tendency, concerning the nearest atomic neighbours of different components, manifests itself to a greater extent in Ni–Fe alloy with more high Ni abundance [71].

The short-range order parameters, $\alpha(\mathbf{r})$, obtained during measurements at 60 K, coincide with those obtained for measurements at 297 K, because of a slow diffusion at so low temperatures.

For comparison, it should also be noted that the detailed investigation of $Ni_3Fe$ alloy, prepared through mechanical alloying (or high-energy compaction), was carried out in [112]. Apparently, the process of atomic ordering is diffusion-based and faster in a nanocrystalline form of $Ni_3Fe$ [112]. It is due to the high diffusive mobility of atoms in nanocrystalline materials.

Within the SCFA, the statistical thermodynamics of the long-range atomic order of f.c.c.-Ni–Fe alloy is completely determined by two energy parameters—$\widetilde{w}_{tot}(\mathbf{k}_X)$ and $\widetilde{w}_{tot}(\mathbf{0})$, where $\mathbf{k}_X = 2\pi(1\,0\,0)$ and $\mathbf{k}_0 = \mathbf{0}$ are the wave vectors that correspond to the high-symmetry reciprocal-space points $X$ and $\Gamma$ of f.c.c. alloy, respectively; $(1\,0\,0)$ is the half-translation of reciprocal b.c.c. lattice (for f.c.c. crystal) lying along $[1\,0\,0]$ direction.

Thus, one can get the configurational free energy of f.c.c.-Ni–Fe alloy *per* site, $f = u - Ts$ ($u$—internal energy *per* site, $s$—entropy *per* site; see (II.2)), and then determine fundamental thermodynamic characteristics of equilibrium alloy, including the jump of long-range atomic-order parameter $\Delta\eta(T_K, c_{Fe})$ ($< 0.47$) (see Ref. [101] and Fig. III.1) at $T_K(c_{Fe})$, the reduced concentration-dependent order($L1_2$)–disorder($A1$) transformation temperature $\tau_K = k_B T_K / \left| \widetilde{w}_{tot}(\mathbf{k}_X) \right|$ (see Figs. III.2 and III.6 below) *etc.*

It is significant that the disorder–order phase transformation into the 'high-symmetry' $L1_2$($Ni_3Fe$)-type state is transition of the 1-st kind. This means that Kurnakov's disorder–order temperature, $T_K$, can be estimated within the SCFA from Eqs. (II.2), (II.3), and the following condition $\Delta f(\Delta\eta, T_K) = 0$, *i.e.*

$$0 = \Delta f(\Delta\eta, T_K) = f(\Delta\eta, T_K) - f(0, T_K) \cong$$

$$\cong \frac{3}{32} \Delta\eta^2 \widetilde{w}_{tot}(\mathbf{k}_X) + \frac{1}{4} k_B T_K \left( 3\left( c_{Fe} - \Delta\eta/4 \right) \ln\left( c_{Fe} - \Delta\eta/4 \right) + 3\left( 1 - c_{Fe} + \Delta\eta/4 \right) \ln\left( 1 - c_{Fe} + \Delta\eta/4 \right) + \right.$$
$$+ \left( c_{Fe} + 3\Delta\eta/4 \right) \ln\left( c_{Fe} + 3\Delta\eta/4 \right) + \left( 1 - c_{Fe} - 3\Delta\eta/4 \right) \ln\left( 1 - c_{Fe} - 3\Delta\eta/4 \right) \right) -$$
$$- k_B T_K \left( c_{Fe} \ln c_{Fe} + \left( 1 - c_{Fe} \right) \ln\left( 1 - c_{Fe} \right) \right).$$

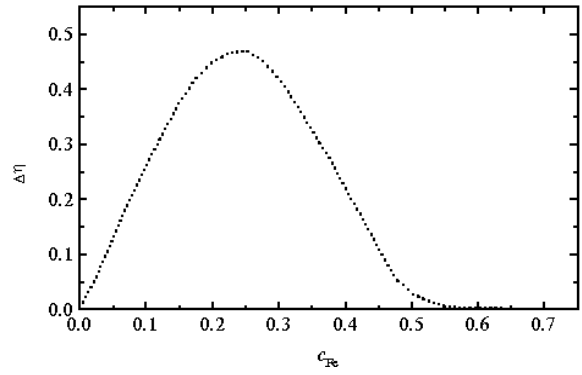

Fig. III.1. Calculated dependency of the jump of long-range order parameter, $\Delta\eta$, at $T_K = T_K(c_{Fe})$ on the Fe content, $c_{Fe}$, for the disordering f.c.c.-Ni–Fe alloy with $L1_2$-type (Ni$_3$Fe) structure.

Thus, to obtain the jump of the long-range order parameter for Ni$_3$Fe-type structure, one has to use the numeric computing to solve the following equation:

$$\frac{3}{32}\Delta\eta \ln\frac{(c_{Fe}-\Delta\eta/4)(1-c_{Fe}-3\Delta\eta/4)}{(c_{Fe}+3\Delta\eta/4)(1-c_{Fe}+\Delta\eta/4)} + \frac{1}{4}\big(3(c_{Fe}-\Delta\eta/4)\ln(c_{Fe}-\Delta\eta/4)+$$

$$+3(1-c_{Fe}+\Delta\eta/4)\ln(1-c_{Fe}+\Delta\eta/4)+(c_{Fe}+3\Delta\eta/4)\ln(c_{Fe}+3\Delta\eta/4))$$

$$+(1-c_{Fe}-3\Delta\eta/4)\ln(1-c_{Fe}-3\Delta\eta/4)\big)-c_{Fe}\ln c_{Fe}-(1-c_{Fe})\ln(1-c_{Fe}) \approx 0$$

and substitute the solution of it (*i.e.* $\Delta\eta(c_{Fe})$ dependence) for the following approximate expression for the reduced Kurnakov's temperature:

$$\tau_K \cong \Delta\eta(c_{Fe})\left(\ln\frac{(c_{Fe}-\Delta\eta(c_{Fe})/4)(1-c_{Fe}-3\Delta\eta(c_{Fe})/4)}{(c_{Fe}+3\Delta\eta(c_{Fe})/4)(1-c_{Fe}+\Delta\eta(c_{Fe})/4)}\right)^{-1}.$$

The comparatively low value of the jump of the long-range order parameter at $T_K$ (*i.e.* $\Delta\eta(T_K, c_{Fe}) < 0.47$) for Ni$_3$Fe-type superstructure is evidence of stability of this long-range ordered phase with respect to the disordered phase (with short-range order only) and justifies, in some way, the neglect of the correlation effects in spatial locations of component atoms (at least in the vicinity of Kurnakov's temperature) at statistical-thermodynamic consideration of this superstructure.

It is significant that the SCFA is asymptotically correct in the limiting cases of high and low temperatures but not correct in the range within the vicinity of the point of the absolute loss of stability of a disordered phase (see Refs in [1, 2]), which is slightly less than the 1-st kind transformation tempera-ture (see Refs in [1, 2, 101, 102, 109]). However, the larger the radius of interatomic interactions, the smaller the width of this range (see Refs in [1, 2, 95, 102, 109]). Thus both long-range action in 'interchange' energy of alloy and decay of correlation effects strengthen non-linear character of dependence $\eta(T/T_K)$ and approach the $L1_2$–$A1$ phase transformation of Ni–Fe alloy, at least, to 'anomalous' phase transition of the 1-st kind [107, 109].

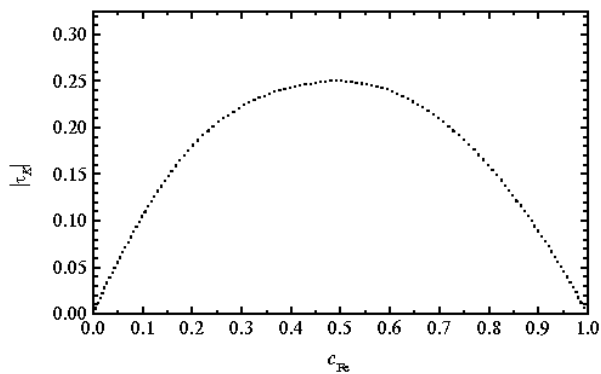

Fig. III.2. Calculated dependence of reduced Kurnakov's temperature (modulo), $|\tau_K|$, for the 1-st kind $A1$–$L1_2$ transformation on the Fe content, $c_{Fe}$, in f.c.c.-Ni–Fe alloy.

Within the framework of the SCFA, respective curves $\Delta\eta(T_K, c_{Fe})$ and $\tau_K(c_{Fe})$ for the 1-st kind order($L1_0$)–disorder($A1$) transformation (close to the 'critical' phase transition) of f.c.c.-Ni–Fe alloy have more trivial approximate forms—$\Delta\eta(T_K, c_{Fe}) \cong 0$ and $\tau_K(c_{Fe}) \cong c_{Fe}(1 - c_{Fe})$ [1, 2, 101].

The temperature-dependence plots of the equilibrium long-range order parameter $\eta(T, c_{Fe})$ in the vicinity of order–disorder transformation point of f.c.c.-Ni–Fe alloy with allowance for possible effect of small C additions on them may also be calculated and are presented in subsection 5.2 (see Fig. III.7).

**4.2. Kinetics of $L1_2$-Type Long-Range Atomic Ordering During Annealing of Ni–Fe Alloys.** The advantages of Khachaturyan–Cook microscopic approach [1] can be considered as relating also to the annealing-time dependence of the long-range order on the data of measurements of time dependence of radiation intensity (x-rays, thermal neutrons *etc.*) on diffraction pattern for a solid solution (if such data would certainly be available). The kinetics of a relaxation of above-mentioned coherent-scattering intensity for a binary f.c.c.-Ni–Fe solution, where $L1_2$-type ordering is generated with the three-ray star $\{1\,0\,0\}$ ($L1_0$-type ordering is generated with only one superlattice vector that enters this star), can be described with the long-range order kinetic approximation based on the Önsager kinetic equations for discrete diffusion relaxation of non-equilibrium single-site occupation probabilities of finding a Fe atom (at lattice sites $\{\mathbf{R}\}$ and any moment of time $t$),

$$P_{Fe}(\mathbf{R}, t) = c_{Fe} + \frac{1}{4}\eta(t)(e^{i2\pi\mathbf{a}_1^*\cdot\mathbf{R}} + e^{i2\pi\mathbf{a}_2^*\cdot\mathbf{R}} + e^{i2\pi\mathbf{a}_3^*\cdot\mathbf{R}}), \qquad (III.3)$$

which corresponds to the superlattice wave vector $\mathbf{k}_X = 2\pi(1\,0\,0) = 2\pi\mathbf{a}_1^*$ [1, 2]. In that case, the derivative of free-energy $\delta F_{conf}/\delta P_{Fe}(\mathbf{R}, t)$ as well as the $\Delta P_{Fe}(\mathbf{R}, t) = \frac{1}{4}\eta(t)(e^{i2\pi\mathbf{a}_1^*\cdot\mathbf{R}} + e^{i2\pi\mathbf{a}_2^*\cdot\mathbf{R}} + e^{i2\pi\mathbf{a}_3^*\cdot\mathbf{R}})$ may be represented as a superposition of the concentration waves:

$$\delta F_{conf}/\delta P_{Fe}(\mathbf{R}, t) = \hat{c}(c_{Fe}, \eta(t)) + \hat{\eta}(c_{Fe}, \eta(t))E(\mathbf{R}), \quad P_{Fe}(\mathbf{R}, t) = c_{Fe} + \Delta P_{Fe}(\mathbf{R}, t) = c_{Fe} + \eta(t)E(\mathbf{R}) \quad (III.4)$$

where $E(\mathbf{R}) = \frac{1}{4}(e^{i2\pi\mathbf{a}_1^*\cdot\mathbf{R}} + e^{i2\pi\mathbf{a}_2^*\cdot\mathbf{R}} + e^{i2\pi\mathbf{a}_3^*\cdot\mathbf{R}})$, and the time-dependent parameters $\hat{c}(c_{Fe}, \eta)$ and $\hat{\eta}(c_{Fe}, \eta)$ can be expressed with actual long-range order parameter $\eta(t)$ and composition $c_{Fe}$ by substituting the last expression (III.4) for the left-hand side of previous equality and comparing the result with its right-hand side, as the $\delta F_{conf}/\delta P_{Fe}(\mathbf{R}, t)$ value possesses the same crystallographic symmetry as function $P_{Fe}(\mathbf{R}, t)$ [1, 2] for the superstructure involved.

Substitution of the last relations (III.4) into the Önsager kinetic equations for single-site probability [1], followed by the Fourier transformation, gives the equation in terms of the long-range order parameter:

$$\frac{d\eta}{dt} \approx -\frac{\tilde{L}_0(\mathbf{k}_X)c_{Fe}(1 - c_{Fe})}{k_B T}\hat{\eta}(c_{Fe}, \eta(t)), \qquad (III.5)$$

where

$$\tilde{L}_0(\mathbf{k}_X) = -2\sum_{\mathbf{r}}'L_0(\mathbf{r})\sin^2(\mathbf{k}_X \cdot \mathbf{r}/2), \qquad (III.6)$$

and $\{-L_0(\mathbf{R} - \mathbf{R}')\}$ is a probability of an elementary jump between the sites $\mathbf{R}$ and $\mathbf{R}' = \mathbf{R} - \mathbf{r}$ at the time-unit interval for the 'exchange' diffusion mechanism in an ordered binary solution f.c.c.-Ni–Fe (without vacancies, when $\forall t\, P_{Ni}(\mathbf{R}, t) = 1 - P_{Fe}(\mathbf{R}, t)$). (The prime symbol in (III.6) means that the term $\mathbf{r} = 0$ is omitted.)

This equation (III.5) describes long-range ordering kinetics, if the temperature is below the absolute instability limit $T_c^-$ ($< T_K$) of the disordered phase, and there is a spinodal ordering, *i.e.* if the nucleation-and-growth mechanism (that corresponds to the 1-st kind of atomic disorder–order phase transformation) does not take place.

By using the SCFA [1, 2] for a calculation of a configurational free energy, $F_{conf} = U_{conf} - TS_{conf}$, one has the following:

$$U_{conf} \cong \frac{1}{2} \sum_{\mathbf{R}} \sum_{\mathbf{R}'} w_{tot}(\mathbf{R} - \mathbf{R}') P_{Fe}(\mathbf{R}) P_{Fe}(\mathbf{R}'), \ S_{conf} \cong -k_B \sum_{\mathbf{R}} \{ P_{Fe}(\mathbf{R}) \ln P_{Fe}(\mathbf{R}) + [1 - P_{Fe}(\mathbf{R})] \ln [1 - P_{Fe}(\mathbf{R})] \},$$

$$\frac{\delta F_{conf}}{\delta P_{Fe}(\mathbf{R})} \cong \sum_{\mathbf{R}'} w_{tot}(\mathbf{R} - \mathbf{R}') P_{Fe}(\mathbf{R}') + k_B T \ln \frac{P_{Fe}(\mathbf{R})}{1 - P_{Fe}(\mathbf{R})}. \tag{III.7}$$

For instance, as revealed for $L1_2$-type ordered f.c.c.-Ni–Fe alloy of non-stoichiometric composition [123], the substitution of $P_{Fe}(\mathbf{R}, t) = c_{Fe} + \eta(t)E(\mathbf{R})$ into the last expression results in

$$\frac{\delta F_{conf}}{\delta P_{Fe}(\mathbf{R})} \cong c_{Fe} \tilde{w}_{tot}(\mathbf{0}) + \frac{k_B T}{4} \ln \frac{\left( c_{Fe} + \dfrac{3\eta}{4} \right) \left( c_{Fe} - \dfrac{\eta}{4} \right)^3}{\left( 1 - c_{Fe} - \dfrac{3\eta}{4} \right) \left( 1 - c_{Fe} + \dfrac{\eta}{4} \right)^3} +$$

$$+ \left[ \eta \tilde{w}_{tot}(\mathbf{k}_X) + k_B T \ln \frac{\left( c_{Fe} + \dfrac{3\eta}{4} \right) \left( 1 - c_{Fe} + \dfrac{\eta}{4} \right)}{\left( 1 - c_{Fe} - \dfrac{3\eta}{4} \right) \left( c_{Fe} - \dfrac{\eta}{4} \right)} \right] E(\mathbf{R}). \tag{III.8}$$

Hence,

$$\hat{\eta}(c_{Fe}, \eta) \approx \eta \tilde{w}_{tot}(\mathbf{k}_X) + k_B T \ln \frac{\left( c_{Fe} + \dfrac{3\eta}{4} \right) \left( 1 - c_{Fe} + \dfrac{\eta}{4} \right)}{\left( 1 - c_{Fe} - \dfrac{3\eta}{4} \right) \left( c_{Fe} - \dfrac{\eta}{4} \right)}. \tag{III.9}$$

Substitution of this expression into the kinetic equation for $\eta(t)$ gives the non-linear differential relation:

$$\frac{d\eta}{dt} \approx -c_{Fe}(1 - c_{Fe}) \frac{\tilde{L}_0(\mathbf{k}_X)}{k_B T} \left[ \eta \tilde{w}_{tot}(\mathbf{k}_X) + k_B T \ln \frac{\left( c_{Fe} + \dfrac{3\eta}{4} \right) \left( 1 - c_{Fe} + \dfrac{\eta}{4} \right)}{\left( 1 - c_{Fe} - \dfrac{3\eta}{4} \right) \left( c_{Fe} - \dfrac{\eta}{4} \right)} \right]. \tag{III.10}$$

The solution of it may be given in another form through the implicitly-defined function:

$$\int_{\eta_0}^{\eta} \left[ \frac{\eta' \tilde{w}_{tot}(\mathbf{k}_X)}{k_B T} + \ln \frac{\left( c_{Fe} + \dfrac{3\eta'}{4} \right) \left( 1 - c_{Fe} + \dfrac{\eta'}{4} \right)}{\left( 1 - c_{Fe} - \dfrac{3\eta'}{4} \right) \left( c_{Fe} - \dfrac{\eta'}{4} \right)} \right]^{-1} d\eta' \approx -c_{Fe}(1 - c_{Fe}) \tilde{L}_0(\mathbf{k}_X) t, \tag{III.11}$$

where $\eta_0 = \eta(t = 0)$ is the magnitude of the initial long-range order parameter at

$$T < T_c^- \approx -c_{\text{Fe}}(1 - c_{\text{Fe}})\frac{\tilde{w}_{\text{tot}}(\mathbf{k}_X)}{k_B} \quad \left( = -0.1875\frac{\tilde{w}_{\text{tot}}(\mathbf{k}_X)}{k_B} < T_K \approx -0.205\frac{\tilde{w}_{\text{tot}}(\mathbf{k}_X)}{k_B} \text{ for } c_{\text{Fe}} = \frac{1}{4} \right) \quad \text{(III.12)}$$

(see above-mentioned Fig. III.2 and Refs [2, 101]).

It is necessary to emphasize that, in the general case, the intensities of the radiation coherent-scattering for superstructural reflections are proportional to the squared modulus of $\Delta\tilde{P}(\mathbf{k}, t)$:

$$I_L(\mathbf{k}, t) \propto |\Delta\tilde{P}(\mathbf{k}, t)|^2 \propto \{\eta(t)\}^2. \quad \text{(III.13)}$$

Within the scope of the linear approximation of Önsager equation (III.5) (or (III.10)) with right-hand expression (III.4) (or (III.8)) linearized with respect to $\eta$ in the small long-range order stage of relaxation, there is the exponential change (decay or growth) of amplitudes of the respective concentration waves. Relationship (III.13) provides the following explicit $\ln\{I_L(\mathbf{k}, t)/I_L(\mathbf{k}, 0)\}$–time dependence:

$$\ln\left(\frac{I_L(\mathbf{k}, t)}{I_L(\mathbf{k}, 0)}\right) \cong -2\tilde{L}_0(\mathbf{k})\left[c_{\text{Fe}}(1 - c_{\text{Fe}})\frac{\tilde{w}_{\text{tot}}(\mathbf{k})}{k_B T} + 1\right]t. \quad \text{(III.14)}$$

In principle, Eq. (III.14) enables to determine theoretically the diffusivities, which enter in the expression similar to (III.6) and/or $\tilde{w}_{\text{tot}}(\mathbf{k})$ value from the time dependences of x-ray or neutron diffraction intensities, if any measured at various temperatures as it was carried out, for example, in [85, 107].

Because the characteristic kinetic function $\tilde{L}_0(\mathbf{k})$ involves the probabilities of diffusion jumps, the method described above and based on measurements of time dependences of the reflection-point intensities of radiation (x-rays, neutrons *etc.*) at various temperatures, $T (< T_c^- < T_K)$, provides unique possibilities for the direct determination of kinetics parameters of elementary diffusion events. This can be done if the 'modulation' period of a periodic concentration profile is commensurate with the intersite distance. Accordingly, regarding the diffusivity $D^0$, one makes assumption that, for instance, elementary diffusion jumps of substitutional atoms are permitted only between the nearest-neighbour sites in f.c.c. lattice. Then one obtains

$$\tilde{L}_0(\mathbf{k}) \cong 2(D^0/a^2)\sum_{\mathbf{r}_l}\sin^2(\mathbf{k}\cdot\mathbf{r}_l/2) \quad \text{(III.15)}$$

where the summation is performed over twelve $\{\mathbf{r}_l\}$ sites of the 1-st co-ordination shell in the f.c.c. crystal lattice with parameter $a$. If one considers a one-dimensional 'modulation' along the [1 0 0] direction, $\mathbf{k} = (k_x\, 0\, 0)$, the summation over $\{\mathbf{r}_l\}$ yields

$$\tilde{L}_0(\mathbf{k}) \cong 8(D^0/a^2)\{1 - \cos(k_x d)\} \quad \text{(III.16)}$$

where $d$ is the (100) interplanar distance [1].

The scheme of computation, introduced in this subsection, can hereinafter bring to the fruitful applications for investigation of the f.c.c.-Ni–Fe alloy.

Now let us consider the case of 'exchange' mechanism governing the diffusion in Ni$_3$Fe alloy (which is similar by the composition to disordered f.c.c.-$^{62}$Ni$_{0.765}$Fe$_{0.235}$ solution) at temperatures *below* the temperature of order($L1_2$)–disorder phase transformation. Atomic distribution function in this ordering alloy can be represented as a superposition of the static concentration waves (III.3) [1]. Using concentration-waves' approach along with SCFA, one can also apply the differential kinetic equation (III.10) for relaxation of long-range order parameter $\eta$ [1] of $L1_2$-type Ni$_3$Fe; as mentioned above, in this equation, $\mathbf{k}_X = 2\pi(1\, 0\, 0)$—wave vector (in a first Brillouin zone of f.c.c. lattice), which generates the $L1_2$-type superstructure, $\tilde{w}_{\text{tot}}(\mathbf{k}_X)$—Fourier component of pairwise-interaction 'mixing' energy for Ni and Fe atoms, $\tilde{L}_0(\mathbf{k}_X)$—Fourier transform of Önsager kinetic coefficients; algebraic magnitudes of $\widetilde{w}_{\text{tot}}(\mathbf{k}_X)$ for Ni–Fe alloy at different temperatures were estimated earlier (see Table II.3 and Eq. (II.8)). An implicit solution $t = t(\eta)$ of Eq. (III.10) is given by the Eq. (III.11).

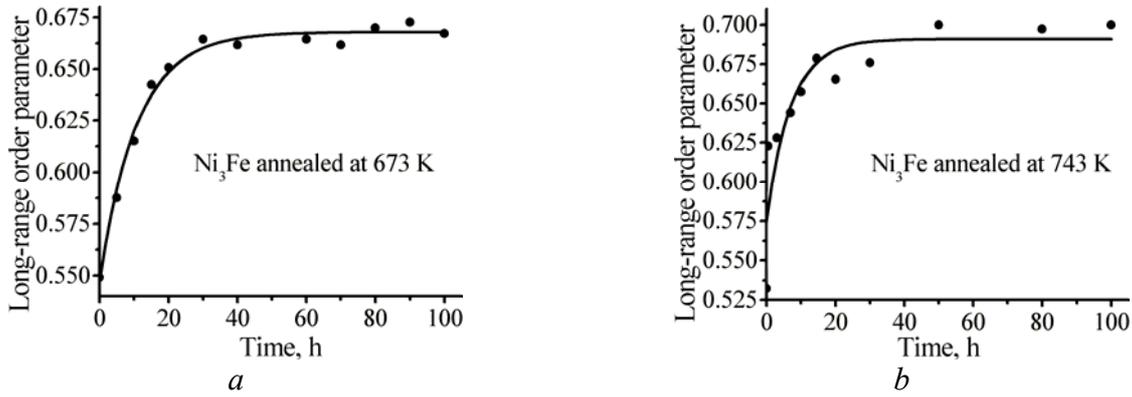

Fig. III.3. Time dependence of long-range order parameter for $L1_2$-type Ni$_3$Fe alloy; ●—experimental data [107] for different temperatures: 673 K (a), 743 K (b).

Assuming the atomic jumps between nearest-neighbour sites only and taking into consideration the condition that the total number of atoms in a system is conserved, for f.c.c. lattice, we can write Eq. (III.6) as follows:

$$\widetilde{L}_0(\mathbf{k}) \approx -4L_0(r_1)[3 - \cos(\pi h_x)\cos(\pi h_y) - \cos(\pi h_y)\cos(\pi h_z) - \cos(\pi h_z)\cos(\pi h_x)]$$

for $\mathbf{k} = 2\pi(h_x\,h_y\,h_z)$, where $\{-L_0(r_1)\}$ is proportional to the circular-jump probability of a pair of atoms at nearest-neighbour sites $\mathbf{R}$ and $\mathbf{R}'$ ($r_1 = |\mathbf{R} - \mathbf{R}'|$) per unit time. Using experimental data from Ref. [107] (see Fig. III.3), Eq. (III.11), and the last-named expression, we obtain roughly averaged values: $\langle -L_0(r_1)\rangle \approx 4 \cdot 10^{-7}$ s$^{-1}$ for $T = 673$ K and $\langle -L_0(r_1)\rangle \approx 6 \cdot 10^{-6}$ s$^{-1}$ for $T = 743$ K [123]. Substituting $L_0(r_1)$ (and 0 instead of $L_0(r_n)$ with $n \geq$ II) into expression similar to

$$D^0 \approx -(1/6)\sum_{n=1}^{\infty} L_0(r_n)r_n^2 Z_n$$

($Z_n$—co-ordination number for $n$-th co-ordination shell) [104], 'exchange' diffusion mobilities of Fe and Ni atoms in their pairs within the $L1_2$-type Ni$_3$Fe alloy may be roughly estimated by averaging experimental data over annealing times: $\langle D^0\rangle \approx 1.1 \cdot 10^{-21}$ cm$^2$/s for $T = 673$ K and $\langle D^0\rangle \approx 1.5 \cdot 10^{-20}$ cm$^2$/s for $T = 743$ K [123]. (It is important that diffusion migration of the 'slowest' particles is that process, which finally forms an equilibrium state of ordered system.)

According to the Arrhenius formula, estimated diffusion-migration activation energy for Fe (and Ni) atoms is approximately 1.6 eV.

For disordered f.c.c.-$^{62}$Ni$_{0.765}$Fe$_{0.235}$-type solution, probabilities of 'random-walk' jumps of both Fe and Ni atoms into a given site of the nearest-neighbour shell around a 'zero'-site heterogeneity or into the next-nearest-neighbour sites were simulated in [104, 123]. Obtained microparameters of diffusion were estimated on the basis of available experimental data [85] concerning neutron diffuse-scattering evolution kinetics caused by the short-range order relaxation of the solid solution. Respective 'microdiffusivities' determine the short-range order change and give information about its kinetics. This stands for the predominance of atomic jumps within the 1-st co-ordination shell, which are mainly governed by the vacancy-controlled mechanism of diffusion within the commonly accepted interpretation.

Table III.2. Vacancy-controlled diffusion ($D_{\text{Fe}}$), self-diffusion ($D_{\text{Fe}}^*$), and 'interdiffusion' ($D_0$) coefficients for disordered f.c.c.-$^{62}$Ni$_{0.765}$Fe$_{0.235}$ solution.

| $T$ [K] | $D_{\text{Fe}}$ [cm$^2$/s] [104, 123] | $D_{\text{Fe}}^*$ [cm$^2$/s] [104, 123] | $D_0$ [cm$^2$/sec] [85] |
|---|---|---|---|
| 776 | $4.5 \cdot 10^{-17}$ | $1.8 \cdot 10^{-17}$ | $2.5 \cdot 10^{-18}$ |
| 783 | $6.9 \cdot 10^{-17}$ | $2.6 \cdot 10^{-17}$ | $3.6 \cdot 10^{-18}$ |

Such calculated diffusion and self-diffusion coefficients for Fe atoms in f.c.c.-$^{62}$Ni$_{0.765}$Fe$_{0.235}$ solution are listed in Table III.2. In the last column, 'interdiffusion' coefficients extrapolated in [85] from the high-temperature measurements are also presented.

The total activation energies of vacancy-mediated diffusion and self-diffusion of 'slow' Fe atoms are estimated approximately as 3.2 and 2.6 eV, respectively [123]. These energies for 'slow' (Fe) atoms in disordered $^{62}$Ni$_{0.765}$Fe$_{0.235}$ alloy are many higher than the above-mentioned estimated 'exchange'-diffusion migration energy of corresponding atoms in pairs in long-range ordered Ni$_3$Fe (1.6 eV) because the latter does not involve the energy of vacancy formation (migration energy in ordered $L1_2$-Ni$_3$Fe is evaluated within the alloy model without vacancies). Therefore, vacancy formation energy in a former model is 38–50% [123].

'Exchange' diffusion mobilities of atoms in pairs within the $L1_2$-type ordered Ni$_3$Fe alloy (for different temperatures) are lower than diffusivities in Table III.2 because of some reasons. Firstly, because of temperature- and concentration-dependent statistical-thermodynamic factors of correlated diffusion of the interacting atoms. (For instance, a mixing energy is dependent on both $T$ and $c_{Fe}$.) Secondly, below the order–disorder transformation temperature, the diffusion mechanism in long-range ordered alloys may be modified, and this will affect the magnitude of diffusion coefficient in the direction of observed variation; in fact, the probability of 'exchange' ('ring') mechanism of diffusion is small. (It is proved by the magnitude of Önsager kinetics coefficient.) Thirdly, because of the long-range ordering phase is formed at more low temperatures as compared with a disordered phase.

## 5. Estimations of Temperatures of the Structural and Magnetic Transformations in Alloys

### 5.1. In Addition to the Ni–Fe Phase Diagrams.
As stated above, within the framework of the SCFA, the statistical thermodynamics of the long-range atomic order of f.c.c.-Ni–Fe alloy is completely determined by two energy parameters—$\tilde{w}_{tot}(\mathbf{k}_X)$ and $\tilde{w}_{tot}(\mathbf{0})$, where $\mathbf{k}_X = 2\pi(1\,0\,0)$ and $\mathbf{k_0} = \mathbf{0}$ are the wave vectors corresponding to the high-symmetry reciprocal-space points $X$ and $\Gamma$ of this f.c.c. alloy, respectively; $(1\,0\,0)$ is the half-translation of reciprocal b.c.c. lattice (of f.c.c. crystal) lying along the $[1\,0\,0]$ direction.

So, one can obtain the configurational free energy $f$ of f.c.c.-Ni–Fe alloy *per* site (see (II.2)), and then determine fundamental thermodynamical characteristics of equilibrium state, including the (reduced) concentration-dependent order($L1_2$)–disorder($A1$) transformation (Kurnakov's) temperature $\tau_K = k_B T_K / |\tilde{w}_{tot}(\mathbf{k}_X)|$ (see Fig. III.2). Besides, within the SCFA, the curves $\tau_K(c_{Fe})$ and $\Delta\eta(T_K, c_{Fe})$ for the 1-st kind order($L1_0$)–disorder($A1$) transformation (in the vicinity of the critical phase transition) have more trivial approximate forms—$\tau_K(c_{Fe}) \cong c_{Fe}(1 - c_{Fe})$ and $\Delta\eta(T_K, c_{Fe}) \cong 0$ [1, 2, 101].

The apparent asymmetry of the curves for *absolute* Kurnakov's temperature $T_K = T_K(c_{Fe})$ (as well as Curie temperature $T_C = T_C(c_{Fe})$; see Figs. II.2 and III.4) is introduced by magnetic and correlation effects, particularly asymmetrical magnetic contribution of both Ni and Fe atoms with different spins to the total 'interchange' energies $\{\tilde{w}_{tot}(\mathbf{k})\}$, which is dependent on temperature and concentration (see (II.6), (II.8), and Fig. III.5):

$$T_K(c_{Fe}) = \{\tilde{w}_{prm}(\mathbf{k}) + \tilde{J}_{FeFe}(\mathbf{k})s_{Fe}^2\sigma_{Fe}^2 + \tilde{J}_{NiNi}(\mathbf{k})s_{Ni}^2\sigma_{Ni}^2 - 2\tilde{J}_{NiFe}(\mathbf{k})s_{Ni}s_{Fe}\sigma_{Ni}\sigma_{Fe}\}\tau_K(c_{Fe})\,. \tag{III.17}$$

### 5.2. Detailing of Binary Sections of the F.C.C.-Ni–Fe–C Phase Diagram.
The influence of C atoms on Kurnakov's temperature, $T_K$, of the 1-st kind of atomic ordering in substitutional f.c.c.-Ni–Fe alloy for superstructural $L1_2$ or $L1_0$ types (which are generated with a star of a wave vector $\mathbf{k}_X = 2\pi(1\,0\,0)$ on the basis of reciprocal lattice of f.c.c. crystal) may be analytically studied (allowing for long-range interatomic interactions).

In particular, in linear on $c_C$ approximation (if $c_C \ll 1$), the temperature of structural disorder–order phase transformation [3],

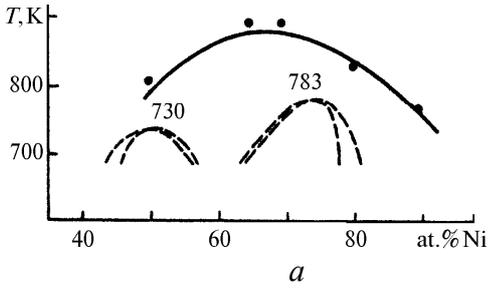

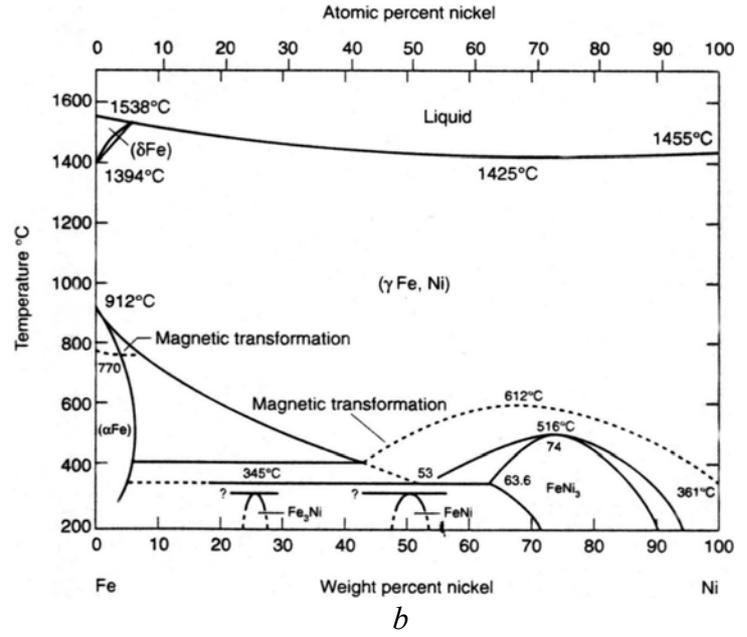

Fig. III.4. (*a*) Schematic plots of Curie (solid line) and Kurnakov's (dashed line) temperatures dependent on Ni content, which are numerically estimated (with cluster variation method) for f.c.c.-Ni–Fe alloys; black circles—experimental points (see Refs in [60, 116]). (*b*) Ni–Fe alloy phase diagram which incorporates information obtained in the studies of slowly cooled Fe–Ni meteorites and neutron- and electron-irradiated samples to produce the low-temperature NiFe ($L1_0$) and NiFe$_3$ ($L1_2$) ordered phases unattainable within the laboratory experiments (see also Refs in [71, 84, 86, 98, 111, 120–122]).

$$T_K(c_{Fe}, c_C) \cong -\frac{c_{Fe}(1-c_{Fe})}{k_B}\tilde{w}_{tot}(\mathbf{k}_X)(1+\tau_{(100)}) - \frac{(1+3\tau_{(100)})}{k_B}\frac{\omega^2_{FeNiC}(\mathbf{k}_X)}{\tilde{w}_{tot}(\mathbf{k}_X)}c_C, \qquad (III.18)$$

within the certain Fe concentration range ($c_{Fe} \in (0, 1)$), can be increased by addition of C. Here (in the designations introduced above), the Fourier component of mixing energy is as follows:

$$\tilde{w}_{tot}(\mathbf{k}_X) = \tilde{w}_{prm}(\mathbf{k}_X) + \tilde{J}_{FeFe}(\mathbf{k}_X)s^2_{Fe}\sigma^2_{Fe} + \tilde{J}_{NiNi}(\mathbf{k}_X)s^2_{Ni}\sigma^2_{Ni} - 2\tilde{J}_{NiFe}(\mathbf{k}_X)s_{Ni}s_{Fe}\sigma_{Ni}\sigma_{Fe}, \qquad (III.19)$$

$$\omega_{FeNiC}(\mathbf{k}_X) \approx \tilde{W}^{CFe}(\mathbf{k}_X) - \tilde{W}^{CNi}(\mathbf{k}_X),$$

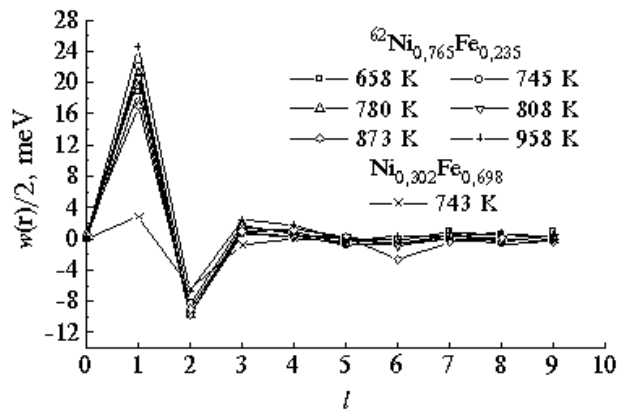

Fig. III.5. Variation of 'mixing' energy $w_{tot}(\mathbf{r})$ with a number of co-ordination shell $l$ for different annealing temperatures in f.c.c.-Ni–Fe alloys (within the postulated calibration equation $w_{tot}(\mathbf{r})|_{\mathbf{r}=0} = 0$, and without explicit consideration of the strain-induced effects) [100].

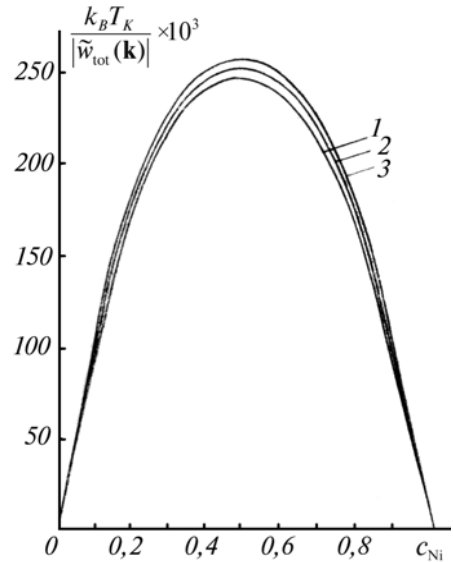

Fig. III.6. Kurnakov's point dependence on $c_{Ni}$ for substitutional f.c.c.-Ni–Fe subsystem with different values of $c_C$ and for $\mathbf{k} = \mathbf{k}_X$ where $|\omega_{FeNiC}(\mathbf{k}_X)/\tilde{w}_{tot}(\mathbf{k}_X)| \approx 0.8$ ($1$—$c_C = 0$, $2$—$c_C = 0.05$, $3$—$c_C = 0.1$).

and

$$\tau_{(100)} \approx \begin{cases} \left\{ \dfrac{21\left[1 - 3c_{Fe}\left(1 - c_{Fe}\right)\right]}{2\left(1 - 2c_{Fe}\right)^2} - 2 \right\}^{-1} & (L1_2), \\[2mm] 0 & (L1_0). \end{cases} \qquad (III.20)$$

For $c_{Fe} \in (0.686, 0.785)$, $\tilde{w}_{tot}(\mathbf{k}_X) \cong -0.282$ eV, $\omega_{FeNiC}(\mathbf{k}_X) \cong -0.240$ eV. In this way, the curves of the Kurnakov's order–disorder phase transformations over the binary sections of the phase diagram of f.c.c.-Ni–Fe–C alloys (see Fig. III.6) are prognosticated as well as the dependence of the long-range order parameter of substitutional f.c.c.-Ni–Fe subsystem on the temperature, $T$, in the vicinity of $T_K$ at different $c_C$ values (see Fig. III.7).

On the other hand, by using the known energies of exchange interactions, the temperatures of magnetic transitions in a weak interstitial–substitutional f.c.c.-Ni–Fe–(0–0.0007 C) solutions in a disordered state with the zero parameter of the long-range order may be calculated, and the diagram of magnetic states of substitutional Ni–Fe-subsystem doped with interstitial C may be plotted numerically

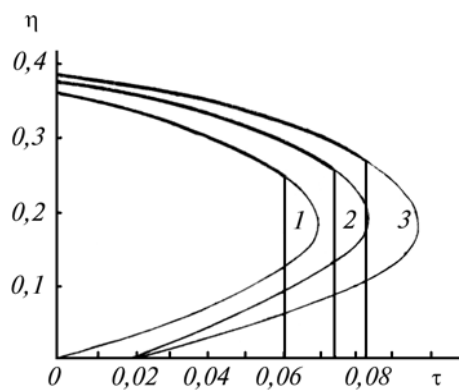

Fig. III.7. Dependence of the long-range order parameter, $\eta$, of substitutional f.c.c.-Ni–Fe subsystem on the temperature deviation, $\tau \equiv (T - T_c^-)/T_c^-$ (at different values of $c_C$: $1$—$c_C = 0$, $2$—$c_C = 0.03$, $3$—$c_C = 0.05$); $|\omega_{FeNiC}(\mathbf{k}_X)/\tilde{w}_{tot}(\mathbf{k}_X)| \approx 0.8$. The thick line marks the order-parameter values within the range of thermodynamically-equilibrium ordered alloy ($T_c^- = -c_{Fe}(1 - c_{Fe})\tilde{w}_{tot}(\mathbf{k}_X)/k_B \leq T \leq T_K$).

(for this concentration range) by way of illustration, as provided by the following forecasting formula:

$$T_C(c_{Fe}, c_C) \cong -\frac{[(-1893.11 + 385395 c_C) c_{Ni} + (3144.01 + 11661918 c_C) c_{Fe}]}{6} +$$

$$+ \frac{1}{6}\sqrt{[(-1893.11 + 385395 c_C) c_{Ni} - (3144.01 + 11661918 c_C) c_{Fe}]^2 + 4(-4580.33 - 2164050 c_C)^2 c_{Fe} c_{Ni}}$$

(III.21)

where $T_C$ is measured in [K], and $c_{Ni} = 1 - c_{Fe}$ [113].

## Conclusions

The results of performed numerical calculations of strain-induced ('elastic') pairwise interaction energies of interstitial–interstitial, interstitial–substitutional, and substitutional–substitutional atoms in α-Ni and γ-Fe were presented. The microscopic theory applied in this paper enables to take into account the atomic structure of solid solutions. The elasticity moduli, lattice spacings, and Born–von Kármán parameters of the host lattice are the input parameters used for the computations of the numerical coefficients $\{A^{\alpha\beta}(\mathbf{R} - \mathbf{R}')\}$, which are material parameters of α-Ni and γ-Fe. These coefficients enable to calculate directly the strain-induced pairwise interaction energies between any dissolved atoms within many co-ordination shells provided that the coefficients of the concentration expansion of the host lattice (related to the relevant substitutional and interstitial atoms) are known.

The strain-induced interaction in some α-Ni- and γ-Fe-base solid solutions is generally strong and must necessarily be taken into account for the analysis of structure and properties of solid solutions.

The calculation of the C activity and C distribution parameters yielded with the use of Mössbauer spectroscopy shows that the strain-induced interaction model is applicable, for instance, to Fe–C austenite, but it must be supplemented with additional 'electrochemical' C–C repulsion at least in the first co-ordination shell.

In other words, the applied procedure of Monte Carlo computer simulation of the constitution of nanoscale Fe–C-austenitic crystallite is based on the analysis of both the dependences of relative fractions of any type of atomic configurations on interatomic C–C interaction energies and the revealed correlation between the potential energy of such modelling system and numbers of iterations (as well as Monte Carlo steps) in the approach to constrained equilibrium.

As shown, for adequate reproduction of experimental thermodynamic C-activity data in austenite, one needs to consider a simulating crystallite of size not less than 52.5×52.5×52.5 in units of parameter of f.c.c. conventional unit cell. For the relaxation, it is enough to pass no more than 10 (for instance, 5) Monte Carlo steps.

In computations, the 'electrochemical' (direct) and strain-induced (indirect) contributions to C–C interaction are taken into account. The sets of energies of this (total) interaction, $\{W_n^{CC}\}$, within the first six interstitial co-ordination shells ($n = 1, ..., 6$) with rated radii up to 6.35 Å are estimated, and optimal set ($W_1^{CC} \approx 0.078$ eV, $W_2^{CC} \approx 0.144$ eV, $W_3^{CC} \approx -0.039$ eV, $W_4^{CC} \approx 0.003$ eV, $W_5^{CC} \approx 0.0025$ eV, $W_6^{CC} \approx 0.0197$ eV) is selected. The last set optimally corresponds to experimental concentration and temperature dependences of C activity and Mössbauer-spectroscopy data on the nearest neighbourhood of Fe atoms with C octahedral interstitials. The abundances of the different relative positional atomic Fe–C and C–C clusters (depending on C–C interaction energies) are determined.

The increase of C content in f.c.c.-Ni–Fe–C alloys results in deviation of the dependence of the relative saturation magnetization on reduced temperature, by means of decrease its convex camber, with respect to the well-known Brillouin function for pure Ni. Similar effect is redoubled for compositions that approach to the standard Invar one. This effect is accountable for the increase of the number of atoms with spin 3/2 ($s_{Fe}$) at the expense of carriers with smaller spin 1/2 ($s_{Ni}$) and supported with the calculation results, which use the obtained exchange interaction parameters.

The exchange interaction parameters and the Fourier components of exchange interaction energies of atomic 'spins' for disordered f.c.c.-Ni–Fe alloy were fitted, and the different signs for nearest neighbours were revealed that points to presence of Fe–Fe antiferromagnetic and both Ni–Fe and Ni–Ni ferromagnetic interactions in the involved alloys.

The optimal set is as follows (depending on the spin states [50, 58, 98, 99] of constituents): $J_{NiNi}(r_1) \approx -0.018$ eV, if Ni-atom spin is $s_{Ni} = 1/2$, and $J_{FeFe}(r_1) \approx +0.006$ eV, $J_{NiFe}(r_1) \approx -0.020$ eV for Fe spin $s_{Fe} = 3/2$ (or $J_{FeFe}(r_1) \approx +0.030$ eV, $J_{NiFe}(r_1) \approx -0.044$ eV for Fe spin $s_{Fe} = 1/2$) for binary f.c.c.-Ni–Fe alloy. As shown for the first time, a small C amount addition to the f.c.c.-Ni–Fe alloy increases the antiferromagnetic Fe–Fe component as well as ferromagnetic Ni–Fe one and decreases the ferromagnetic Ni–Ni component of exchange interactions.

Increasing C content in the f.c.c.-Ni–Fe–C alloys results in elevation of the (reduced) concentration-dependent 1-st kind order ($L1_2$ or $L1_0$)–disorder($A1$) transformation temperature. Effect of C alloying on the Curie temperature is crucially dependent on both C and Fe concentrations.

The Khachaturyan–Cook microscopic approach at issue relates the time dependences of the long-range or short-range orders to atomic diffusion. It enables to use the data of measurements of time dependence of radiation diffraction or diffuse-scattering intensity for a crystal of binary substitutional solid solution Ni–Fe for calculation of both probabilities of elementary atomic-diffusion jumps to different lattice sites per unit time and 'exchange' or vacancy-controlled diffusion coefficients, respectively.

For instance, for disordered f.c.c.-Ni–Fe alloy, total activation energies of vacancy-controlled diffusion and self-diffusion of 'slow' (Fe) atoms are approximately 3.2 and 2.6 eV, respectively, and the 'exchange'-diffusion migration energy of these atoms in atomic pairs in long-range ordered $Ni_3Fe$ Permalloy is estimated as 1.6 eV.

## Acknowledgements

We are grateful to Prof. D.G. Rancourt from University of Ottawa (Canada), Dr. G.E. Ice from Oak Ridge National Laboratory (U.S.A.), Dr. F. Bley from Laboratoire de Thermodynamique et Physico-Chimie Métallurgiques, ENSEEG (Saint Martin D'Heres, France), Prof. J. Desimoni from Facultad de Ciencias Exactas, UNLP, IFLP-CONICET (La Plata, Argentina), Prof. V.E. Antonov from Institute of Solid State Physics, R.A.S. (Chernogolovka, Russia), Dr. A.E. Krasovskii from Institute of Magnetism, N.A.S.&M.E.S. of the Ukraine (Kyyiv), and especially to Prof. M.S. Blanter from the Moscow State Academy of Instrumental Engineering and Information Science (Russia) as well as Mr. D. Pavlyuchkov from Forschungszentrum Juelich GmbH (Germany) for communicating their applicable results and/or important references. The support by Dr. S.M. Bugaychuk from G.V. Kurdyumov Institute for Metal Physics, N.A.S. of the Ukraine (Kyyiv) in correcting the original English translation of a given article is gratefully acknowledged. The grant #28/08-H (06) from the N.A.S. of the Ukraine is gratefully acknowledged for the support in part. We are grateful to Dr. H.M. Zapolsky and Prof. D. Blavette from Group of Material Physics, UMR CNRS 6634, University of Rouen (St. Etienne du Rouvray, France) for discussion of the results obtained within the framework of the given work. One of the authors (S.M.B.) acknowledges the financial assistance within the frameworks of the collaboration project under financial support of Group of Material Physics by the Ministere de l'Enseignement Supérior et de la Recherche, France, and else one of the authors (Y.B.P.) acknowledges the financial assistance he has received from the Centre for Advanced Materials Processing (CAMP) of the 21st Century Frontier R&D Program funded by the Ministry of Science and Technology, Korea.

**Defects and Diffusion in Ceramics X**

**Semi-Empirical Parameterization of Interatomic Interactions and Kinetics of the Atomic Ordering in Ni-Fe-C Permalloys and Elinvars**